\documentclass[11pt]{article}

\usepackage{mathtools}
\usepackage{tikz}
\usetikzlibrary{decorations.markings}

\usepackage{setspace}
\usepackage{lipsum}
\usepackage{multirow}
\usepackage{fullpage}
\usepackage{amsmath}
\usepackage{amssymb}
\usepackage{setspace}
\usepackage{bbm}
\usepackage{dsfont}
\usepackage{graphics}
\usepackage{longtable}
\usepackage[font=footnotesize,labelfont=bf,justification=centerlast,width=.9\textwidth]{caption}
\usepackage{color}
\usepackage{etoolbox}

\usepackage{array}
\usepackage{hyperref}

\usepackage{caption}
\hypersetup{
	bookmarks=true,%
	colorlinks,%
	citecolor=blue,%
	filecolor=black,%
	linkcolor=black,%
	urlcolor=blue
}

\def \F {\mathcal{F}}
\def \N {\mathcal{N}}

\def \CFT {\text{CFT}}
\def \D {\text{D}}
\def \M {\text{M}}
\def \aD {\overline{\D}}
\def \aM {\overline{\M}}
\begin{document}
	
\numberwithin{equation}{section}

\begin{titlepage}
\begin{center}
			
			\hfill \\
			\hfill \\
			\vskip 0.45in
			
			{\Large \bf  Quantum Black Hole Entropy, Localization\\ and the Stringy Exclusion Principle }\\
			
			\vskip 0.4in
			
			{\large  Jo\~{a}o Gomes}\\
			\vskip 0.3in

			{\it Institute of Physics, University of Amsterdam,
				Science Park 904, Postbus 94485, 1090 GL Amsterdam, The Netherlands} \vskip .5mm
			{\it Institute for Theoretical Physics, University of Utrecht, Princetonplein  3584 CC Utrecht, The Netherlands} \vskip .4mm
				\texttt{j.m.vieiragomes@uva.nl}	
			
\end{center}
		
		\vskip 0.35in
		
		\begin{center} {\bf ABSTRACT } \end{center} Supersymmetric localization  has lead to remarkable progress in computing quantum corrections to BPS black hole entropy. The program has been successful especially for computing perturbative corrections to the Bekenstein-Hawking area formula.  In this work, we consider non-perturbative corrections related to polar states in the Rademacher expansion, which describes the entropy in the microcanonical ensemble. We propose that these non-perturbative effects can be identified  with a new family of saddles in the localization of the quantum entropy path integral. We argue that these saddles, which are euclidean $AdS_2\times S^1\times S^2$ geometries, arise after turning on singular fluxes in M-theory on a Calabi-Yau. They cease to exist after a certain amount of flux, resulting in a finite number of geometries; the bound on that number is in precise agreement with the stringy exclusion principle. Localization of supergravity on these backgrounds gives rise to a finite tail of Bessel functions in agreement with the Rademacher expansion. As a check of our proposal, we test our results against well known microscopic formulas for one-eighth and one-quarter BPS black holes in $\mathcal{N}=8$ and $\mathcal{N}=4$ string theory respectively, finding agreement. Our method breaks down precisely when mock-modular effects are expected in the entropy of one-quarter BPS dyons and we comment upon this. Furthermore, we mention possible applications of these results, including an exact formula for the entropy of four dimensional $\mathcal{N}=2$ black holes.
		
		\vfill
		
		\noindent \today

	\end{titlepage}
	
	\newpage
	
	\tableofcontents

\section{\label{sec1}Introduction}

Supersymmetric localization \cite{Banerjee:2009af,Dabholkar:2010uh} has lead to the possibility of evaluating exactly the $\text{AdS}_2$ path integral that computes the quantum entropy \cite{Sen:2008vm} of BPS black holes. This technique has been particularly successful for computing perturbative quantum corrections to the Bekenstein-Hawking entropy in toroidal compactifications , where an almost exact matching with the microscopic theory was obtained \cite{Dabholkar:2011ec}. 

The goal of this work is to address instead non-perturbative corrections to black hole entropy related to polar states in the microscopic theory. We want to understand the origin of these effects, perhaps as new saddle points of the path integral, and if so how to compute quantum corrections around each saddle. In  toroidal compactifications, such non-perturbative effects are not present, which in a way is what explains the simplicity of the microscopic formulas. Nevertheless, for $\N=4$ and $\N=2$ compactifications, these non-perturbative effects are crucial to understand black hole entropy in the limit of very large central charge, which is where the four dimensional semiclassical description holds. Though exponentially subleading, these non-perturbative effects can become relevant  when the number of polar states grows exponentially, which is the case of large central charge.

Recent attempts to compute exactly the quantum entropy, rely on the four dimensional effective action, which includes instanton effects in the prepotential of supergravity.  In order to address the problem of non-perturbative corrections within the context of supersymmetric localization, we need to first understand the UV dynamics that are responsible for those effects. With this in mind, the following question arises: can we rely on effective field theory such as supergravity or do we need the full string theory? 

Another issue is of concern. Since localization is an off-shell computation and as such does not depend on the values of the coupling constants, it is valid at strong and weak coupling. Translating to supersymmetric black holes and their quantum entropy, this means that the localization computation should hold for any value of charges. The reason is that, in string theory, the scalar fields, which  play the role of the coupling constants, become functions solely of the charges at the black hole attractor. However, in view of the AdS/CFT correspondence and the fact that we are working in the microcanonical ensemble, this raises many issues related to the validity of effective field theory.  For example, the characteristic length scale of the geometry is a function of the charges, and so by scaling these it is possible that a particular dimensional point of view is more appropriate than other. Conversely, we may ask which microscopic Lagrangian are we localizing?

To better understand  these issues, we take a pedestrian approach. We start by recalling the original localization computation of  \cite{Dabholkar:2010uh} in four dimensional supergravity and discuss its validity using effective field theory. Then in section \S\ref{sec 1.1} we consider the five dimensional point of view. We argue this is more appropriate to describe the physics of the Rademacher expansion. In section \S\ref{sec 1.2}, we discuss the connection between non-perturbative effects in the black hole entropy and the counting of polar states. Along the way, we present the main lines of our solution, which, in essence, is a reformulation of the OSV formula. We require this formula to be compatible with the Rademacher expansion and to reproduce at the same time the counting of polar states.

\subsection{The four dimensional point of view and the OSV formula}

 In \cite{Dabholkar:2010uh},  it is shown that the path integral of $\mathcal{N}=2$ supergravity \footnote{The Lagrangian is based on the off-shell superconformal formalism and it is thus related to the Wilsonian rather than the 1PI effective action of string theory.}on $AdS_2\times S^2$, which computes the entropy of a BPS black hole, reduces to a finite dimensional integral by means of supersymmetric localization.  The answer for the black hole degeneracy $d(q,p)$ is  schematically of the form
\begin{equation}\label{OSV}
d(q,p)\sim \int d\phi\, e^{-\pi q\phi+4\pi\text{Im}F(\phi+ip)},
\end{equation}
where $F(X)$ is the four dimensional holomorphic prepotential that encodes the couplings of the vectors to the Weyl multiplet, and $q,p$ are respectively the electric and magnetic charges; the integration variables $\phi$ correspond to normalizable modes that are left unfixed by localization. 

The result (\ref{OSV}) is a reincarnation of the conjectured OSV formula \cite{Ooguri:2004zv}, which  relates the black hole quantum entropy to the topological string partition function. In the original formulation \cite{Ooguri:2004zv}, the reason this happens is because the supergravity prepotential $F(X)$ in (\ref{OSV}) is directly related to the perturbative free energy $F(g_{top},t)$ of the topological string on the  Calabi-Yau compactification manifold. To be more precise, the topological string computes four dimensional higher derivative terms, also called F-type terms, of the form $F_{g}(t)R_{-}^2T_{-}^{2g-2}+\text{h.c.}$ ($g>0$) with $R_{-}$ being the anti-self-dual curvature two-form and $T$ the graviphoton field \cite{Gopakumar:1998ii,Gopakumar:1998jq}.
 The $F_g(t)$ \footnote{In the present discussion, we keep the free energies $F_g(t)$ holomorphic, which is appropriate for the Wilsonian point of view of the entropy function and hence also of the localization computation. We will comment on possible non-holomorphic corrections later in section \S \ref{sec effective action}.  } are defined in a perturbative expansion in powers of the topological string coupling constant $g_{\text{top}}$, that is,
\begin{equation}\label{top free energy}
F(g_{top},t)=\sum_{g=0}g_{\text{top}}^{2g-2}F_{g}(t).
\end{equation}
For $t\gg 1$, the tree level $g=0$ term can be approximated by $F_0(t)\simeq D_{abc}t^at^bt^c/g_{top}^2$
with $D_{abc}$ the intersection matrix of the Calabi-Yau threefold, while the function $F_1(t)$ is the one-loop correction which approximates to $F_1(t)\simeq \frac{c_{2a}t^a}{24}+\mathcal{O}(e^{-t})$. The corrections of order $e^{-t}$ are due to worldsheet instantons, while the parameter $c_{2a}$ can be identified with the second Chern-class $c_2(X)$ (tangent bundle) of the Calabi-Yau.  The map between the topological string variables and the supergravity fields is the following: the complexified K\"{a}hler parameter  $t^a=X^a/X^0$ and $g_{top}=1/X^0$, where $X^0$ is the dilaton that sits in the supergravity multiplet and $X^a$ are the vectormultiplet complex scalar fields.  \footnote{We are omitting minor details regarding the map between the complex scalar fields $X^a$ and the topological string variables. These details will become clear later on.}.

For a general Calabi-Yau, the function (\ref{top free energy}) is known only as an asymptotic expansion in $g_{top}$. If one tries to use localization at the level of the four dimensional effective Lagrangian as in \cite{Dabholkar:2010uh} we run into serious problems; not only we have to constrain the scalar $X^0\sim 1/g_{top}$ to very large values, but we also need to include an arbitrary number of higher derivative corrections. The best we can do is to compute order by order in the inverse of the charges using perturbative methods. 

Nonetheless, in  compactifications that preserve more supersymmetry, like in  $\mathcal{N}=8$ and $\mathcal{N}=4$,  the prepotential (\ref{top free energy}) is one-loop exact, that is, all $F_{g>1}(t)$ vanish. In this case, the tree level free energy is given exactly by $F_0(t)=D_{abc}t^at^bt^c$ and the one-loop contribution $F_1(t)=\ln g(t)$, with $g(t)$ being a worldsheet instanton partition function. Since the prepotential is now one-loop exact one might expect to be able to use supersymmetric localization at the level of the four dimensional effective action.  As a matter of fact, previous work shows that in the $\N=8$ theory it is possible to reproduce exactly all the perturbative corrections to the area formula \cite{Dabholkar:2011ec} including non-perturbative corrections, related to orbifold geometries, that depend on intricate Kloosterman sums \cite{Dabholkar:2014ema}. In the $\N=4$ case, however, the situation is not so satisfactory because the localization integral (\ref{OSV}) is not able to reproduce all the features predicted by the microscopic theory. In particular, it fails to reproduce the measure that is known from microscopics even after taking into account the one-loop determinants \cite{Murthy:2015yfa,Gupta:2015gga}. In a way, this is partially justified from the microscopic studies \cite{Shih:2005he,Gomes:2015xcf,Murthy:2015zzy}. These studies predict a measure $\mathcal{M}(\phi,p)$ to (\ref{OSV}) that depends strongly on the worldsheet instantons. The precise dependence is of the form
\begin{equation}\label{worldsheet corrections}
\mathcal{M}(\phi,p)\sim \pi p^2-\frac{\partial}{\partial_{\text{Im}\tau}} \ln |g(\tau)|^2,\;\;\tau=\frac{\phi^1+ip^1}{\phi^0},
\end{equation}
where $g(\tau)$ can be identified with a worldsheet instanton partition function \cite{Harvey:1996ir} that appears in topological string free energy, and $p^2$ is a quadratic magnetic charge invariant.  Since the instanton corrections carry non-trivial information about the Calabi-Yau manifold, related to Gromov-Witten invariants, it would be puzzling if the four dimensional localization computation, including the one-loop determinants, could explain those corrections. In general the determinants are given in terms of equivariant indices of the four dimensional background with no connection to the Calabi-Yau invariants. Instead, one needs to understand the dynamics that are responsible for the quantum corrections that one observes at the level of the microscopic measure.  

The near-horizon geometry can help clarify some of these issues by drawing a clear contrast between four and five dimensional physics. Lets consider the attractor geometry of the $\D0-\D4$ black hole in IIA and uplift to M-theory. The near horizon geometry \cite{Beasley:2006us} is spherically symmetric and contains a local $AdS_3$ factor which consists of the M-theory circle fibered over $AdS_2$, that is,
\begin{equation}\label{AdS3}
ds^2=\vartheta(p)\left[ds^2_{AdS_2}+\frac{1}{(\phi^0)^2}(du+A)^2+ds^2_{S^2}\right]+ds^2_{CY}.
\end{equation}
Here $u$ parametrizes the circle, $A$ is the Kaluza-Klein gauge field, $\vartheta(p)$ is the physical size which can be taken to be large, and $ds^2_{CY}$ is the Calabi-Yau metric. Both $AdS_2$ and $S^2$ factors inside the brackets have unit size in string units. When reducing to four dimensional IIA string theory the radius of the circle becomes the scalar $1/X^0$ in (\ref{top free energy}). Given the map between the supergravity varibles and the topological string, finite radius means finite topological string coupling constant. So for finite radius, the Kaluza-Klein modes, that one obtains after compactification on the circle, have masses which are comparable to the $AdS_2$ inverse size and thus the solution is best described in five dimensional supergravity. In contrast to four dimensions,  the part of five dimensional $\N=2$  Lagrangian that contains the couplings of the vectors to the Weyl multiplet, is completely determined by  the parameters $D_{abc}$  and $c_{2a}$, which appear in the topological string free energy. Therefore, at the level of the Lagrangian that one obtains after dimensional reduction on the Calabi-Yau, there seems to be no information about the worldsheet instantons.

\subsection{The five dimensional point of view and the Rademacher expansion}\label{sec 1.1}

The four dimensional problem just described holds in the regime for which the radius of the circle is parametrically smaller than the size of $AdS_2$, or equivalently in the regime of weak topological string coupling constant. However, if
supersymmetric localization should hold for any value of the charges \footnote{In the black hole problem, charges play the role of the coupling constants in quantum field theory. When computing perturbative corrections to black hole entropy, we expand in inverse powers of the charges.}, the regime of $g_{top}\sim 1$ \footnote{The scalar fields are functions of the charges in the attractor geometry and so is $g_{top}$ too.} should be equally well valid,  but this corresponds to take the five dimensional point of view.  In the following, we shall argue that the microscopic Rademacher expansion is more appropriate to describe the five dimensional physics, and we build our solution based on this idea. 

As we explain later in more detail, localization at the level of the five dimensional theory, initiated in \cite{Gomes:2015xcf,Gomes:2013cca}, gives instead the finite dimensional integral 
\begin{equation}\label{5d localization int}
d(q,p)=\int \prod_{a=0}^{n_V}d\phi^a\ \frac{\vartheta(p)}{\phi^0} e^{-\pi q_b\phi^b+4\pi\text{Im}F_{cl}(\phi+ip)},
\end{equation}
with $F_{cl}(X)$ the "classical" prepotential that we define as
\begin{equation}\label{classical prepotential}
F_{cl}(X)=\frac{1}{6g_{\text{top}}^2}D_{abc}t^at^bt^c+\frac{c_{2a}t^a}{24}.
\end{equation}
This is the one-loop approximation of the prepotential (\ref{top free energy}) without the worldsheet instanton corrections. It is thus clear that the dependence on the worldsheet corrections (\ref{worldsheet corrections}) must arise from other effects. In contrast, the localization integral captures only perturbative quantum corrections around the attractor background (\ref{AdS3}), as an expansion in the area.  

We can check that the integral (\ref{5d localization int}) matches the expectations from the microscopic theory. Performing the various integrals,  it is found  that the final result matches the microscopic degeneracy - a Bessel function, valid precisely for large $g_{top}$ \cite{Gomes:2015xcf}, including the measure factor in $\N=8$ as well as in all $\N=4$ CHL examples in both $K3\times T^2$ and $T^4\times T^2$ compactifications. For this reason, we consider the five dimensional point of view  to be a step closer to understanding the quantum measure and the role of the worldsheet instantons.  

Besides the leading Bessel, the microscopic $\N=4$ answer contains a series of subleading Bessel contributions. Schematically, they arise after expanding the functions $g(\tau)$ (\ref{worldsheet corrections}) as instanton sums and then performing appropriate integrals \cite{Gomes:2015xcf,Murthy:2015zzy}. As a result, non-perturbative corrections to black hole entropy are generated. Remarkably, this series of Bessel contributions can be matched, to a certain extent, to the polar state contributions in a mock Jacobi Rademacher expansion \cite{Murthy:2015zzy}.   

The main goal of this work is to clarify the origin of the subleading Bessel functions, in a way consistent with the quantum entropy functional. Even though they are non-perturbative for large $g_{\text{top}}$, they can become relevant in the opposite regime of $g_{\text{top}}\ll 1$, which occurs for large central charge, because the number of Bessel functions can increase exponentially. According to the supergravity/topological string map, that regime corresponds to the four dimensional point of view, which is why it is crucial to understand the origin of the subleading Bessels. Preliminary steps in this direction were already taken in \cite{Gomes:2015xcf}, where it was suggested that the subleading Bessel contributions could arise from additional configurations of the full string theory path integral. To better understand the claim lets look in more detail to the Rademacher expansion \cite{Rademacher:1964ra}, which is an exact formula for the Fourier coefficients of modular forms- the black hole microscopic degeneracy \footnote{To be more precise one needs to consider also Mock modular forms \cite{Dabholkar:2012nd}. The Rademacher expansion suffers some modifications but these are not relevant for the present discussion. }. Schematically one has
 \begin{equation}\label{Rademacher intro}
 d(\Delta)=\sum_{\Delta_{\text{polar}}>0}^{\text{Max}}\Omega(\Delta_{\text{polar}})\sum_{c=1}^{\infty}\frac{1}{c}Kl(\Delta,\Delta_{\text{polar}},c)\int_{\epsilon-i\infty}^{\epsilon+i\infty}\frac{dt}{t^{\nu+1}}\exp{\left(\frac{\Delta}{4t c}+\frac{\Delta_{\text{polar}}t}{c}\right)},
 \end{equation}
where $d(\Delta)$ is the black hole degeneracy, which is a function of the charge combination $\Delta(q,p)$, $\Omega(\Delta_{\text{polar}})$ is the degeneracy associated with the polar terms and $Kl(\Delta,\Delta_{\text{polar}},c)$ are Kloosterman sums; each of the integrals are modified Bessel functions of the first kind. The microscopic $\N=4$ answer derived in \cite{Murthy:2015zzy} has precisely this form after neglecting the $c>1$ terms. Also, in this work we will only be considering the terms with $c=1$, usually referred as polar Bessels. For $\Delta\gg 1$ with $\Delta_{\text{polar}}$ fixed, the Bessel functions have saddle points at
\begin{equation}\label{saddles intro}
t\sim \sqrt{\frac{\Delta}{\Delta_{\text{polar}}}}\gg 1,
\end{equation}
and growth
\begin{equation}
\exp{\left(\sqrt{\Delta\Delta_{\text{polar}}}\right)}\gg 1.
\end{equation}
The leading contribution in (\ref{Rademacher intro}) therefore comes from the term with maximal $\Delta_{\text{polar}}$, with $\Delta_{\text{polar}}$  the polarity. In terms of the bulk physics, this is precisely the leading Bessel function that one obtains by evaluating the localization integral (\ref{5d localization int}), with $\text{Max}(\Delta_{\text{polar}})$ given by the charge combination $D_{abc}p^ap^bp^c+c_{2a}p^a$. The terms with $\Delta_{\text{polar}}<\text{Max}$ generate exponentially suppressed corrections and hence are non-perturbative. Furthermore, the value of $t$ can be identified with the topological string coupling constant $1/\phi^0$, and so we see that the saddles (\ref{saddles intro}) lie at large values of $1/\phi^0$, when the five dimensional point of view makes sense.  In this regime of charges it is thus pertinent to ask to what the non-perturbative saddles correspond from the five dimensional point of view. This is  puzzling because, given the localization computation (\ref{5d localization int}), there seems to be no room for any other additional contribution to the path integral. Though, it is possible that these saddles arise from other configurations in the full M-theory path integral. From which ones and how do they contribute? These are some of the questions that we want to address.

Our approach is mainly heuristic. In essence, we propose that the full path integral of M-theory receives the contribution of a new family of configurations which are euclidean geometries of the type $AdS_2\times S^1\times S^2$. The $AdS_2\times S^1$ factor is a local $AdS_3$ geometry, such as (\ref{AdS3}), with euclidean time contractible, and  guarantees that after reduction on the circle, one obtains the four dimensional $AdS_2\times S^2$  attractor background. This also follows from the fact that the four dimensional localization equations fix the geometry to be exactly $AdS_2\times S^2$ \cite{Gupta:2012cy}. Therefore we see that from a five dimensional point of view there is not much room for the space of allowed geometries, except that it must have a circle fibered over the attractor geometry.

To be consistent with the path integral and the localization computation, we argue that the new configurations are exact solutions of different five dimensional Lagrangians that we see as effective descriptions. The difference between Lagrangians is a finite renormalization of the parameters that define the theory such as $c_{2a}$ (\ref{classical prepotential}), which is the gauge-gravitational Chern-Simons coupling in five dimensions.  The supersymmetric localization computation at the level of the five dimensional theory reproduces the tail of polar Bessel functions observed in the microscopic answers, including the  exact spectrum of $\Delta_{\text{polar}}$. That is, for each euclidean $AdS_2\times S^1\times S^2$ geometry we find a Bessel function with index and argument given as in (\ref{Rademacher intro}), thus explaining the origin of the non-perturbative effects from a five dimensional point of view.

The renormalization of $c_{2a}$ has an additional effect. It corrects the physical size of the $AdS_2\times S^1\times S^2$ geometry in such a way that it can become zero, thus imposing a physical condition on the number of geometries. We find that this bound is in perfect agreement with the stringy exclusion principle \cite{Maldacena:1998bw}. The bound on the number of possible geometries is essentially the reason why there is only a finite number of polar Bessel functions in the Rademacher expansion. In the semiclassical limit, that is, when the central charge is very large, the number of geometries close to maximal polarity is dense which allows for a saddle point approximation. The result of this can be identified with the dilute gas approximation of the $AdS_3$ path integral as in \cite{Gaiotto:2006ns}, and the non-perturbative corrections around that saddle correspond to excitations of the fields dual to the chiral primary states.

It is also instructive to compare the above proposal with the effective field theory computation in $\mathbb{R}^4\times S^1$, which is the setup considered in \cite{Gopakumar:1998ii,Gopakumar:1998jq} and revisited in \cite{Dedushenko:2014nya}, for deriving the Gopakumar-Vafa formula. They consider a one-loop computation  for the Kaluza-Klein modes of vectors and hypermultiplets on the circle $S^1$, in the background of a self-dual graviphoton field. The result of this computation is the four dimensional higher derivative terms proportional to the topological string free energies $F_g(t)$ (\ref{top free energy}). At the on-shell level we do not expect the $\mathbb{R}^4$ and $AdS_2\times S^2$ computations to differ much when the curvatures are very small. So computing the instanton contributions, in $F_g(t)$, to the on-shell entropy function, we can generate non-perturbative corrections to the entropy area formula \cite{Murthy:2015zzy}. Nevertheless, at the quantum level, placing the theory on the $AdS_2\times S^1\times S^2$ background  (\ref{AdS3}), leads to problems related to the stringy exclusion principle \cite{Maldacena:1998bw}. The path integral of the reduced theory \footnote{By the AdS/CFT we keep track of all the Kaluza-Klein modes.} on $AdS_2\times S^1$ \footnote{To be more precise on thermal $AdS_3$, which, in a way, is a modular transformed version of $AdS_2\times S^1$. }  contains fluctuations that are not unitary  and hence are expected to backreact on the background solution \cite{Castro:2011ui}. The role of the exclusion principle is to artificially truncate the perturbative spectrum of fluctuations in agreement with the dual field theory. The exclusion principle is more relevant for small central charge which makes it a non-perturbative effect. The way we circumvent this problem is by considering the full M-theory path integral, instead of using the effective five dimensional Lagrangian with the massive hypermultiplets that are needed to obtain the Gopakumar-Vafa formula.

In fact, we show that in the limit of charges for which the circle becomes parametrically smaller than the size of $AdS_2$, while keeping the curvature small, we recover the perturbative partition function, in an expansion in the charges, as determined by the four dimensional effective action. This regime of charges is obtained by scaling $\text{Max}(\Delta_{\text{polar}})$ faster than $\Delta$ such that we have $1/\phi^0\ll 1$ at the saddle point. In this regime, we shall recover the Gopakumar-Vafa corrections to black hole entropy.   We explain, in addition, how the on-shell logarithmic corrections computed in \cite{Banerjee:2010qc,Banerjee:2011jp} arise from our formalism. 

To put it more explicitly, we provide with a non-perturbative formula for black hole entropy that correctly interpolates between the five and four dimensional physics. For small central charge $c$, one has only a small number of geometries and thus also a small number of Bessels.  Schematically we have  the gravitational answer
\begin{equation}
d_{\text{grav}}(\Delta)\simeq \int \frac{dt}{t^{\nu+1}}\exp{\left[\frac{\Delta}{4t}+c\,t\right]},\qquad c\sim 1
\end{equation}
which is the Bessel function, in agreement with the Rademacher expansion. Whereas for large central charge the high density of geometries, and so of Bessels, allows for a saddle point approximation. For the $\N=8$ and $\N=4$ models, we recover partially the OSV formula, that is, the holomorphic part, with corrections that we can systematically compute,
\begin{equation}
d_{\text{grav}}(\Delta)\sim \int d\phi\, e^{\Delta \phi}|Z_{\text{top}}(\phi,p)|^2+\ldots,\qquad c\propto p^3\gg 1
\end{equation}
Here $Z_{\text{top}}(\phi,p)$, which encodes the holomorphic free energies, can be seen as  a canonical partition function for the non-perturbative effects. These effects can then be related to the Gromov-Witten worldsheet instantons.
 
 \subsection{The polar state side of the story}\label{sec 1.2}

So far we have described the problem from the black hole point of view. However, there is another side to this story, which is not directly connected to black holes. This is the context of polar states and its relation to chiral primary states. We will study these states, which can be seen as $\D6-\aD6$ bound states, and we shall argue that the proposed $AdS_2\times S^1\times S^2$ geometries are the bulk duals of these microscopic configurations, after a modular transformation.

Polar states are characterized by having negative  charge discrimant $-\Delta_{\text{polar}}<0$ in the R sector of the CFT. Since black holes have necessarily positive charge discriminant, polar states must correspond in the bulk to configurations that do not form single center black holes. However, the reason why the  information about polar states enters in the black hole counting formula (\ref{Rademacher intro}) is due to the  modular properties of the CFT
partition function. In fact, knowing the spectrum of polar states defines 
completely the spectrum of non-polar states, and so using modularity we can relate one to the other.

There is an extensive literature on the problem of determining the spectrum of polar states and then use modularity to study corrections to black hole entropy 
\cite{Gaiotto:2006ns,Gaiotto:2006wm,Denef:2007vg}. One of the most complete of such studies is the work of Denef and Moore. Succintly, they perform an extensive study of polar multi-center black hole solutions in four dimensional $\N=2$ supergravity, with the goal of determining their contribution to the spacetime index. The main ingredients used are attractor flow trees \cite{Denef:2000nb} and the wall-crossing phenomena. They find that at large topological string coupling, the main contribution to the index comes from two center black hole solutions, corresponding to a configuration of a single $\D6$ and a single anti-$\D6$ ($\overline{\D}6$) with worldvolume fluxes, located at different positions in $\mathbb{R}^3$. The fluxes considered contain, besides the smooth part, a singular component, which is represented by ideal sheaves. Their contribution to the index gives rise to a refined version of the OSV answer, which includes a measure of the sort described by (\ref{worldsheet corrections}). 

The multi-center black hole solutions studied by Denef and Moore, admit a decoupling limit after an uplift to five dimensions \cite{Denef:2007yt,deBoer:2008fk}. In particular for the two center solution, the region near the core, where the $\D6$ and $\aD6$ sit, is zoomed in, and the decoupled geometry becomes asymptotically $AdS_3\times S^2$ with no black holes inside \cite{deBoer:2008fk}. It can be shown that these solutions carry Virasoro charges 
consistent with the values expected for $\Delta_{\text{polar}}$. Nevertheless, this result holds only for $\Delta_{\text{polar}}$ very close to its maximal value. We revisit this construction and establish a parallelism with our solutions. 

In all the works on black hole entropy through polar state counting,  one uses the $\CFT_2$ as an intermediate step. First, we build a partition function for the 
polar states $Z_{\text{polar}}(\tau,z^i)$, with $\tau$  the complex structure of 
the torus where the $\text{CFT}_2$ 
lives, and $z^i$ are chemical potentials. Then, we use modularity to 
construct the black hole partition function \cite{deBoer:2006vg} as
\begin{equation}\label{Zpolar}
Z_{\text{BH}}\simeq Z_{\text{polar}}(-1/\tau,z^i/\tau).
\end{equation}
This is only an approximate equality because we are not including the contribution due to other elements in the  $SL(2,\mathbb{Z})$ modular group. Nevertheless, for the purpose of studying non-perturbative effects due to the polar contributions, it is enough to consider only the modular transformation $\tau\rightarrow -1/\tau$. 

From the $\text{CFT}$ point of view,  $Z_{\text{polar}}$ naturally receives the contribution from only a finite number of states, those with negative discriminant. Nevertheless, from the bulk, one has to truncate artificially the perturbative 
spectrum of Kaluza-Klein fields on $AdS_3$, which are the fields dual to chiral primary states (polar states in the R sector). The truncation is known as the stringy exclusion principle \cite{Maldacena:1998bw} and asserts that quantum gravity in $AdS_3$ is inherently non-perturbative. 

The solution that we propose in this work is greatly inspired by the polar geometries studied in \cite{deBoer:2008fk}. The asymptotically  $AdS_3\times S^2$ polar configurations have a complicated geometry, but for large central charge we can write the metric as a background global $AdS_3\times S^2$ geometry plus corrections proportional to the singular fluxes, which are of the order of the inverse of central charge. A modular transformation makes the euclidean time circle contractible giving rise to asymptotically black hole like $AdS_2\times S^1\times S^2$ geometries \cite{Maldacena:1998bw,Dijkgraaf:2000fq,Murthy:2009dq}. However, in view of the localization computation that we want to perform, these solutions are not satisfactory because they do not have an exact $AdS_2\times S^2$ factor \cite{Gupta:2012cy,Gomes:2013cca}.  In a sense, which we would like to understand in more detail, our solutions are the non-perturbative analog, when taking into account the full string theory, of these modular transformed polar configurations. Conversely, we expect the fully quantum corrected polar configuration to have an  exact global $AdS_3\times S^2$ factor. 

To build intuition about the quantum corrected polar configurations we proceed as follows. The approach described in \cite{Denef:2007vg,deBoer:2008fk} considers first the backreaction of a two center $\D0-\D2-\D4-\D6$ configuration in four dimensions and then its uplift to M-theory. Equivalently, we can think of the same bound state as a $\D6-\aD6$ brane configuration with worldvolume fluxes $F$. It is well known that such configuration uplifts in M-theory to a Taub-Nut/anti-Taub-Nut geometry with fluxes $G\sim \omega\wedge F$, with $G$  the field strength of the M-theory three-form and $\omega$ is a normalizable two form of the Taub-Nut geometry. Therefore, fluxes on the $\D6$ branes map to fluxes in M-theory. If the worldvolume fluxes are ideal sheaves \cite{Denef:2007vg} we can generate arbitrary $\D0-\D2$ charges while keeping fixed the $\D4$ charge. We argue that the presence of such fluxes on the Calabi-Yau  can induce corrections on the five dimensional Lagrangian after reduction. Then, solving the full five dimensional equations of motion we find instead the black hole  $AdS_2\times S^1\times S^2$ geometry without corrections, but with the physical size $\vartheta$ (\ref{AdS3}), and the attractor values of the scalar fields depending explicitly on the fluxes. Localization on these backgrounds reproduces the finite tail of polar Bessel functions in the Rademacher expansion, thus setting the stage for a possible derivation of the solutions that we propose.  The presence of singular M-theory fluxes can be understood as quantum fluctuations of the K\"{a}hler class of the Calabi-Yau, which allows us to make a connection with the quantum foam picture of topological strings studied in \cite{Iqbal:2003ds}.

To guide the construction of our solution, we will revisit the counting of chiral primary states on $AdS_3$ and its relation to $Z_{\text{polar}}$ (\ref{Zpolar}) following \cite{Gaiotto:2006ns,Gaiotto:2006wm}.  Since our goal is to interpret the quantum black hole entropy as a partition function of M-theory, we will want to reproduce the counting of chiral primaries purely in terms of the eleven dimensional M-theory fields, and this will lead us inevitably to the polar $\D6-\aD6$ configurations with singular fluxes. The counting consists essentially in building multi-particle states on top of the vacuum $AdS_3$ by acting with the quanta that we obtain from the fields dual to the chiral primary states \cite{deBoer:1998ip,deBoer:1998us}. To do that we need to analyze the Kaluza-Klein tower of fields on $AdS_3$ coming from the supergravity fields and the $\M2$, and (anti)-$\aM2$ branes, wrapping cycles on the Calabi-Yau. Contrary to \cite{Gaiotto:2006ns}, which works in the dilute gas approximation, we will reconsider the same counting but for finite central charge. Imposing the stringy exclusion principle and spectral flow symmetry will enable us to reproduce the non-perturbative corrections induced by the polar Bessels in the Rademacher expansion, including the polar coefficients $\Omega(\Delta_{\text{polar}})$.

\subsection{Outline}

The plan of the paper is as follows. In section \S \ref{sec quantum saddles}, we start by describing in more detail 
the Rademacher expansion and connect it to previous work on black hole entropy and localization in supergravity. Then we review the microscopic formula for the entropy of one-quarter BPS black holes, which includes the effect of the subleading Bessel contributions. We use this formula as a check of our proposal and later make comments on $\N=2$ black hole entropy.  Before moving to the discussion about the $\D6-\aD6$ configurations, in section \S \ref{sec polar states} we review the problem of counting chiral primaries on $AdS_3$, which includes M2 and anti-M2-branes wrapping holomorphic cycles of the Calabi-Yau \cite{Maldacena:1998bw,Gaiotto:2006ns}. Taking into account the stringy exclusion principle and spectral flow symmetry we obtain a formula that agrees precisely with the microscopic $\N=4$ answer at finite charges; this formula will serve as a guide for the solution that we propose. Then in section \S \ref{sec bh bound states}, we review the $\D6-\aD6$ configurations with worldvolume fluxes and their decoupling limit. We  argue for the existence of a family of $AdS_2\times S^1\times S^2$ configurations and then in section \S  \ref{sec Loc} we compute the partition function using localization. The result of this is a finite sum over Bessel functions, whose spectrum is in agreement with the spectrum of polar states of a Jacobi form. Finally in section \S \ref{sec Foam} we discuss a connection between our solutions and the quantum foam picture of non-perturbative topological string. 

\section{Quantum saddle points from Rademacher expansion}\label{sec quantum saddles}

The Fourier coefficients of Jacobi forms of non-positive weight admit an exact expansion in terms of an infinite sum of Bessel functions. This expansion is known as Rademacher expansion \cite{Dijkgraaf:2000fq} and provides with a simple way to address the asymptotic behaviour of the integer Fourier coefficients. We review this expansion and connect to previous work on black hole entropy corrections.

Consider a Jacobi form $J_{k,\omega}$ of level $k$ and negative weight $\omega$, with 
 Fourier expansion 
\begin{equation}
J_{k,\omega}(\tau,z)=\sum_{n\geq 0,l}c(n,l)q^ny^l,\quad q=e^{2\pi 
i\tau},\,y=e^{2\pi i z}.
\end{equation}
The coefficients $c(n,l)$ with non-negative discriminant $\Delta\equiv n-l^2/(4k)\geq 0$, which are known as non-polar coefficients, admit an exact expansion in terms of an infinite sum of Bessel functions. Known as Rademacher expansion 
\cite{10.2307/2371313,Dijkgraaf:2000fq} it has the form
\begin{equation}\label{rademacher}
c(n,l)=\sum_{(m,s)\in\text{polar}}c(m,s)\sum_{c=1}^{\infty}\frac{1}{c}Kl(n,l;m,s;c)\,\int_{\epsilon-i\infty}^{\epsilon+i\infty}\frac{du}{u^{5/2-\omega}}\exp{\left[2\pi
 \frac{\Delta}{cu}-2\pi\frac{\Delta_p u}{c}\right]}.
\end{equation}
The coefficients $c(m,s)$ have negative discriminant, or polarity, $\Delta_p\equiv m-s^2/(4k)<0$, and are thus the polar coefficients, and $Kl(n,l;m,s;c)$ are Kloosterman sums \cite{Kloos}. The structure in (\ref{rademacher}) is completely fixed by modularity except for the knowledge of the polar coefficients.

One of the great advantages of the expansion (\ref{rademacher}), is that it is very appropriate to the study of asymptotics. For $\Delta\gg1$ with finite $\Delta_p$ each of the Bessel functions have a saddle point at
\begin{equation}\label{saddles up}
u_{p}=\sqrt{\frac{\Delta}{|\Delta_p|}}\gg 1.
\end{equation} 
Around each saddle we can expand perturbatively in powers of $\sqrt{\Delta|\Delta_p|}\gg 1$ such that
\begin{eqnarray}\label{saddle expansion}
c(n,l)\simeq&& \sum _{\Delta_{\text{min}}\leq\Delta_{p}\leq\Delta_{\text{max}}} e^{4\pi \sqrt{\Delta|\Delta_{p}|}}\left(1+\ldots\right)+\sum _{\Delta_{\text{min}}\leq\Delta_{p}\leq\Delta_{\text{max}}} \sum_{c>1}e^{4\pi \sqrt{\Delta|\Delta_p|}/c}\left(1+\ldots\right),
\end{eqnarray}
where the $\ldots$ denote perturbative corrections in powers of $1/\Delta$ around each of the saddles $u_p$, we are ignoring a normalization factor for each of the perturbative series. Therefore we see that the sum of polar contributions results in a tail of exponentially suppressed terms relative to the term of maximal polarity. 

For holomorphic Jacobi forms\footnote{The discussion for nearly-holomorphic Jacobi forms is very similar.}, the leading term in the expansion (\ref{saddle expansion}) comes from the most polar term, which has $\Delta_p=-k/4$. From the bulk physics point of view, we can identify the leading exponential growth with the black hole entropy area formula, since we have $4\pi \sqrt{\Delta|\Delta_p|} =A/4$, where $A$ is the area of the black hole horizon. Similarly, in the near-horizon attractor geometry (\ref{AdS3}) the saddle value of $u_p$ is identified with $1/\phi^0$. 

In addition, we can compute quantum perturbative corrections to the leading saddle using localization and a connection to Chern-Simons theory \cite{Gomes:2015xcf}.  The result is the finite dimensional integral (\ref{5d localization int}), which we review in section \S\ref{sec 5.1} using localization at the level of five dimensional supergravity. The idea of \cite{Gomes:2015xcf} is roughly the following. We start with the four dimensional localization integral (\ref{OSV}) and approximate the prepotential $F(X)$ according to the regime $\phi^0\ll 1$, where the leading saddle lives.  Indeed, the on-shell complexified K\"{a}hler class becomes large, that is, $t\gg 1$ and the one-loop topological string free energy approximates to $F_1(t)\simeq c_{2a}t^a/24$ leading to the classical prepotential (\ref{classical prepotential}).   The quantum measure, on the other hand, is fixed by a zero mode argument using the Chern-Simons formulation.  Though this formulation is well justified in the regime of $\phi^0\ll 1$, it is argued in \cite{Gomes:2015xcf} that the zero mode argument can be extrapolated also for the regime $\phi^0\gg 1$, which allows to define a quantum measure. 

The subleading saddle points in (\ref{saddle expansion}), corresponding to the polar terms with $|\Delta_p|<k/4$, lead to exponentially suppressed  corrections of the form
\begin{equation}\label{subleading saddles}
c(n,l)\sim e^{\frac{A}{4}}+\sum_{\Delta_p<\Delta_{\text{max}}}e^{4\pi \sqrt{\Delta|\Delta_p|}}+\ldots,
\end{equation}
with $4\pi \sqrt{\Delta|\Delta_p|}<A/4$. Given what we know already for the leading Bessel function in terms of bulk physics, it becomes pertinent to understand what is the origin of the subleading saddles from the quantum entropy functional. In fact, there is partial understanding for the leading saddles in the $c>1$ tails (\ref{saddle expansion}),
\begin{equation}\label{orbifold saddles}
c(n,l)\sim e^{\frac{A}{4}}+\ldots \sum_{c>1} e^{\frac{A}{4 c}}+\ldots,
\end{equation}
at the level of the quantum entropy path integral \cite{Sen:2009vz,Dabholkar:2014ema}. In this case, the subleading terms that grow as $\exp{A/4 c}$ arise after including in the path integral $\mathbb{Z}_c$ orbifolds of locally $AdS_3$ geometries \cite{Murthy:2009dq}. The orbifold explains the exponential growth $\sim \exp{A/4c}$ that characterizes them, because the area is reduced by a factor of $1/c$ due to the orbifold. 

There is something particular to the subleading polar terms when compared with the orbifold saddles, which is partially the reason why their bulk interpretation is more difficult.  While for the orbifold saddles the values of $u_p$ are consistent with the attractor background, for the subleading polar saddles (\ref{subleading saddles}) the values of $u_p$ (\ref{saddles up}) are quite distinct from the on-shell attractor background values, which can be determined from the leading Bessel;  they differ from finite renormalizations. If these saddles indeed correspond to bulk saddle configurations, then they can not be solutions of five dimensional supergravity that one obtains after compactification on the Calabi-Yau.

\subsection{Degeneracy from Siegel Modular Forms}\label{deg 1/4 dyons}

In the following we present a study of the microscopic $\N=4$ degeneracy for dyons in $K3\times T^2$ and $T^4\times T^2$ CHL orbifold compactifications \cite{Gomes:2015xcf,Murthy:2015zzy}; we describe in detail the role of the polar contributions. Though, our considerations are valid also for $\N=2$ compactifications, we will use the $\N=4$ answer as a check of our proposal.

The index $d(m,n,l)$ of $1/4$-BPS dyons in four dimensional $\N=4$ compactifications, has generating function the reciprocal of a Siegel modular form $\Phi_{k}$, that is,
\begin{equation}\label{Siegel modular form}
\frac{1}{\Phi_{k}(\rho,\tau,z)}=\sum_{m,n,l}d(m,n,l) e^{2\pi i m\rho}e^{2\pi i n\tau }e^{2\pi i lz}.
\end{equation}
Here $k$ is  the weight of the modular form under a congruence subgroup of $Sp(4,\mathbb{Z})$, and depends on the  orbifold compactification. The integers $m,n,l$ label respectively the T-duality invariants $P^2/2$, $Q^2/2$ and $Q.P$  with electric charges $Q$ and magnetic charges $P$ (in a particular heterotic frame). For further details we point the reader to \cite{Sen:2007qy}.

Conversely we can extract the integers $d(m,n,l)$- the black hole degeneracies, by performing an inverse Fourier transform. The function $1/\Phi_{k}$ contains poles, and thus by deforming the contour, the integral picks the residues at those poles. It turns out that the dominant contribution to the  black hole degeneracy is
\begin{eqnarray}\label{Siegel deg}
d(m,n,l)\simeq(-1)^{l+1}\int_{\mathcal{C}}\frac{d^2 u}{u_2^{k+3}}\left(2\pi m-\partial_{u_2}\Omega(u,\bar{u})\right)\,\exp{\left(\pi\frac{n+|u|^2 m-u_1 l}{u_2}-\Omega(u,\bar{u})\right)},
\end{eqnarray} 
with 
\begin{equation}\label{Omega instanton}
\Omega(u,\bar{u})=\ln g(u)+\ln g(-\bar{u}),\;u=u_1+iu_2.
\end{equation}
The functions $g(u)$ are modular forms of weight $k+2$, with Fourier expansion
\begin{equation}
g(u)=\sum_{n} d(n)e^{2\pi i u n}.
\end{equation} 

Choosing appropriately the contour $\mathcal{C}$  in (\ref{Siegel deg}) \cite{Gomes:2015xcf}, we can rewrite the degeneracy (\ref{Siegel deg}) as a finite sum of integrals of Bessel type, that is,
\begin{eqnarray}
d(m,n,l)\simeq &&(-1)^{l+1}2\pi i\sum_{r=0}^{m+2n_p-1}\Big(m+2n_p-r\Big)\,\nonumber\\
&&\times\sum_{\substack{s\geq 0\\  |r-2s|<m\\ c_q(m,r,s)>0}}^{r}d(r-s)d(s)e^{\pi i (r-2s)\frac{l}{m}}\,\nonumber\\
&&\times\int_{\epsilon-i\infty}^{\epsilon+i\infty}du_2\int_{-i\infty}^{i\infty}du_1 \frac{1}{u_2^{k+3}}\exp{W_{r,s}(u,m,n,l)},\nonumber\\
{}\label{finite sum Bessels}
\end{eqnarray}
with
\begin{eqnarray}\label{W potential}
W_{r,s}(u,m,n,l)=&&\pi\frac{n+|u|^2 m-u_1 l}{u_2}+2\pi (2n_p-r)u_2+2\pi i(r-2s)u_1,
\end{eqnarray}
and
\begin{equation}\label{quantum central charge}
\frac{c_q(m,r,s)}{24}=n_p-s+\frac{(m-r+2s)^2}{4m},
\end{equation}
with $n_p=1,0$ for $K3$ and $T^4$ CHL models respectively. Integrating over $u_1$ we obtain a sum over Bessel functions with the series resembling the Rademacher expansion (\ref{rademacher}). This has led the authors in \cite{Murthy:2015zzy} to test this possibility against an exact mock-Jacobi Rademacher expansion \cite{Dabholkar:2012nd}. 

For $r=s=0$, extremization of (\ref{W potential}) gives  the Cardy formula
\begin{equation}
d(m,n,l)\sim e^{2\pi\sqrt{c_L \Delta/6}},\;\Delta=n-\frac{l^2}{4k}\gg 1,
\end{equation}
where $c_L=6m+24n_p$ can be identified with the  left central charge (of the non-supersymmetric side of the $(0,4)$ $\CFT_2$). The values of $u_1$ and $u_2$ at the saddle point are
\begin{equation}\label{leading saddles n=4}
u_1=\frac{l}{2m},\;u_2=\sqrt{\frac{\Delta}{m+4n_p}}.
\end{equation}
From the bulk physics, $u_1$ and $u_2$ are mapped respectively to the values of the scalar fields $X^1/X^0$ and $1/X^0$ of the four dimensional supergravity.

For $(r,s)\neq (0,0)$ we can proceed similarly. Each term has exponential growth
\begin{equation}
\exp{W_{r,s}(u,m,n,l)}\sim e^{2\pi\sqrt{c_q \Delta/6}},\;\Delta\gg 1,
\end{equation}
and the values of the saddles are 
\begin{equation}
u_1=\frac{l}{2m}-i\frac{(r-2s)}{2m}u_2,\;u_2=\sqrt{\frac{6\Delta}{c_q}}.
\end{equation}
From here we see that these saddles differ from (\ref{leading saddles n=4}) by finite renormalizations parametrized by $r,s$.

\section{Polar states from M2 and anti-M2 branes}\label{sec polar states}

In this section, we review the problem of counting chiral primary states from the bulk theory on $AdS_3\times S^2\times X_{CY}$, with $X_{CY}$ a Calabi-Yau manifold, and how it connects to the study of black hole entropy. This section is essentially a review of  \cite{Gaiotto:2006ns} and companion works \cite{Gaiotto:2006wm,Simons:2004nm}. We develop on their formulas for black hole entropy and provide with corrections, which follow mainly from the stringy exclusion principle. The final result for the degeneracy of one-quarter BPS dyons in $\N=4$ compactifications can be shown to agree with the microscopic formula (\ref{finite sum Bessels}).

The idea of \cite{Gaiotto:2006ns} is to compute the contribution to the elliptic genus coming from the fields on $AdS_3$, which are dual to the chiral primary states of the CFT. The elliptic genus is nevertheless formulated in the R sector, and thus to count primary states, we must first do a spectral flow transformation to the NS sector. This map consists on the identification
\begin{eqnarray}\label{spectral flow map}
&&L_0|_{NS}=L_0|_{R},\\
&&\bar{L}_0|_{NS}=\bar{L}_0|_{R}+J^3_{0}|_{R}+\frac{c_R}{24},\\
&&J^3_0|_{NS}=J^3_{0}|_{R}+\frac{c_R}{12},
\end{eqnarray}
where the $|_{R,NS}$ subscripts denote the R and NS sectors, $L_0,\bar{L}_0$ and $J^3_{0}$ are the Virasoro and R-symmetry generators respectively, and $c_L,c_R$ are the left and right central charges. Under this transformation, polar states essentially map to chiral primary states, which we can count from the spectrum of Kaluza-Klein fields on $AdS_3$ \cite{Maldacena:1998bw,deBoer:1998ip,deBoer:1998us}. These include not only the contribution coming from the supergravity fields but also the contribution of M2 and anti-M2 branes wrapping holomorphic two-cycles on the Calabi-Yau. The black hole partition function is obtained by performing a modular transformation, after flowing to the R sector. 

The main result of \cite{Gaiotto:2006ns} is the factorization of the partition function (index), over the chiral primary states, as the square of the topological string partition function $Z_{\text{top}}$. Essentially the result is
\begin{equation}\label{Ztop^2}
\text{Tr}_{\text{ch.p.}}(-1)^Fe^{2\pi i \tau L_0}e^{2\pi i z^aJ_a}=Z_{\text{top}}(\tau,z^a)\times Z_{\text{top}}(\tau,-z^a),
\end{equation}
where the trace only goes through the chiral primary states (ch.p.); for simplicity we have omitted a $-c_L/24$ factor in $L_0$. In the NS sector, chiral primary states obey the condition $\overline{L}_0=J^3_0$, which maps to the condition $\bar{L}_0-c_R/24=0$ in the R sector (\ref{spectral flow map}). In addition, chiral primary states can carry charges under $U(1)$ currents $J_ a$. In the dilute gas approximation of \cite{Gaiotto:2006ns}, the trace in the bulk is a trace over BPS multi-particle states \cite{deBoer:1998us,deBoer:1998ip,Maldacena:1998bw} with arbitrary spin and occupation numbers. As a consequence, the result of the trace is a formal infinite product over all the quantum numbers, which can be related to $Z_{\text{top}}$ using the Gopakumar-Vafa invariants \cite{Gopakumar:1998ii,Gopakumar:1998jq}. That is, 
\begin{equation}\label{Ztop}
Z_{\text{top}}(\tau,z^a)=\prod_{m_a,n}(1-e^{2\pi i \tau (\frac{1}{2}m_ap^a+n)}e^{2\pi i m_az^a})^{N_{m_a,n}}.
\end{equation}
This is the key step that allows the authors in \cite{Gaiotto:2006ns} to make a connection with the OSV conjecture \cite{Ooguri:2004zv}. Here $p^a$ is the magnetic flux on the sphere $S^2$, induced by the M5-brane wrapping a divisor four cycle in the homology class $P$ Poincare dual to $[P]=p^a\Sigma_a$ with $\Sigma_a\in H^2(X_{CY},\mathbb{Z})$.

The factors $Z_{\text{top}}(\tau,z^a)$ and $Z_{\text{top}}(\tau,-z^a)$ arise essentially from the contributions of respectively M2-branes and anti-M2-branes  wrapping  holomorphic cycles in the Calabi-Yau; there is also a contribution coming from the supergravity fields but they will not be relevant for the discussion of $\N=4$ compactifications, which is our main interest in this section. What allows the sum over arbitrary M2 and anti-M2 charges is the fact that in AdS space, a brane and its anti-brane can preserve mutual supersymmetries. Indeed, a M2-brane wrapping a holomorphic cycle $Q\in H_2(X_{CY})$, siting at the origin of $AdS_3$ and at the north pole of $S^2$ preserves the same set of supersymmetries as an anti-M2-brane wrapping the same cycle $Q$, sitting at the origin of $AdS_3$ but now at the south pole of $S^2$ \cite{Simons:2004nm,Gaiotto:2006ns}. The fact that these are supersymmetric configurations on $AdS_3\times S^2\times X_{CY}$ will play a very important role in the remaining of the letter.

If the theory has $\N=4$ supersymmetry, for example when $X_{CY}=K3\times T^2$, the partition function (\ref{Ztop}) simplifies considerably, that is,
\begin{equation}\label{dedekind function}
Z_{\text{top}}(\tau,z^a)=\prod_{m_1>0}\left(1-e^{2\pi i m_1(\tau\frac{p^1}{2}+z^1)}\right)^{-24}.
\end{equation}
Here $p^1$ parametrizes the class of $K3\in H_4(\mathbb{Z})$, which is Poincare dual to $H^2(T^2)$. The coefficient $24$ in the product is the Euler character of $K3$, which allows for a generalization to other $\N=4$ compactifications. In the case of $\N=8$ supersymmetry this partition function is trivially one.

Formula (\ref{Ztop^2}) is valid only in the limit of very large central charge and for low density of chiral primaries, and so it is not the complete answer. The reason is that it does not take into account the stringy exclusion principle, which puts a bound on the total spin of the multi-particle states, that is,
\begin{equation}\label{exclusion principle bound}
J^3_0|_{NS}\leq \frac{c_R}{12}.
\end{equation}
 It makes sense as a grand canonical partition function valid for infinite central charge, in which case the stringy exclusion principle constraint can be relaxed. The exclusion principle can not be seen in perturbation theory on $AdS_3$, because from the bulk point of view the multi-particle states are free bosonic excitations with no limit in their particle number. Instead, for finite central charge the contribution coming from the perturbative spectrum of Kaluza-Klein fields on $AdS_3$ must be truncated due to the stringy exclusion principle. Since $\bar{L}_0=J^3_0$, by supersymmetry, then the bound on $J^3_0$ imposes a bound on $\bar{L}_0$. Moreover, we have $L_0=\bar{L}_0$ for a static solution and so $L_0$ is also bounded. Therefore, only a finite number of states contribute to (\ref{Ztop^2}). 

Physically, adding $q_a$ M2-branes sitting at the north pole adds non-zero angular momentum
\begin{equation}\label{M2 ang momentum}
J^3_0|_{NS}=\frac{1}{2}q_ap^a,
\end{equation}
much like an electron in a background magnetic field, while the $\bar{q}_a$ anti-M2-branes, because they sit at the south pole,  contribute with the same sign angular momentum, that is, $J^3_0=\bar{q}_ap^a/2$. Therefore, flowing to the R sector, we find that the state carries R-charge
\begin{equation}
J^3_0|_R=-\frac{c_R}{12}+\frac{1}{2}(q_a+\bar{q}_a)p^a.
\end{equation}
We then see that the exclusion principle gives a bound on the number of M2 and anti-M2 branes.

The trace in the R sector must contain only states that do not form black holes, up to a spectral flow transformation. In terms of the Virasoro charges this implies  
\begin{equation}\label{polarity chiral}
L_0-\frac{c_L}{24}+\frac{1}{2}D^{ab}j_aj_b<0,
\end{equation}
in the R sector, where we have reincorporated a $-c_L/24$ factor. Here  $j_a=q_a-\bar{q}_a$ is the total M2-brane charge where $q_a$ and $\bar{q}_a$ are respectively the M2 and anti-M2 charges, and $D_{ab}=D_{abc}p^c$, with $D_{abc}$ the intersection matrix of the Calabi-Yau. Since $j_a\in \mathbb{Z}$ and $D_{ab}$ is not unimodular,  $j^a$ lives on the lattice $\Lambda^*/\Lambda$ with $\Lambda$ the lattice $k^a\in \mathbb{Z}$ and $\Lambda^*$ its dual under the metric $D_{ab}$; the quotient removes spectral flow charges \cite{deBoer:2006vg}.  Since the configuration is static, that is, $L_0=\bar{L}_0$ in the NS sector, the condition becomes
\begin{eqnarray}\label{M2 polarity constraint}
L_0-\frac{c_L}{24}+\frac{1}{2}D^{ab}j_aj_b<0\Leftrightarrow \frac{p^3}{24}+\frac{c_2\cdot p}{24}-\frac{1}{2}(q_a+\bar{q}_a)p^a-\frac{1}{2}D^{ab}(q_a-\bar{q}_a)(q_a-\bar{q}_a)>0\nonumber, \\
{}
\end{eqnarray} 
where we used the fact that $c_L=p^3+c_2\cdot p$, with $p^3=D_{abc}p^ap^bp^c$ and $c_2\cdot p=c_{2a}p^a$ \cite{Maldacena:1997de}. We can show that (\ref{M2 polarity constraint}) is spectral flow invariant. In particular, for $K3\times T^2$ or $T^4\times T^2$ CHL orbifolds  this condition becomes 
\begin{equation}\label{c_L corrected bulk}
n_p-\bar{q}_1+\frac{\left(P^2/2-q_1+\bar{q}_1\right)^2}{2P^2}>0,
\end{equation}
with $n_p=0,1$ for the $T^4,K3$ respectively. Here we have used the fact that the only non-vanishing components of $D_{abc}$ are $D_{1ab}\equiv C_{ab}$ and permutations, together with $c_{2a}\equiv 24 n_p\delta_{a,1}$ and $P^2\equiv C_{ab}p^ap^b$. Setting $P^2=2m$, $q_1=(r-s)$ and $\bar{q}_1=s$  we obtain precisely the effective central charges (\ref{quantum central charge}).

The formula (\ref{Ztop^2}) also misses important degeneracy factors when taking the trace over the chiral primaries. In the limit when the M2-brane charge $q_a,\bar{q}_a$ is parametrically smaller than $p$, which we are taking to be large, these degeneracy factors are irrelevant for the purpose of arriving at (\ref{Ztop^2}). This is the dilute gas approximation of \cite{Gaiotto:2006ns}. However, since our main interest is for finite central charge, we need to take into account those degeneracy factors.  Essentially we follow the discussion  in \cite{Gaiotto:2006wm}. Under spectral flow from NS to R sector the chiral primaries, which are annihilated by $J^{-}_{1}$, flow to lowest $SU(2)_R$ weight states because $J^{-}_{1}$ flows to $J^{-}_{0}$. For example the $AdS_3$ R vacuum corresponds to a lowest weight state with $J^3_0=-c_R/12=-k/2$ with $k$ the $SU(2)_R$ level. Therefore acting with $J^{+}_{0}$ we generate the full multiplet, which leads to a degeneracy of $2|J|+1$ states. In addition, these states have to be tensored with the zero modes $\psi^{+\pm}$ of the centre of mass multiplet\footnote{The $(0,4)$ MSW $\CFT_2$ superconformal algebra \cite{Maldacena:1997de} contains the centre of mass multiplet, besides the small $\N=4$ algebra.} which carry spin $1/2$. The total angular momentum after including the contribution of the M2-branes is
\begin{equation}\label{total angular momentum}
J^3_0=\frac{c_R}{12}-\frac{1}{2}(q_a+\bar{q}_a)p^a-\frac{1}{2},
\end{equation}
which leads to a degeneracy 
\begin{equation}
2J^3+1=\frac{c_R}{6}-(q_a+\bar{q}_a)p^a.
\end{equation} 
Substituting in this expression the  values of $p^a$ and $q_a,\bar{q}_a$ for the $K3\times T^2$ and $T^4\times T^2$ CHL examples, that is,  $q_1=(r-s)$ and $\bar{q}_1=s$, one obtains
precisely 
\begin{equation}
\frac{c_R}{6}-(q_a+\bar{q}_a)p^a=p^1(m+2n_p-r),
\end{equation}
 which we identify with the measure factor in the first line of expression (\ref{finite sum Bessels}). Since degeneracy is always positive we must have 
 \begin{equation}\label{stringy bound}
 m+2n_p-r>0,
 \end{equation}
which is the bound implied by the stringy exclusion principle \cite{Maldacena:1998bw}. In the limit when $p^3\propto m\rightarrow \infty$ this bound can be relaxed which is why one obtains the infinite products (\ref{Ztop}).

In addition to the $SU(2)_R$ degeneracy,  we need to tensor with the states associated with the quantization of fluctuations of the M2 and anti-M2-branes wrapping holomorphic cycles in the Calabi-Yau. For $K3\times T^2$ they can be identified with the degeneracies of $r-s$ M2-branes and $s$ anti-M2-branes wrapping $T^2$, which are given by the dedekind function (\ref{dedekind function}). These explain the factors $d(s)d(r-s)$ in the second line of (\ref{finite sum Bessels}).

Assembling all the factors, we construct, in the R sector, the polar partition function
\begin{equation}
Z_{\text{polar}}(\tau,z^a)=\sum_{L_0,j_a\in \text{ polar}} e^{2\pi i\tau (L_0-\frac{c_L}{24})}e^{2\pi iz^aj_a},
\end{equation}
where the sum is over the states obeying the condition (\ref{c_L corrected bulk}) and (\ref{stringy bound}). The black hole partition function is obtained after a modular transformation \cite{Gaiotto:2006ns}, that is,
\begin{equation}\label{modular transf polar states}
Z_{\text{BH}}(\tau,z^a)\simeq\tau^{-\omega} e^{\pi i\frac{D_{ab} z^az^b}{\tau}}Z_{\text{polar}}(-1/\tau,z^a/\tau).
\end{equation}
There are further corrections to this formula coming from other elements of $SL(2,\mathbb{Z})$; they give contributions of the orbifold type (\ref{orbifold saddles}). The parameter $\omega$ is the weight of the elliptic genus under modular transformations and can be determined as follows. We decompose the elliptic genus in spectral flow sectors as $\chi(\tau,z)=\sum_{\mu}h_{\mu}(\tau)\theta_{\mu}(\tau,z^a)$, with $\theta_{\mu}(\tau,z^a)$ a multidimensional theta function \footnote{The part of the elliptic genus that contains the black hole entropy is the vector-valued modular form $h_{\mu}(\tau)$. Its Fourier coefficients are the quantities that are invariant under U-duality.}. The function $h_{\mu}(\tau)$ contains the information about black hole degeneracy, while the theta functions contains the states related by spectral flow symmetry.
On one hand, from the Siegel modular form (\ref{Siegel modular form}), of weight $k$, one finds Jacobi forms of weight $-k$ and a single chemical potential $z$. This implies that part of the Jacobi form that contains the information about black hole degeneracy, which is a vector valued modular form, must have weight $-k-1/2$, and hence also $h_{\mu}(\tau)$. On the other hand, since $\theta_{\mu}(\tau,z^a)$ has weight $b_2/2$ with $b_2$ the second Betti number of the Calabi-Yau, we find that the weight of the elliptic genus is $\omega=-k-1/2+b_2/2$. For the $K3\times T^2$ compactification we have $b_2=23$ and thus $\omega=1$. Similarly for the other CHL compactifications we have $b_2=2k+2+1$ \cite{Sen:2007qy}, which also gives $\omega=1$!

The black hole degeneracy is computed by an inverse Fourier transform, that is,
\begin{eqnarray}
&&d_{\text{BH}}(n,l_a)\sim\int  \prod_{a=1}^{b_2} dz^a d\tau\, Z_{\text{BH}}(\tau,z^a)e^{-2\pi i\tau n-2\pi i z^al_a}\\
&=&\sum_{\substack{q_a,\bar{q}_a\\ \frac{c_R}{6}-(q_a+\bar{q}_a)p^a>0}}d(q_a)d(\bar{q}_a)\left(\frac{c_R}{6}-(q_a+\bar{q}_a)p^a\right)\times \nonumber\\
&&\int \prod_{a=1}^{b_2} dz^a d\tau \tau^{-\omega} e^{\pi i\frac{D_{ab} z^az^b}{\tau}-\frac{2\pi i}{\tau}z^a(q_a-\bar{q}_a)}e^{\frac{2\pi i}{\tau}\left(\frac{p^3}{24}+\frac{c_2\cdot p}{24}-\frac{1}{2}(q_a+\bar{q}_a)p^a\right)}e^{-2\pi i\tau n-2\pi i z^al_a}\nonumber,\\
{}
\end{eqnarray}
with the additional constraint that the sum obeys (\ref{M2 polarity constraint}). Specializing the various parameters to the $\N=4$ examples and performing the various gaussian integrals in $z$, we obtain almost precisely the one-quarter BPS degeneracy described in section \S\ref{deg 1/4 dyons}. The only difference is the contour. While in the formula above we take $\tau$ over the Fourier contour $]0,1]$, in the Rademacher expansion one has $1/\tau$ running over $]\epsilon -i\infty,\epsilon+i\infty[$. It looks puzzling how to go from one contour to the other without picking additional contributions. Nevertheless, for the purpose of computing saddle point corrections, both integrals are equally valid. As we explain later, one of the great advantages for using localization is that it naturally picks the Rademacher contour, which then acquires a physical interpretation.

\section{Black hole bound states and horizonless geometries}\label{sec bh bound states}

In this section, instead of thinking in terms of M2 and anti-M2 branes wrapping cycles on the Calabi-Yau, we consider an equivalent description in type IIA string theory consisting of a $\D6$ and a $\aD6$ configuration wrapping the Calabi-Yau, and carrying $U(1)$ fluxes $F$  in their worldvolume.  This section is essentially a review of the polar configurations of Denef and Moore \cite{Denef:2007vg} and their decoupling limit \cite{deBoer:2008fk}. The main goal is to find a microscopic description for the family of saddle geometries that we propose. Under certain assumptions, we argue that the quantum entropy path integral should be seen as an M-theory path integral with eleven dimensional instanton solutions.  Then we propose an effective five dimensional description, which is amenable for using localization.

For the charge configuration of interest, the total $\D6$ charge is zero but the presence of fluxes induce lower dimensional charges due to the couplings of the worldvolume fields to the Ramond-Ramond gauge fields $A_3,A_1$, such as
\begin{equation}\label{Ramond couplings}
\int_{\D6}F\wedge F\wedge A_3,\;\int_{\D6}F\wedge F\wedge F\wedge A_1 ,
\end{equation}
which generate $\D2$ and $\D0$ charges respectively. Uplifting to M-theory, the pair $\D6-\aD6$ becomes a Taub-Nut and anti-Taub-Nut configuration, while the $\D2$ and $\aD2$ charges lift to $\M2$ and $\aM2$ charges and the $\D0$ charges become momentum along the M-theory circle.

From the M-theory point of view the fluxes on the $\D6$-brane lift to four fluxes $G=dC_3$ \cite{Sen:1997js} in M-theory, with $C_3$ the three-form that couples to M2-branes, that is,
\begin{equation}
G\propto \omega_{TN}\wedge F.
\end{equation} 
 $\omega_{TN}$ is the self-dual normalizable two form of the Taub-Nut geometry, and $F$ is the total flux in the $\D6$; and similarly for the $\aD6$ brane. Therefore fluxes on the $\D6$ branes map to fluxes in the bulk M-theory. 

To be consistent with the $\M2$ brane picture of the previous section, we want to turn on fluxes that generate arbitrary $\M2\sim \D2$ charges as well as $\D0$ charges, but keep fixed the $\D4$ charge, which lifts to the $\M5$ brane, parametrized by the magnetic charges $p^a$. In order to do that, we parametrize the flux in the form $F=p+\F$, where $p\in H_2(X)$. To keep the $\D4$ charge equal to $p$, we need to impose that the first Chern-class of $\F$ is zero, whereas to generate arbitrary $\D0,\D2$ charges we must keep its higher Chern-classes arbitrary. In other words, this amounts to
\begin{equation}\label{singular fluxes}
\int_{C^a}\F=0,\;\int \F\wedge \F\neq 0,\;\int \F\wedge \F\wedge \F\neq 0,
\end{equation}
for any two cycle $C_a$ in the Calabi-Yau. Such conditions on the fluxes are only possible if the flux has singularities. If the flux was smooth then vanishing of the first Chern-class would imply vanishing of the higher Chern-classes. The way to regularize the singularities is to drop the notion of line bundle and use ideal sheaves, which are torsion free sheaves with vanishing first Chern-class \cite{Denef:2007vg}. The ideal sheaf is a generalization of the notion of line bundle. Usually if the Calabi-Yau is an algebraic variety, then the singularities can be blown up leading to a new space $\hat{X}$ where the torsion free sheaves become line bundles. It was argued in \cite{Iqbal:2003ds} that we should include such configurations in the quantum gravity path integral.  We follow a similar approach and consider the path integral of M-theory in the presence of such configurations. 

The induced four dimensional charges have the form \cite{Denef:2007vg}
\begin{eqnarray}
&&\Gamma_6=e^{p/2}(1-\beta+n\omega)\left(1+\frac{c_2(X)}{24}\right)\nonumber\\
&&=\left(1,\frac{p^a}{2},\frac{p_a}{8}+\frac{c_{2a}}{24}-\beta_{a},\frac{p^3}{48}+\frac{c_{2}\cdot p}{48}-\frac{1}{2}\beta\cdot p+n\right)\label{Gamma6},
\end{eqnarray}
for the first $\D6$ center, where we have defined $p_a=D_{abc}p^cp^b$, while for the second center $\aD6$, we have
\begin{eqnarray}
&&\Gamma_{\bar{6}}=-e^{-p/2}(1-\bar{\beta}+\bar{n}\omega)\left(1+\frac{c_2(X)}{24}\right)\nonumber\\
&&=\left(-1,\frac{p^a}{2},-\frac{p_a}{8}-\frac{c_{2a}}{24}+\bar{\beta}_{a},\frac{p^3}{48}+\frac{c_{2}\cdot p}{48}-\frac{1}{2}\bar{\beta}\cdot p+\bar{n}\right)\label{Gamma6bar}.
\end{eqnarray}
We have denoted $\beta$ and $n$ respectively the second and third Chern-classes of the ideal sheaves, and similarly for $\bar{\beta}$ and $\bar{n}$. We have used the notation in \cite{Denef:2007vg} which assigns charges (D6,D4,D2,D0) as $(p^0,p^a,q_a,q_0)$. We see that the total charges are as follows: the total $\D6$ charge is zero, and the total $\D4$ charge is $p^a$ as required; on the other hand the total $\D2\sim \M2$ charge is $\beta_a-\bar{\beta}_a$ and the $\D0$ charge is $p^3/24+c_2\cdot p/24-(\beta+\bar{\beta})\cdot p/2+n+\bar{n}$.

Once we consider backreaction in four dimensional supergravity this charge configuration gives rise to a two center supersymmetric black hole solution. In fact we can obtain multi-center configurations by considering many other non-local charges \cite{Denef:2000nb}. For a generic charge configuration there can exist both single and multi-center solutions. However, certain configurations have the  property that for the same  total charge, only multi center solutions exist; these are the polar configurations.  A special feature of multi center configurations is that for certain values of the asymptotic moduli the distance between two centers goes to infinity and the bound state leaves the spectrum leading to the phenomenon of wall-crossing. 

The two center black hole solution is a complicated geometry. The details about the metric can be found in \cite{Denef:2000nb,Denef:2007vg}. Nevertheless, the essential feature that we need is that the metric is determined in terms of $\mathbb{R}^3$ harmonic functions  $H(x)$,
\begin{eqnarray}\label{harmonic functions}
&&H^0=\frac{p^0}{R^{1/2}|x-x_{6}|}-\frac{p^0}{R^{1/2}|x-x_{\bar{6}}|}+h^0,\;H^a=\frac{p^a}{2R^{1/2}}\left(\frac{1}{|x-x_{6}|}+\frac{1}{|x-x_{\bar{6}}|}\right)+h^a\\
&&H_a=\frac{1}{R^{1/2}}\left(\frac{q_a}{|x-x_{6}|}-\frac{\tilde{q}_a}{|x-x_{\bar{6}}|}\right)+h_a,\;H_0=\frac{1}{R^{1/2}}\left(\frac{q_0}{|x-x_{6}|}+\frac{\tilde{q}_0}{|x-x_{\bar{6}}|}\right)+h_0\nonumber,
{}
\end{eqnarray}
 where $q_a$ and $\tilde{q}_a$ are the $\D2$ charges induced respectively by the $\D6$ and $\aD6$, and similarly for $q_0,\tilde{q}_0$. The parameter $R$, which is a free parameter, is the asymptotic radius of the M-theory circle. In addition, supersymmetry imposes a certain integrability condition on the harmonic functions, which force the centers to stay at a predetermined distance. This equilibrium distance is a function of the charges and also the asymptotic moduli. The wall-crossing phenomena happens when we tweak the moduli such that this distance goes to infinity in which case the bound state  splits into its constituents. Moreover, the two center solution is not a static geometry and carries angular momentum \cite{Denef:2000nb} 
\begin{equation}\label{total J}
\vec{J}=\frac{1}{2}\langle\Gamma_6,\Gamma_{\bar{6}}\rangle\frac{\vec{x}_{6\bar{6}}}{r_{6\bar{6}}},
\end{equation}
where $\langle\Gamma_1,\Gamma_{2}\rangle$ is the symplectic charge inner product defined as $\langle\Gamma_1,\Gamma_{2}\rangle \equiv -p^0_1q^2_0+p^a_1q^2_a-q^1_ap^a_2+q^1_0p^0_2$.

 The region near the core of the centers admits a decoupling limit \cite{deBoer:2008fk} after an uplift to five dimensions. Before moving to a general discussion about the decoupled two center configurations, first we describe the simplest $\D6-\aD6$ configuration, which corresponds to setting the singular fluxes to zero. Without loosing generality we consider $c_2(X)=0$ for the moment. Therefore, the configuration consists of a $\D6$ at a position $x_6$ and  a $\aD6$ at a position $x_{\bar{6}}$, carrying $U(1)$ fluxes $F=p^a\omega_a/2$, with $\omega_a$ a basis of $H^2(X)$, and $\bar{F}=-p^a\omega_a/2$ for the $\aD6$.  The charge vectors are therefore
 \begin{equation}\label{D6 charges}
 \Gamma_6=e^{\frac{p}{2}},\;\Gamma_{\bar{6}}=-e^{-\frac{p}{2}}.
 \end{equation}
  Each center has non-zero entropy but from asymptotic infinity one finds that a black hole with the same total charges cannot exist because the discriminant function $\hat{q}_0=q_0-D_{ab}q^aq^b/2$ is positive which renders the entropy formula $\sim \sqrt{-\hat{q}_0}$ imaginary. From the M-theory point of view this configuration lifts to a Taub-Nut and anti-Taub-Nut configuration with M-theory three form flux $G\propto \omega_{TN}\wedge p$, with $\omega_{TN}$ the normalizable self-dual two form of the Taub-Nut geometry. This means that from the M-theory point of view the solution is completely smooth with no horizon.
 
The decoupling limit consists effectively in taking the  constants of the harmonic functions to zero with the exception of the component $h_0$ which becomes $-R^{3/2}/4$. This renders a configuration where the centers sit at a fixed distance completely determined by their charges, that is,
\begin{equation}\label{r66}
r_{6\bar{6}}=\frac{4\langle \Gamma_6,\Gamma_{\bar{6}}\rangle}{p^0_6 R^2},
\end{equation}
where $\langle \Gamma_6,\Gamma_{\bar{6}}\rangle$ is the symplectic charge inner product. For the charge configuration (\ref{D6 charges}) we find $\langle \Gamma_6,\Gamma_{\bar{6}}\rangle =p^3/6$.

The decoupled geometry based on the harmonic functions is still a complicated solution. Nevertheless, we can follow the observation made in \cite{Denef:2007yt}, and use oblate-spheroidal coordinates defined as
\begin{eqnarray}\label{oblate coords}
|x-x_{6}|=\frac{r_{6\bar{6}}}{2}(\cosh(2\eta)+\cos(\theta)),\\
|x-x_{\bar{6}}|=\frac{r_{6\bar{6}}}{2}(\cosh(2\eta)-\cos(\theta)),
\end{eqnarray}
to simplify considerably the problem. The five dimensional metric in the new coordinates becomes precisely the global $AdS_3\times S^2$ geometry
\begin{eqnarray}\label{purely fluxed ads3}
ds^2=4U^{2/3}\left(-\cosh^2\eta d\tau^2+d\eta^2+\sinh^2\eta d\sigma^2\right)+U^{2/3}(d\theta^2+\sin^2\theta (d\phi+A)^2),
\end{eqnarray}
where $U=p^3/6$. Furthermore, the attractor equations fix the five dimensional vector-multiplet scalars $M$ to constants $M^a=U^{1/3}p^a$ while the gauge fields have constant flux on the sphere $F^a= p^a e_2$, with $e_2$ the volume form. Furthermore, the sphere is twisted by the gauge field
\begin{equation}\label{R gauge field}
A=d\sigma -d\tau,
\end{equation}
which is flat everywhere except at the origin where it has a delta function singularity; this follows from the fact that four dimensional solution carries angular momentum. From the CFT point of view, the Wilson line $\oint A$ around the boundary circle $\sigma$, which is contractible in the full geometry,  is necessary to make the fermions periodic, as expected for the R vacuum. In other words, the Wilson line generates a spectral flow transformation which takes the NS vacum with $L_0=\bar{L}_0=0$ and $J^3_0=0$ to the R vacuum with $L_0=0$ and $\bar{L}_0=p^3/24$ and maximal $J^3_0=-p^3/12$ \cite{Kraus:2006nb,deBoer:2008fk}. Finally note that according to the change of coordinates (\ref{oblate coords}), the positions of the $\D6$ and $\aD6$ map to the center of $AdS_3$ and respectively to the north and south of $S^2$, which is consistent with the M2 branes description. 

The geometry described above corresponds to a particular case of what is known as ambi-polar Eguchi-Hanson metric with Gibbons-Hawking (GH) charges $q=1$ and $-q$ \cite{Bena:2010gg}. For general $q$ this is a $\mathbb{Z}_q$ quotient of global $AdS_3\times S^2$. We review this construction following \cite{Bena:2010gg}. We use cylindrical polar coordinates $(z,\rho,\phi)$ on $\mathbb{R}^3$, and consider GH charges located on the $z$-axis at $z=\pm a$, and define
\begin{equation}
r_{\pm}=\sqrt{\rho^2+(z\mp a)^2}.
\end{equation}
The five dimensional geometry is entirely determined by the harmonic functions
\begin{eqnarray}\label{Eguchi-Hanson}
&&V^0=q\left(\frac{1}{r_+}-\frac{1}{r_-}\right),\; V^1=k\left(\frac{1}{r_+}+\frac{1}{r_-}\right),\nonumber\\
&&V_{1}=-\frac{k}{q}\left(\frac{1}{r_+}-\frac{1}{r_-}\right),\; V_0=-\frac{2k^2}{a q^2}+\frac{k^2}{2q^2}\left(\frac{1}{r_+}+\frac{1}{r_-}\right).\nonumber
\end{eqnarray}
Using the oblate spheroidal coordinates we can map the GH space to global $AdS_3\times S^2$ (\ref{purely fluxed ads3}) with size
\begin{equation}
L^2=(k^2)^{2/3}.
\end{equation}
So to agree with (\ref{purely fluxed ads3}) we need $k^2=U$ or $k=(p^3)^{1/2}$.

Lets now consider the case of a general two center charge configuration and its decoupling limit \cite{deBoer:2008fk}. In this case, the geometry is only asymptotically  $AdS_3\times S^2$ 
\begin{eqnarray}\label{approximate AdS3}
ds^2\simeq U^{2/3}\left(d\eta^2+e^{2\eta}(-d\tau^2+d\sigma^2)\right)+U^{2/3}(d\theta^2+\sin^2\theta (d\phi+\tilde{A})^2),\;\eta\gg 1
\end{eqnarray}
with $U=p^3/6$ and
\begin{equation}\label{twisting gauge field}
\tilde{A}=\frac{J}{J_{\text{max}}}(d\tau-d\sigma),\; J_{\text{max}}=\frac{p^3}{12}.
\end{equation}
Here  $J=\langle\Gamma_6,\Gamma_{\bar{6}}\rangle /2$ is the total angular momentum of the two center configuration (\ref{total J}). The remaining field configuration consists of five dimensional gauge fields $A^{a}$ and scalars $M$, which have the attractor solution
\begin{eqnarray}\label{attractor fields}
&&A^{a}_{5D}\simeq -p^a \cos\theta (d\phi+\tilde{A})+2D^{ab}q_b (d\sigma +d\tau),\\
&&M^a\simeq U^{-1/3} p^a,
\end{eqnarray}
for $\eta\gg 1$, with $q_a$ the total $\D2$ charge. 

We can use the expansion of the metric and the gauge fields at infinity to compute the Virasoro charges. After removing the contribution of the $U(1)$ and $SU(2)_R$ currents to the stress tensor \cite{deBoer:2008fk}, we find 
\begin{eqnarray}
\bar{L}_0-\frac{c_R}{24}&=&0,\\
L_0-\frac{c_L}{24}&=&-(q_0-\frac{1}{2}D^{ab}q_aq_b),
\end{eqnarray}
with $q_0,q_a$ the total charges. These represent the contributions to the stress tensor coming purely from the gravitational sector. For the charge configuration (\ref{Gamma6}) and (\ref{Gamma6bar}) this gives
\begin{equation}
q_0-\frac{1}{2}D^{ab}q_aq_b=\frac{p^3}{24}+\frac{c_2\cdot p}{24}-\frac{p\cdot(\beta+\bar{\beta})}{2}+n+\bar{n}-\frac{1}{2}D^{ab}\Delta\beta_a\Delta\beta_b,
\end{equation}
with $\Delta\beta_a=\beta_a-\bar{\beta}_a$. Since $q_0-\frac{1}{2}D^{ab}q_aq_b>0$ is the polarity, we see that these configurations indeed map to the polar states in the $\CFT_2$.

Let us now consider the limit of charges $p^a\rightarrow \lambda p^a$ with large $\lambda$ while keeping fixed the fluxes $\beta,\bar{\beta}$ and $n,\bar{n}$ in (\ref{Gamma6}) and (\ref{Gamma6bar}). In particular, this ensures that the individual center black hole charges are kept large, which is necessary for the supergravity solution to be valid.  The harmonic functions (\ref{harmonic functions}) split into a  term proportional to the smooth fluxes $p$ and a term coming from the contribution of the ideal sheaves fluxes $\beta,\bar{\beta}$ and $n,\bar{n}$, which are parametrically smaller of order $1/\lambda^2, 1/\lambda^3$ respectively. Since in the absence of the singular fluxes the geometry is global $AdS_3\times S^2$, we can write the full geometry as global $AdS_3\times S^2$ plus corrections, that is,
\begin{equation}
ds^2=ds^2_{AdS_3\times S^2}+\delta g_{\mu\nu},
\end{equation} 
where $\delta g_{\mu\nu}$  is a function of the singular fluxes $\beta,\bar{\beta}$ and $n,\bar{n}$ and thus of order $ \mathcal{O}(1/\lambda^3,1/\lambda^2)$. We can proceed similarly for the gauge fields and scalars
\begin{eqnarray}
&&A^{a}_{5D}= -p^a \cos\theta (d\phi+\tilde{A})+\delta A^a,\\
&&M^a= U^{-1/3} p^a+\delta \sigma^a,
\end{eqnarray}
with $\delta A^a$ and $\delta M^a$ of order $\mathcal{O}(1/\lambda,1/\lambda^2)$. Moreover, near the boundary $r=e^{\eta}\sim \infty$ the perturbations $\delta g_{\mu\nu}$ and $\delta A^a$ are of order $\mathcal{O}(r^0)$ while $\delta M^a$ is of order $\mathcal{O}(1/r)$ and thus they are normalizable. We can identify these perturbations as coming from the backreaction of the fields dual to the chiral primary states described in section \S\ref{sec polar states}.

\subsection{Purely fluxed solutions from Gauge-Gravitational Chern-Simons}

 In this section, we set the singular fluxes to zero and consider the case of charges induced by  mixed gauge-gravitational Chern-Simons terms, which are proportional to $c_2$, the second Chern-class of the Calabi-Yau. We saw previously  that the decoupling of the two-center solution, after uplift to five dimensions,  gave a geometry that was asymptotically $AdS_3\times S^2$. In this section we try a different approach. Instead of solving for the backreacted four dimensional solution and then uplift, we consider  the problem directly in five dimensional supergravity in the presence of higher derivatives terms, given by the supersymmetrization of gauge-gravitational Chern-Simons terms. We find that the theory admits global $AdS_3\times S^2$ with same quantum numbers as the asymptotically AdS solutions.

The Ramond couplings (\ref{Ramond couplings}) must be supplemented with the terms $\int R\wedge R\wedge A_3$ and $\int R\wedge R\wedge F\wedge A_1 $, where $R$ is the curvature two-form. The two center charge configuration in the absence of the singular fluxes is
\begin{eqnarray}
\Gamma_6=&&e^{p/2}(1+c_2(X)/24)\nonumber\\
=&&\left(1,\frac{p^a}{2},\frac{1}{8}D_{abc}p^bp^c+\frac{c_{2a}}{24}, \frac{p^3}{48}+\frac{c_{2}\cdot p}{48}\right),\\
{}\nonumber\\
\Gamma_{\bar{6}}=&&-e^{-p/2}(1+c_2(X)/24)\nonumber\\
=&&\left(-1,\frac{p^a}{2},-\frac{1}{8}D_{abc}p^bp^c-\frac{c_{2a}}{24}, \frac{p^3}{48}+\frac{c_{2}\cdot p}{48}\right),
\end{eqnarray}
 with $c_2(X)$ the second Chern-class (tangent bundle) of Calabi-Yau $X$.  As explained in the previous section, the solution that we find in two derivative supergravity after decoupling limit, is a geometry which is asymptotically $AdS_3\times S^2$, and carries total angular momentum
\begin{equation}\label{total ang mom higher derivative}
J=\frac{1}{2}\langle \Gamma_6,\Gamma_{\bar{6}}\rangle=\frac{p^3}{12}+\frac{c_{2}\cdot p}{24}.
\end{equation}

The decoupled geometry is not a solution of the full equations of motion because the four dimensional multi-center geometries, described in \cite{deBoer:2008fk,Denef:2000nb}, are solutions of two derivative supergravity and thus  higher derivative corrections are not taken into account. To correctly describe the exact solution we need to consider the problem in the presence of the higher derivatives terms, which arise from the reduction of the eighth-derivative $C\wedge I_8(R)$ term in M-theory. After compactification on the Calabi-Yau this gives rise to gauge-gravitational Chern-Simons terms plus their supersymmetric completion. This includes terms such as
\begin{equation}
c_{2a}\int A^a\wedge R\wedge R ,\;c_{2a}\int M^a R\wedge\star R,
\end{equation}
with $A^a$ the five dimensional gauge field and $M^a$ the real scalar field that sits in the vector multiplet.
This problem was studied in \cite{Castro:2008ne,Castro:2007hc} by considering five dimensional off-shell supergravity with a mixed gauge-gravitational Chern-Simons term. The solution found is in fact the near-horizon geometry of a black ring, which has the form $AdS_2\times S^1\times S^2$. But after a simple analytic continuation, that we describe in further detail in section \S \ref{sec subleading Bessels}, we can bring the metric to  global $AdS_3\times S^2$. A few properties of the solution are the following. The physical size $L^2$ is given by
\begin{equation}\label{U with c_2 correction}
L^2=\left(\frac{p^3}{6}+\frac{c_2\cdot p}{12}\right)^{2/3},
\end{equation}
in contrast with the two derivative result $L^2=(p^3/6)^{2/3}$ (\ref{approximate AdS3}). The difference is nevertheless negligible in the limit of $p\gg 1$, which is when the two derivative solution is justified. Note that in this case $L^2$ agrees precisely with the decoupling two center distance $r_{6\bar{6}}$ (\ref{r66}); this fact will become relevant later. Furthermore, the attractor equations for the $U(1)$ gauge fields and the scalar fields are exactly
  \begin{eqnarray}
  &&A^{a}_{5D}= -p^a \cos\theta d\phi,\\
  &&M^a= U^{-1/3} p^a,
  \end{eqnarray}
which contrasts with the approximate solutions in the two derivative theory.

So far the sphere is not twisted and thus the gravitini are antiperiodic along the spatial circle, which makes this a solution in the NS sector. Since our interest is in the R sector we need to turn on a non-trivial connection on the sphere, such as (\ref{R gauge field}), so that its holonomy around the contractible cycle effectively changes the gravitini periodicities. This has an effect on the total angular momentum carried by the solution. This is easier to see from an holographic point of view.  To do that, we reduce the theory on the sphere keeping its isometries gauged, which gives rise to three dimensional $SU(2)_R$ Chern-Simons terms \cite{Hansen:2006wu} with level $k_R$ \cite{Kraus:2005vz}
\begin{equation}
k_R=\frac{p^3}{6}+\frac{c_2\cdot p}{12}.
\end{equation}
From the three dimensional point of view the twisting connection $A=d\sigma-d\tau=d\bar{z}$ on the sphere, with $z$ a right-moving coordinate, induces a current $J_{R\bar{z}}=ik_R A_{\bar{z}}/2$ \cite{Hansen:2006wu}. The angular momentum is the $R$-charge of the solution, that is,
\begin{equation}
J_{R}^0=-\oint \frac{d\bar{z}}{2\pi i}J_{R\bar{z}}=-\frac{k_R}{2},
\end{equation}
in precise agreement with (\ref{total J}). Note also that the solution corresponds to a lowest $SU(2)_R$ weight state. In this case the ratio $J/J_{\text{max}}=1$ in (\ref{twisting gauge field}). The key difference is on the $SU(2)_R$ level, which suffers a correction due to the mixed Chern-Simons terms.

\subsection{Purely fluxed solutions from Ideal Sheaves}\label{sec Ideal sheaves}

In this section, we want to consider the purely fluxed $\D6-\aD6$ configurations directly from the M-theory point of view, without resorting to the four dimensional charge configuration and its supergravity solution. In particular we want to follow the intuition from the previous section and search for exact solutions to the full equations of motion that carry the same charges. Our goal in this section is to reproduce the sum over $\M2/\aM2$-branes on $\text{AdS}_3$ of the section \S \ref{sec polar states}, but now in terms of the eleven dimensional M-theory fields, such that we can interpret the degeneracy as an M-theory partition function. Since our main interest is the $\N=4$ theory we will only consider ideal sheaves with second Chern-classes. 

In section \S\ref{sec polar states}  we considered  configurations of M2 and anti-M2 branes wrapping holomorphic cycles on the Calabi-Yau and sitting at the origin of $AdS_3$ and at the north and south poles of $S^2$ respectively. The configuration carries M-theory flux $G\propto e_2\wedge p$ where $p$ is the flux along the Calabi-Yau and $e_2$ is the volume form of the sphere. The map between the fluxes $\beta,\bar{\beta}$ and the number of M2 and anti-M2 branes living on $AdS_3\times S^2\times X_{CY}$ is
\begin{equation}
\beta_{a}=q_a,\;\bar{\beta}_{a}=\bar{q}_a.
\end{equation}
It follows that the positions of the M2 and anti-M2 branes on $AdS_3\times S^2$ are consistent with the positions of the $\D6$ and $\aD6$ branes (\ref{oblate coords}).  Indeed, in the absence of M2 branes the geometry corresponds to the decoupling limit of a $\D6$ and $\aD6$ configuration with worldvolume flux $p$ after uplift to M-theory.

We have explained that turning fluxes on the $\D6$ brane is equivalent to turning on fluxes in M-theory. The question we want to answer is how the singular M-theory fluxes affect the full eleven dimensional geometry. At this point we should proceed carefully because we do not really know how to deal with singular gauge field configurations in the path integral. Therefore, our approach is mainly heuristic. A crucial aspect in the M2 brane construction was that the configuration on global $AdS_3\times S^2$ preserved the same set of supersymmetries independently on the number of M2 branes.  Inspired by this result we consider an ansatz for the exact geometry which we take to be global $AdS_3\times S^2$.

We start from the $AdS_3\times S^2$ ansatz and write the metric using the ambi-polar Eguchi-Hanson coordinates defined in (\ref{Eguchi-Hanson}), which depend on the parameters $q,a,k$. In this work we consider only smooth geometries and so we set $q=1$; $q$ maps to the number of $\D6$ branes, which is consistent with our problem. The parameter $a$ is the physical distance between the centers while $k$ parametrizes the harmonic functions. From the decoupling limit of  the multi-center geometry, the distance between the centers (\ref{r66}) is proportional to the symplectic charge inner product, which has the value
\begin{equation}\label{quantum size}
r_{6\bar{6}}= \frac{4}{R^2}\left(\frac{p^3}{6}+\frac{c_2\cdot p}{12}-(\beta+\bar{\beta})\cdot p\right).
\end{equation}
This is the charge product $\langle\Gamma_{6},\Gamma_{\bar{6}}\rangle$ for the charges (\ref{Gamma6}) and (\ref{Gamma6bar}).
Since we are turning on singular fluxes in M-theory, we can expect corrections to the harmonic functions (\ref{harmonic functions}) and also to the distance formula (\ref{r66}). Nevertheless the charge combination $\langle \Gamma_6,\Gamma_{\bar{6}}\rangle$ is a topological invariant and hence it is integer quantized. In view of this, it seems natural to assume that the distance between the centers remains unchanged.  Moreover, we expect the geometry to asymptote to the perturbative geometry (\ref{purely fluxed ads3}) when we take the fluxes $\beta,\bar{\beta}$ to be parametrically smaller than $p$. This means that the constant term in the harmonic function $V_0$ (\ref{Eguchi-Hanson}) must equal the corresponding term $h_0=R^{3/2}/4$ in the formula (\ref{harmonic functions}), so we have 
\begin{equation}
\frac{2 k^2}{a}=\frac{R^{3/2}}{4}.
\end{equation}
 Hence, using that $a=r_{6\bar{6}}$, we find
\begin{equation}\label{quantum corrected size}
k^2=\frac{1}{2 R^{1/2}}\langle \Gamma_6,\Gamma_{\bar{6}}\rangle.
\end{equation}
We thus see that both the parameters $a$ and $k$, that parametrize the full solution, depend only on the combination
\begin{equation}\label{c2 renormalization}
\hat{c}_2=c_2-12(\beta+\bar{\beta}),
\end{equation}
that appears in (\ref{quantum size}). This suggests that the effect of the singular fluxes is to renormalize the second Chern-class $c_2$ by a shift $-12(\beta+\bar{\beta})$. Assuming this renormalization we can easily determine other parameters of the theory such as the central charges and angular momentum. What we need to do is to reconsider the problem studied in the previous section but with the renormalized second Chern-class $\hat{c}_2$.

For example the effective three dimensional Chern-Simons theory that we obtain after reduction on the sphere contains  $SL(2,\mathbb{R})_L\times SL(2,\mathbb{R})_R\times SU(2)_R$ and abelian Chern-Simons terms. The levels for each gauge group are respectively $\tilde{k}_L,\tilde{k}_R$ and $k_R$. Furthermore, by supersymmetry we must have $\tilde{k}_R=k_R$.  Since we have $c_L=6\tilde{k}_L$ and $c_R=6\tilde{k}_R$,  using the values of the central charges with the renormalized $c_2$,  we find
\begin{equation}\label{quantum kr level}
\tilde{k}_R=k_R=\frac{p^3}{6}+\frac{\hat{c}_2\cdot p}{12},
\end{equation}
and
\begin{equation}
\tilde{k}_L=\frac{p^3}{6}+\frac{\hat{c}_2\cdot p}{6}.
\end{equation}
As explained in a previous occasion, to describe the R sector of the theory, the sphere must be twisted by the Wilson line (\ref{R gauge field}), to impose the correct boundary conditions on the gravitini. Following the derivation presented in the previous section, the geometry acquires angular momentum
\begin{equation}
J=\frac{k_R}{2}=\frac{p^3}{12}+\frac{\hat{c}_2\cdot p}{24}.
\end{equation}
This is in perfect agreement with the angular momentum formula (\ref{total J}) for the two center bound state, and it also agrees with the total angular momentum contribution due to the M2-branes (\ref{total angular momentum}) as described in section \S\ref{sec polar states}. 

Given that the size (\ref{quantum corrected size}) must be positive for the geometry to make sense, we must have 
\begin{equation}\label{condition on the size}
\frac{p^3}{6}+\frac{c_2\cdot p}{12}-(\beta+\bar{\beta})\cdot p >0.
\end{equation}
As we discuss later, restricting the fluxes to $\beta,\bar{\beta}\geq 0$, guarantees that the partition function agrees with the Cardy limit of the CFT. If it was not the case then there would be contributions to the path integral overwhelming the area formula predicted from microscopics (\ref{subleading saddles}).   Therefore for $\beta,\bar{\beta}>0$ we obtain an upper bound on the possible amount of fluxes, which in turn leads to a \emph{finite number of geometries}. This bound on the spectrum was also observed in a similar context \cite{deBoer:2009un}.

The bound on the number of geometries is precisely the bound imposed by the stringy exclusion principle \cite{Maldacena:1998bw}.  The principle was introduced in order for the number of chiral primaries in the $\CFT_2$ to match the spectrum of Kaluza-Klein fields on $AdS_3$. The reason is that while from the $\CFT$ the number of chiral primaries follows from fermi statistics and is therefore finite, from the bulk point of view the Kaluza-Klein fields have free bosonic excitations and thus with no limit in their particle number. The exclusion principle gives an unitarity condition that is non-perturbative in nature.  In terms of quantum numbers, the exclusion principle  translates into a bound on the R-charge carried by the fields excitations on top of $AdS_3$. For example, in the $\M2$ brane picture it implies that $\frac{1}{2}(q+\bar{q})\cdot p<\frac{p^3}{12}+\frac{c_2\cdot p}{24}$ where $q,\bar{q}$ are the number of $\M2$ and $\aM2$ respectively, in agreement with the bound (\ref{condition on the size}). 

At this point the renormalization (\ref{c2 renormalization}) is only a conjecture, which is very difficult to show given the nature of the solution. Nevertheless, we can already provide with preliminary evidence,  by giving an alternative derivation for the $SU(2)_R$ level (\ref{quantum kr level}), and then of $\tilde{k}_R$ by supersymmetry. The idea is to determine the coefficient of the $SU(2)_R$ Chern-Simons terms in three dimensions starting directly from the eleven dimensional action. We follow closely \cite{Hansen:2006wu,Freed:1998tg}. We write the M-theory four form flux as 
\begin{equation}
G=e_2(A)\wedge F,
\end{equation} 
where $e_2$ is the volume form of the sphere and contains the effect of gauging the isometries, that is, it depends explicitly on the $SU(2)_R$ connections $A$, which have legs on the $AdS_3$ directions. We decompose the flux $F$ in the  smooth component $p=p^a\omega_a$ with $\omega_a\in H^2(X,\mathbb{Z})$ and the singular term $\mathcal{F}$ , that is $F=p+\F$. The ideal sheaf flux $\F$ has zero first Chern-class and
\begin{eqnarray}\label{Chern classes}
c_2(\F)=\frac{1}{2}\int_{\alpha^a} \F\wedge \F=-(\beta_a+\bar{\beta}_a),\;\;c_3(\F)=\frac{1}{6}\int_X \F\wedge \F\wedge \F=n+\bar{n},
\end{eqnarray}
where $\alpha^a\in H_4(X,\mathbb{Z})$ and $\beta_a,\bar{\beta}_a,n,\bar{n}\in \mathbb{Z}$. The contributions $\beta,n$ and $\bar{\beta},\bar{n}$ are due to the $\D6$ and $\aD6$ respectively. The expressions (\ref{Chern classes}) have to be taken with care since the fluxes $\F$ are singular and require an appropriate regularization. For our purpose the Chern-classes $c_2(\F)$ and $c_3(\F)$ are well defined and given by the values (\ref{Chern classes}). The $SU(2)_R$ Chern-Simons coupling is implicitly related to a lack of gauge invariance of the action and thus we can focus only on the Chern-Simons terms in M-theory. We compute
\begin{equation}
\frac{1}{6}\int C\wedge G\wedge G=\frac{1}{6}\int_{AdS_3\times S^2} e^{(0)}_1\wedge e_2\wedge e_2\int_X F\wedge F\wedge F,
\end{equation}
 with $e^{(0)}_1$ defined locally by the equation $de^{(0)}_1 =e_2$. The first integral on the RHS gives the "descent" of the Pontryagin class of the sphere bundle, that is, 
 \begin{equation}
 \int_{S^2} e^{(0)}_1\wedge e_2\wedge e_2=-\frac{1}{2(2\pi)^2} \text{Tr}\left(AdA+\frac{2}{3}A^3\right),
 \end{equation}
 with $A$ the $SU(2)_R$ connection.  In addition one has
 \begin{eqnarray}
 \frac{1}{6}\int_X F\wedge F\wedge F&=&\frac{p^3}{6}+p^a\int \omega_a\wedge c_2(\F)+\int c_3(\F)\\
 &=&\frac{p^3}{6}-p\cdot (\beta +\bar{\beta})+n+\bar{n}.
 \end{eqnarray}
  Furthermore we have the contribution from the eighth derivative term $ C\wedge I_8(R)$ in M-theory. This term is easier to compute. Since it depends linearly on $C$ only the first Chern-class of $\F$ can contribute but that is zero by definition. Though, it  contributes with a term proportional to  $c_2\cdot p$ \cite{Kraus:2005vz}. The final contribution is therefore
 \begin{eqnarray}
 &&\frac{1}{6}\int C\wedge G\wedge G+\int C\wedge I_8(R)\propto\frac{1}{4\pi}\left(\frac{p^3}{6}+\frac{c_2\cdot p}{12}-p\cdot (\beta+ \bar{\beta})+n+\bar{n}\right) \int \text{Tr}(AdA+\frac{2}{3}A^3)\nonumber.
  \end{eqnarray}
The overall normalization is fixed by setting the fluxes to zero. The coefficient of the Chern-Simons term agrees precisely with the level $k_R$ (\ref{quantum kr level}). We have kept the dependence on the fluxes $n,\bar{n}$ arbitrary to note that the Chern-Simons coefficient is proportional to the symplectic charge product $\langle \Gamma_6,\Gamma_{\bar{6}}\rangle$. This is in agreement with our expectations since as we have shown, the R-charge, and thus the angular momentum, is proportional to the Chern-Simons level.

 We now return to the backreacted geometry. There is an important comment regarding the definition of the five and four dimensional Newton's constants, which will be important later. 
   So far we have been following the conventions used in \cite{deBoer:2008fk} which fix the five dimensional Einstein-Hilbert (EH) term as
 \begin{equation}\label{EH 5d}
 \int d^5x\sqrt{g_5} R^{(5)},
 \end{equation} 
 where $g_5$ and $R^{(5)}$ correspond to the five dimensional metric and Ricci scalars respectively. 
 In the five dimensional off-shell theory, the EH term contains, in contrast to the on-shell version (\ref{EH 5d}), a conformal coupling to vector-multiplet scalars $M$ as
 \begin{equation}\label{conformal coupling}
 \int d^5x\sqrt{g_5} D_{abc}M^aM^bM^c R^{(5)}.
 \end{equation}
  The attractor equations impose $LM^a=p^a$, with $L^2$ the conformal factor of the metric. After reducing on the circle with radius $L/\phi^0$, we obtain the four dimensional EH term $\int d^4x \sqrt{g_4}\frac{L^{-2}p^3}{\phi^0}R$. Instead we want to have 
   \begin{equation}\label{EH choice}
  \int d^4x \sqrt{g_4}\frac{1}{\phi^0}R,
  \end{equation}
in four dimensions. We keep the factor $1/\phi^0$ such that the conformal factor of the four dimensional metric remains constant in the problem; thus we need $L^2\propto p^3$. If we include higher derivatives then we must impose $L^2\propto p^3/6+c_2\cdot p/12$, which is the result found in \cite{Castro:2008ne}. 
 
 In order to establish a map between the on-shell and the off-shell theory, we write the five dimensional EH in terms of the unit size metric $g^{(0)}_5$, which gives $\sqrt{g^{(0)}_5}  \langle\Gamma_{6},\Gamma_{\bar{6}}\rangle R^{(0)}$. Following the same logic outlined above, we find by dimensional reduction that the  size $L^2$ is precisely $\langle\Gamma_{6},\Gamma_{\bar{6}}\rangle$. 
 
 We can also show that this is consistent with the four dimensional off-shell superconformal gravity and the renormalization of $c_2$.  Following \cite{Mohaupt:2000mj}, the EH term is given by
 \begin{equation}
 \int d^4x \sqrt{g_4}i(X^I\bar{F}_I-\bar{X}^IF_I)R,
 \end{equation}
 where the combination $i(X^I\bar{F}_I-\bar{X}^IF_I)$ is the Kahler potential $e^{-K}$. Assuming the renormalization induced by the fluxes, the prepotential $F(X)$ is
 \begin{equation}
 F(X)=-\frac{1}{6}D_{abc}\frac{X^aX^bX^c}{X^0}+\frac{\hat{c}_{2a}}{24}\frac{X^a}{X^0},
 \end{equation}
 with $\hat{c}_2$  (\ref{c2 renormalization}). Using the on-shell  attractor solutions $LX^0=\phi^0$ and $LX^a=\phi^a+ip^a$, the Kahler potential becomes
 \begin{equation}
 e^{-K}=\frac{L^{-2}}{\phi^0}\left(\frac{p^3}{6}+\frac{\hat{c}_2\cdot p}{12}\right).
 \end{equation}
Therefore to obey the choice (\ref{EH choice}) we must have 
 \begin{equation}
 L^2=\frac{p^3}{6}+\frac{\hat{c}_2\cdot p}{12}.
 \end{equation}

 \section{Localization and Non-perturbative Corrections}\label{sec Loc}
 
 In this section, we consider supersymmetric localization at the level of the five dimensional theory. We follow closely the four dimensional solution studied in \cite{Dabholkar:2010uh,Gupta:2012cy}. Different aspects of this computation such as boundary conditions or the choice of localization supercharge were discussed previously in the works \cite{Gomes:2013cca,Dabholkar:2014ema}, which we review along the way. 

The relation between the quantum entropy functional on $AdS_2$ and the partition function on $AdS_3\simeq AdS_2\times S^1$ was discussed in \cite{Gupta:2008ki}. The essential observation is that the ground states of the conformal quantum mechanics dual to the theory on $AdS_2$ map to a chiral half of the $\CFT_2$, which is dual to the theory on $AdS_3$. This gives a simple way to relate the microscopic index computations to the black hole degeneracy \cite{Sen:2009vz, Dabholkar:2010rm}.

According to our proposal, the full partition function will be the sum of different contributions, each coming from a solution parametrized by $\beta,\bar{\beta}$ , that is
\begin{equation}
Z_{AdS_2\times S^1\times S^2}=\sum_{\beta,\bar{\beta}}\int D[\Phi]e^{-S_{\text{E}}[\Phi,\beta,\bar{\beta}]},
\end{equation}
where $D[\Phi]$ denotes a measure for all the fields in five dimensional supergravity, and $S_{\text{E}}[\Phi,\beta,\bar{\beta}]$ is the euclidean action whose Lagrangian depends explicitly on the values of the fluxes $\beta,\bar{\beta}$. For each of the geometries parametrized by $\beta,\bar{\beta}$, we perform localization. 

As we explain shortly, one also needs to consider the contribution of $U(1)$ connections that have a delta function singularity at the origin of $AdS_2\times S^1$. This was also pointed out in \cite{Dijkgraaf:2000fq}. The field is pure gauge but it is not well defined everywhere.  The reason to include them is motivated from the fact that in the decoupled geometry, the M2 brane total charge gives rise to a gauge transformation $\sim D^{ab}(\beta-\bar{\beta})_b dy$ in the five dimensional gauge field (\ref{attractor fields}), with $y$ parameterizing the spatial circle that is contractible. From a physical point of view, this gauge transformation generates a spectral flow transformation for the $U(1)$ currents in the CFT. In the path integral we use a regularization scheme to avoid the singularity at the origin, and show that is consistent with the localization procedure. The computation ends up depending only on the abelian Chern-Simons terms, as expected for the contribution of a large gauge transformation.

 \subsection{5D Supersymmetric Localization}\label{sec 5.1}
 
 We start by reviewing the localization computation of $\N=2$ supergravity on $AdS_2\times S^2$  \cite{Dabholkar:2010uh}. The four dimensional Lagrangian is constructed using the off-shell superconformal formalism, for which a good review is \cite{Mohaupt:2000mj}. The part of the Lagrangian that is most relevant for the computation is based on the holomorphic prepotential
 \begin{equation}\label{N2 prepotential}
 F(X,\hat{A})=\frac{1}{6}D_{abc}\frac{X^aX^bX^c}{X^0}+g\left(\frac{X}{X^0}\right)\hat{A},
 \end{equation}
 where $X$ are the complex scalar fields in the vector-multiplets and $\hat{A}=(T^{-})^2$ is the bottom component of the chiral multiplet $\mathbf{W}^2$, with $\mathbf{W}$ the Weyl superfield; in the on-shell theory $T^{-}$ becomes the graviphoton field. Besides the usual Einstein-Hilbert and Maxwell terms, the Lagrangian contains in addition higher derivative terms parametrized by the function $g(X/X^0)$. It determines the coupling of the vector-multiplets to the square of the Weyl tensor as
 \begin{equation}
 \sim g\left(\frac{X}{X^0}\right)C_{abcd}C^{abcd}+\text{h.c.},
 \end{equation}
 with $C_{abcd}$ the Weyl tensor. Therefore at the on-shell level, we have two derivative supergravity with Weyl square higher derivative corrections. Later we show how to introduce Gauss-Bonnet type of corrections, which are known to contribute to black hole entropy \cite{Sen:2005iz}.
 
 The original localization solutions of \cite{Dabholkar:2010uh}, solve the BPS equation $Q\Psi=0$, with $Q$ a real supercharge that squares to a self-dual\footnote{On $AdS_2\times S^2$ we have isometries $L$ and $J$, respectively rotations on $AdS_2$ and $S^2$. The supercharge $Q$ obeys $Q^2=L-J$. } $U(1)$ isometry of $AdS_2\times S^2$, and $\Psi$ are the vector-multiplet fermions. The remaining equations for the other fields, including the Weyl multiplet, were solved in \cite{Gupta:2012cy}\footnote{There is an important caveat in this computation (which is not fault of the authors). The reason is that in supergravity we can not really define a supercharge as we usually do in supersymmetric field theory. So the construction of the localization deformation of the sort $QV$, which includes all the supergravity fields and is invariant under local supersymmetry, is still unknown. Part of this work provides steps in that direction. }. In particular the localization equations for the Weyl multiplet imply that the four dimensional metric must be of the form $AdS_2\times S^2$, with equal sizes for $AdS_2$ and the sphere. Since the theory is off-shell the solutions are universal and thus independent of the particular details of the Lagrangian. The crucial result of the localization computation is that the vector-multiplet scalars have non-trivial radial profiles, in contrast to the constant on-shell values. In fact, the solution does not obey the equations of motion. This happens because the localization action contains flat directions, which allows the scalar fields $X$ to go off-shell at the expense of turning on the auxiliary fields $Y$. More precisely, the solution is 
 \begin{equation}\label{localization solutions 4D}
 X=X^*+\frac{C}{r},\;Y=\frac{C}{r^2},
 \end{equation}
 where $r\in [1,\infty[$ is the radial coordinate of $AdS_2$ and $X^*$ is the attractor value of the scalar, which is constant. All the other fields remain fixed to their attractor background. Moreover, the solutions are  parametrized by constants $C^0,C^a$, with $a=1\ldots n_V$, with $n_V$ the number of vector multiplets \footnote{The Weyl multiplet does not contain the graviphoton gauge field, which requires to add the compensator vector-multiplet, which contains the scalar $X^0$.}.  
 
 Given the solutions to the localization equations we need to determine their contribution to the physical action. After removing IR divergences \footnote{Integrating on $AdS_2$ leads to infinite volume divergences, which requires introducing a cutoff at finite radius.}, following the prescription in \cite{Sen:2008vm}, we obtain the renormalized action
 \begin{equation}\label{ren action 5D}
 \text{Ren}(S)=-\pi q_I\phi^I+4\pi \text{Im}F(\phi+ip),
 \end{equation}
where $\phi+ip$ is the value of the scalar $X$ at the origin of $AdS_2$, with $\phi\sim \text{Re}(X^*)+C$, which is free to fluctuate. The function $F$ is the holomorphic prepotential (\ref{N2 prepotential}). To arrive at the expression (\ref{ren action 5D}) one needs to use, at an intermediate step, the on-shell equations of motion
  \begin{equation}\label{attractor equations}
  q_I=4\text{Im}\left(\frac{\partial F}{\partial X^I}\right)|_{X^*}.
  \end{equation}
It ensures that the saddle point equations that we obtain from varying the renormalized action against $\phi$ are consistent with the attractor equations of motion. The localization integral is thus given by
\begin{equation}\label{4D loc integral}
Z_{AdS_2\times S^2}\sim \int \prod_{I=0}^{n_V} d\phi^I\exp{\left[-\pi q_I\phi^I+4\pi \text{Im}F(\phi+ip)\right]}.
\end{equation}
The symbol $\sim$ means that we are ignoring a measure factor. A measure  was proposed in \cite{Murthy:2015yfa-1,Gupta:2015gga}, which includes the contribution of the localization one-loop determinants.

As an example, lets evaluate the integral (\ref{4D loc integral}) for the theory on $K3\times T^2$. The prepotential is
\begin{equation}
F(X,\hat{A})=\frac{X^1}{X^0}C_{ab}X^aX^b+\ln \eta^{24}\left(\frac{X^1}{X^0}\right)\hat{A},
\end{equation} 
with $\eta^{24}(t)$ the worldsheet instanton partition function. We have used the fact that $D_{1ab}=C_{ab}$ with the other components zero. The integral (\ref{4D loc integral}) becomes
\begin{equation}
Z_{AdS_2\times S^2}\sim \int  d\tau_1d\tau_2 \exp{\left[\pi \frac{|Q+\tau P|^2}{\tau_2}-\ln |\eta^{24}(\tau_1+i\tau_2)|^2\right]},\;\tau_1+i\tau_2=\frac{\phi^1+ip^1}{\phi^0},
\end{equation}
 after evaluating the $\phi^a$ integrals, with $a=2\ldots n_V$, which are gaussian. We have defined $|Q+\tau P|^2\equiv Q^2+2\tau_1 Q.P+ |\tau|P^2$, with $Q^2,P^2,Q.P$ the T-duality invariants, which are quadratic in the charges $q,p$. Comparing with the microscopic answer (\ref{Siegel deg}) described in section \S\ref{deg 1/4 dyons}, we find agreement up to the measure factor that contains a derivative of $\ln |\eta^{24}|^2$.
 
Now we turn gears to the localization computation in five dimensions. In contrast to four dimensions, the coupling between the Weyl multiplet and the vector-multiplets are determined completely by the constants $D_{abc}$ and $c_2a$. The first parametrizes the different couplings in the two derivative Lagrangian, while the second constant parametrizes the supersymmetrization of the gauge-gravitational Chern-Simons term
\begin{equation}
c_{2a}\int A^a\wedge R\wedge R,
\end{equation}
where $A^a$ is the gauge field in the vector-multiplet, and $R$ is the curvature two form.

 The Lagrangian contains the following terms
\begin{equation}\label{5D lagrangian}
\mathcal{L}_{5D}=\mathcal{L}_{VVV}+\mathcal{L}_{\text{hyper}}+\mathcal{L}_{VWW}.
\end{equation}
 $\mathcal{L}_{VVV}$ is the two derivative  Lagrangian, which is cubic in the vector-multiplets. It contains the coupling of the vectors to the Weyl-multiplet, in particular to the Einstein-Hilbert term, and abelian Chern-Simons terms of the form $\int D_{abc}A^a\wedge F^b\wedge F^c$. $\mathcal{L}_{\text{hyper}}$ is the hypermultiplet lagrangian and $\mathcal{L}_{VWW}$ contains the supersymmetrization of the gauge-gravitational Chern-Simons term, which is linear in the vector-multiplet fields and quadratic in the Weyl multiplet fields.

Supersymmetric localization of the five dimensional theory on $AdS_2\times S^1\times S^2$ goes along the lines described in \cite{Gomes:2013cca}. The off-shell reduction described in \cite{Banerjee:2011ts} plays a very important role and we use it extensively here. We can show that the five dimensional localization equations \cite{Gomes:2013cca} do not allow the fields to depend on the circle coordinate. Therefore, we can write the five dimensional equations in terms of the four dimensional ones, using the off-shell reduction of \cite{Banerjee:2011ts}. As a consequence, the five dimensional solutions are an uplift of the four dimensional.

The uplift goes as follows. Since the four dimensional scalar $X^0$ maps to the radius of the circle and the four dimensional metric is fixed by the localization equations to be of the form $AdS_2\times S^2$, the five dimensional metric becomes 
\begin{equation}\label{metric localization}
ds^2=\vartheta\left[(r^2-1)d\tau^2+\frac{dr^2}{r^2-1}+\frac{1}{((\phi^0)^*+C^0/r)^2}(du+i(\phi^0)^*(r-1)d\tau)^2\right] +\vartheta ds^2_{S^2},
\end{equation}
where $(\phi^0)^*=\vartheta^{1/2}\text{Re}(X^0)^*$ is the on-shell value of $\phi^0$. The factor $\vartheta$ is a constant free parameter, since the theory is Weyl invariant by construction. Physically we need to use a gauge fixing condition, which makes $\vartheta$ a function of the charges. This was explained at the end of section  \S\ref{sec Ideal sheaves}. Similarly, the five dimensional gauge fields are not fixed by the localization equations. The scalars $X^a$ map to the Wilson lines of the five dimensional gauge field, and so we obtain
\begin{equation}\label{localization 5D gauge field}
A_{5D}^a=-2\frac{(\phi^a)^*+\frac{C^a}{r}}{(\phi^0)^*+\frac{C^0}{r}}\left(du+i(\phi^0)^*(r-1)d\tau\right)+(A^a)^*_{4D},
\end{equation}
where $(\phi^a)^*=\vartheta^{1/2}\text{Re}(X^a)^*$ are the on-shell values of $\phi^a$ and $(A^a)^*_{4D}$ is the on-shell four dimensional gauge field. In addition, there is a map between the four and five dimensional auxiliary fields, which we refer the reader to \cite{Banerjee:2011ts,Gomes:2013cca} for more details. For $C^0=C^a=0$ the metric becomes the locally $AdS_3$ metric (\ref{AdS3}) and the five dimensional gauge fields become flat, in agreement with the five dimensional attractor equations \cite{deWit:2009de}.

Before moving to the computation of the renormalized action, we discuss the boundary terms. These are necessary to ensure a well defined variational problem. At the boundary, the gauge fields have the form
\begin{equation}
A_{5D}^a\simeq A_f^a+p^aA_{\text{Dirac}},\;r\sim \infty,
\end{equation} 
with $A_f$ a flat connection on $AdS_2\times S^1$ and $A_{\text{Dirac}}$ is the Dirac monopole gauge field, which is defined locally. In order to compute the boundary terms it is enough to consider the abelian Chern-Simons terms in $\mathcal{L}_{VVV}$. In contrast, the Maxwell terms give a contribution proportional to $dA_f$ that vanishes at the boundary. The boundary terms are thus of Chern-Simons type as discussed in \cite{Elitzur:1989nr}. Moreover, we need to include a Wilson line for the Kaluza-Klein gauge field, since we keep fixed the electric fields in the microcanonical ensemble \cite{Sen:2008vm}. Details about their computation can be found in \cite{Dabholkar:2014ema}. We will denote these boundary terms generically by $S_{\text{Bnd}}$.

The renormalized action in five dimensions consists of the following terms
\begin{equation}
\text{Ren }(S_{5D})=S_{\text{Bnd}}\left(A_{5D}^a(C^a,C^0)\right)+S_{\text{bulk}}\left(g_{\mu\nu}(C^0),A_{5D}^a(C^a,C^0)\right)+S_{\text{ct}}.
\end{equation}
 $S_{\text{bulk}}$ is the bulk action based on the Lagrangian (\ref{5D lagrangian}) and $S_{\text{ct}}$ are boundary counter terms necessary to remove IR divergences. Computing  the action above is a very complicated task, because the various fields, including the metric, have non-trivial radial profiles. Instead of performing directly the five dimensional computation, we can simplify the problem by reducing the different terms to four dimensions and then use the results of \cite{Dabholkar:2010uh} described in the beginning of this section. The reduction is possible because the fields do not carry any dependence on the circle coordinate.  Nevertheless, this is still a complicated task because the five dimensional Lagrangian contains a large amount of terms. Fortunately, such problem has been object of study in recent years \cite{Butter:2013lta,Butter:2014iwa,Banerjee:2011ts}. The main conclusion is that  under the reduction, the different four dimensional terms can be assembled in four dimensional supersymmetric invariants. 
 
 The way the reduction works is succinctly the following. The two derivative Lagrangian $\mathcal{L}_{VVV}$ gives rise to the four dimensional Lagrangian based on the holomorphic prepotential
\begin{equation}
F(X)=\frac{1}{6}D_{abc}\frac{X^aX^bX^c}{X^0}.
\end{equation}
On the other hand, the reduction of the higher derivative $\mathcal{L}_{VWW}$, gives rise to two different supersymmetric invariants. The first is the supersymmetrization of the  Weyl squared tensor term, which together with the reduction of $\mathcal{L}_{VVV}$, can both  be written in terms of a single  supersymmetric invariant based on the holomorphic prepotential
\begin{equation}\label{classical prepotential reduction}
F(X,\hat{A})=D_{abc}\frac{X^aX^bX^c}{X^0}+c_{2a}\frac{X^a}{X^0}\hat{A}.
\end{equation} 
This is the one-loop $\N=2$ prepotential (\ref{top free energy}), after neglecting the contribution of the world-sheet instantons. Since it depends only on the geometry we call it classical prepotential. The second set of terms can be written in terms of a chiral superspace integral based on the non-linear Kinetic multiplet $\mathbb{T}(\ln \mathbf{X}^0)$ \cite{Butter:2013lta}. In superspace it has the form
\begin{equation}
ic_{2a}\int d^4x d^4\theta \frac{\mathbf{X}^a}{\mathbf{X}^0}\mathbb{T}(\ln \bar{\mathbf{X}}^0)\;+\text{h.c.},
\end{equation}
which is a particular case of the more general type of supersymmetric invariants
\begin{equation}\label{NL Kinetic multiplet}
\int d^4x d^4\theta \Phi'\mathbb{T}(\ln\bar{\Phi}_{\omega}),
\end{equation}
with  $\Phi'$ a chiral superfield of weyl weight zero and $\Phi_{\omega}$ a chiral superfield of weyl weight $\omega$. The Kinetic chiral multiplet  $\mathbb{T}(\ln\bar{\Phi})$, has non-linear supersymmetry transformations due to the anomalous transformation  of $\ln\bar{\Phi}_{\omega}$ under Weyl transformations. For $\omega=0$, $\ln \bar{\Phi}_{\omega=0}$ is a chiral superfield with well defined Weyl transformations. In this case the invariant (\ref{NL Kinetic multiplet}) falls under the category of supersymmetric invariants studied in \cite{deWit:2010za}. 

Following \cite{Butter:2013lta,Butter:2014iwa}, we can develop in components the Lagrangian density $\mathcal{L}_{NL}$ of the invariant (\ref{NL Kinetic multiplet}) as
\begin{align}
\label{NL action Components}
e^{-1} \mathcal{L}_{NL} =&\,
4\,\mathcal{D}^2 A'\,\mathcal{D}^2 \hat{\bar A}
+ 8\,\mathcal{D}^a A'\, \big[\mathcal{R}_{ab}
-\frac{1}{3} \mathcal R \,\eta_{ab}\big]\mathcal{D}^b \hat{\bar A}
+ C'\,\hat{\bar C}
\nonumber \\[.1ex]
&\,
- \mathcal{D}^\mu B'_{ij} \,\mathcal{D}_\mu \hat B^{ij}
+ (\frac{1}{6} \mathcal{R} +2\,D) \, B'_{ij} \hat B^{ij}
\nonumber\\[.1ex]
&\,
- \big[\varepsilon^{ik}\,B'_{ij} \,\hat F^{+\mu\nu} \,
R(\mathcal{V})_{\mu\nu}{}^{j}{}_{k}
+\varepsilon_{ik}\,\hat B^{ij}\,F'^{-\mu\nu} R(\mathcal{V})_{\mu\nu j}{}^k \big]
\nonumber\\[.1ex]
&\,
-8\, D\, \mathcal{D}^\mu A'\, \mathcal{D}_\mu \hat{\bar A}
+ \big(8\, \mathrm{i}\, R(A)_{\mu\nu}
+ 2\, T_\mu{}^{cij}\, T_{\nu cij}\big)
\mathcal{D}^\mu A' \,\mathcal{D}^\nu \hat{\bar A}  \nonumber\\[.1ex]
&\,
-\big[\varepsilon^{ij} \mathcal{D}^\mu T_{bc ij}
\mathcal{D}_\mu A' \,\hat F^{+bc}
+ \varepsilon_{ij} \mathcal{D}^\mu T_{bc}{}^{ij}
\mathcal{D}_\mu \hat{\bar A}\,F'^{-bc}\big] \nonumber\\[.1ex]
&\,
-4\big[\varepsilon^{ij} T^{\mu b}{}_{ij}\,\mathcal{D}_\mu A'
\,\mathcal{D}^c \hat F^{+}_{cb}
+ \varepsilon_{ij} T^{\mu bij}\,\mathcal{D}_\mu \hat{\bar A}
\,\mathcal{D}^c F'^{-}_{cb}\big] 
\nonumber\\[.1ex]
&\,
+ 8\, \mathcal{D}_a F'^{-ab}\, \mathcal{D}^c \hat F^+_{cb}
+ 4\,F'^{-ac}\, \hat F^+_{bc}\, \mathcal R_a{}^b
+\tfrac1{4} T_{ab}{}^{ij} \,T_{cdij} F'^{-ab} \hat F^{+cd} 
\nonumber\\[.1ex]
&\,
+\omega\,\Big\{ - \frac{2}{3}  \mathcal{D}^a A' \,\mathcal{D}_a
\mathcal{R} + 4  \mathcal{D}^a A'\, \mathcal{D}_a D 
-  T^{acij} T_{bc ij}\, \mathcal{D}^b \mathcal{D}_a A'
\nonumber\\[.1ex]
&\quad \qquad
- 2 \mathcal{D}^a F_{ab}'^- \,\mathcal{D}_c T^{cb}{}^{ij} \varepsilon_{ij}
+ \mathrm{i}\,  F'^{-ab} R(A)_{ad}^- \,T_b{}^{dij} \varepsilon_{ij}
+ F_{ab}^- T^{ab ij} \varepsilon_{ij} (\frac{1}{12} \mathcal{R} - \frac{1}{2} D)
\nonumber\\[.1ex]
&\quad\qquad
+ A' \,\big[\tfrac{2}{3} \mathcal{R}^2 - 2\, \mathcal{R}^{ab} \mathcal{R}_{ab} - 6\, D^2
+ 2 \, R(A)^{ab} R(A)_{ab} -  R(\mathcal{V})^{+ab}{}^i{}_j\, R(\mathcal{V})^+_{ab}{}^j{}_i
\nonumber \\
& \quad \qquad\qquad
+ \frac{1}{128}  T^{ab ij} T_{ab}{}^{kl} T^{cd}{}_{ij} T_{cd kl}
+  T^{ac ij} D_a D^b T_{bc ij}\big]\Big\} \,~ +\text{total derivatives},
\end{align}
where we used the notation $A,\Psi_i,B_{ij},F_{ab}^{-},\Lambda_i,C$ for the components of a chiral superfield $\Phi$. The prime variables correspond to the components of the chiral multiplet $\Phi'$ and the hat variables correspond to the components of $\ln\bar{\Phi}_{\omega}$; $e$ is the volume element. The remaining fields $D,T_{ab ij}, A_a,\mathcal{V}^i_{\mu j}$ are respectively the auxiliary scalar, antisymmetric tensor and R-symmetry vector field respectively, while $\mathcal{R}_{ab}$ is the Ricci tensor. These fields sit in the Weyl multiplet.

  For our problem we have $A'=A|_{X^a/X^0}$ and $\hat{\bar A}=A|_{\ln \bar{X}^0}$ and similarly for all the other components of the chiral multiplets. Developing the components of the chiral multiplets in terms of the vector-multiplet fields, we obtain a density $\mathcal{L}_{NL}$ that has a very complicated dependence on the fields. To compute its contribution on the localization solution we proceed as follows.  First we note that in the four dimensional attractor background we have \cite{Mohaupt:2000mj,Dabholkar:2010uh}
\begin{eqnarray}
D=0,\;f_{\mu}^{a}=0,\;b_{\mu}=0,\;A_a=0,\;\mathcal{V}_{\mu i}^{ j}=0\\
\mathcal{R}_a^{b}=\frac{1}{16}T^{-}_{ac}T^{+cb},\;\mathcal{R}_a^a=0,\;\mathcal{D}_cT_{ab}^{-}=\mathcal{D}_cT_{ab}^{+}=0,
\end{eqnarray}  
with all the other fermionic fields in the Weyl multiplet set to zero. This implies that the covariant derivative $\mathcal{D}_a$ and the superconformal invariant derivative $D_a$ become the usual covariant derivative with no dependence on the weight $\omega$. This is enough to show that the last two lines of (\ref{NL action Components}) vanish identically as noticed in \cite{Butter:2013lta}. Furthermore, the two lines
\begin{eqnarray}
&&\omega \Big\{- \frac{2}{3}  \mathcal{D}^a A' \,\mathcal{D}_a
\mathcal{R} + 4  \mathcal{D}^a A'\, \mathcal{D}_a D 
-  T^{acij} T_{bc ij}\, \mathcal{D}^b \mathcal{D}_a A'
\nonumber\\[.1ex]
&&\quad \qquad
- 2 \mathcal{D}^a F_{ab}'^- \,\mathcal{D}_c T^{cb}{}^{ij} \varepsilon_{ij}
+ \mathrm{i}\,  F'^{-ab} R(A)_{ad}^- \,T_b{}^{dij} \varepsilon_{ij}
+ F_{ab}^- T^{ab ij} \varepsilon_{ij} (\frac{1}{12} \mathcal{R} - \frac{1}{2} D)\Big\},
\nonumber
\end{eqnarray}
also vanish, except for the term $T^{acij} T_{bc ij}\, \mathcal{D}^b \mathcal{D}_a A'$ since $A'=X^a/X^0$ is not constant in the localization background. Nevertheless, that term can be replaced by $\mathcal{D}^b(T^{acij} T_{bc ij})\, \mathcal{D}_a A'$ after an integration by parts, which vanishes on the solution. Note that we are not taking into account the total derivatives in (\ref{NL action Components}), and so in practice that term is ambiguous \footnote{Total derivatives can nevertheless contribute to the renormalized action. However such contributions must be consistent with the non-renormalization theorems of \cite{deWit:2010za,Butter:2014iwa}, which states that the either the on-shell values of the BPS black hole entropy and the definition of the electric charges remain unaffected by adding the supersymmetric invariants based on the Kinetic multiplet. }.  On the other hand, the remaining  lines in (\ref{NL action Components}), those which do not come multiplied by $\omega$, give rise to terms that fall in the category of the D-type terms studied in \cite{deWit:2010za}. A few characteristic terms are \cite{Banerjee:2011ts}
\begin{align}
\label{D-term Kahler potential}
&\,
\frac{1}{4}\,\mathcal{H}_{IJ\bar K \bar L}
\big( F_{ab}^-{}^I\, F^{-ab\,J}
-\frac{1}{2} Y_{ij}{}^I\, Y^{ijJ} \big)
\big(F_{ab}^+{}^K \, F^{+ab\,L} -\frac{1}{2} Y^{ijK}\,
Y_{ij}{}^L  \big)
\nonumber\\[.5ex]
&\,-\Big\{ \mathcal{H}_{IJ\bar K}\big(
F^{-ab\,I}\, F_{ab}^{-\,J} -\frac{1}{2} Y^I_{ij}\, Y^{Jij})
\big( \Box_\mathrm{c} X^K
+ \frac{1}{8} F^{-\,K}_{ab}\, T^{ab kl}  \varepsilon_{kl}\big)
+\mathrm{h.c.}\Big\}   \displaybreak[0] \nonumber\\[.5ex]
&\,+\mathcal{H}_{I\bar J}\Big[ 4\big( \Box_\mathrm{c} \bar X^I + \frac{1}{8}
F_{ab}^{+\,I}\, T^{ab}{}_{ij} \varepsilon^{ij}\big)
\big( \Box_\mathrm{c}  X^J + \frac{1}{8} F_{ab}^{-\,J}\, T^{abij}
\varepsilon_{ij}\big) \nonumber\\
& \qquad\qquad +8\,\mathcal{D}_{a}F^{-\,abI\,}\,
\mathcal{D}_cF^{+c}{}_{b}{}^J   - \mathcal{D}_a Y_{ij}{}^I\,
\mathcal{D}^a Y^{ij\,J}
\nonumber\\
&\qquad\qquad + 8\,\mathcal{R}^{\mu\nu}\, \mathcal{D}_\mu X^I
\,\mathcal{D}_\nu \bar X^J \nonumber\\
&\qquad\qquad -\big[\varepsilon^{ik}\, Y_{ij}{}^I\,(F^{+ab\,J}
-\frac{1}{4} X^J T^{ab}{}_{lm}\varepsilon^{lm} )\, 
R(\mathcal{V})_{ab}{}^j{}_k +[\mathrm{h.c.}; I\leftrightarrow J]
\big]  \Big]\nonumber \\
&\, +\cdots \,,
\end{align}
with
\begin{equation}
\mathcal{H}(X,\bar{X})=\frac{i}{384}c_a\left(\frac{X^a}{X^0}\ln \bar{X}^0-\frac{\bar{X}^a}{\bar{X}^0}\ln X^0\right),
\end{equation}
the K\"{a}hler potential and $\mathcal{H}_{IJ\ldots}$ its derivatives. Plugging in the localization solution we still obtain a very complicated expression. Fortunately, this is precisely the problem studied in \cite{Murthy:2013xpa}, where it is shown that D-type terms of the form (\ref{D-term Kahler potential}) vanish identically on the localization solution.

In conclusion, only the reduction of $\mathcal{L}_{VVV}$ and the part of $\mathcal{L}_{VWW}$ that contains a coupling to a Weyl squared term survive once evaluated on the localization solution.  The resulting Lagrangian density is based on the holomorphic prepotential (\ref{classical prepotential reduction}). Therefore, the renormalized action on $AdS_2\times S^1\times S^2$ reduces exactly to the four dimensional one based on the classical prepotential. That is,
\begin{equation}
\text{Ren}(S_{5D})=-q_I\phi^I+4\pi\text{Im}F_{cl}(\phi+ip),
\end{equation}
with $F(X)_{cl}$ the classical prepotential (\ref{classical prepotential reduction}). Developing further this expression we obtain
\begin{equation}\label{Ren 5D}
\text{Ren}(S_{5D})=-\pi \hat{q}_0\phi^0 +\frac{\pi}{6}\frac{p^3+c_2\cdot p}{\phi^0}-\frac{\pi}{2\phi^0}D_{ab}(\phi^a+q^a\phi^0)(\phi^b+q^b\phi^0),
\end{equation} 
with $D_{ab}=D_{abc}p^c$, $c_2\cdot p= c_{2a}p^a$, and $\hat{q}_0=q_0-D_{ab}q^aq^b/2$. The parameter $p^3+c_2\cdot p$ can be identified with the central charge of the dual CFT \cite{Maldacena:1997de}, which is also proportional to the $SL(2,\mathbb{R})_L$ Chern-Simons level of the three dimensional effective theory; similarly, $D_{ab}$ parametrizes the Chern-Simons couplings of the abelian gauge fields \cite{Dabholkar:2014ema,Gomes:2015xcf}. This gives further support in view of the observations made in \cite{Gomes:2015xcf}, which relates the quantum black hole entropy (\ref{Ren 5D}) to the Chern-Simons path integral. In particular, $-\pi \hat{q}_0\phi^0 +\frac{\pi}{6}(p^3+c_2\cdot p)/\phi^0$
can be identified with the Chern-Simons action of a flat $SL(2,\mathbb{R})_L$ connection. The quadratic term proportional to $D_{ab}$ can be identified with the contribution of zero modes of the abelian Chern-Simons terms.

To compute the one-loop determinants we follow essentially the discussion in \cite{Gomes:2015xcf}, which provides with a simple derivation of the measure using a connection to supersymmetric Chern-Simons theory on $AdS_2\times S^1$. The derivation of the measure relies on the fact that the structure of perturbative corrections in the Cardy limit, which is given by the most polar Bessel function, in essence, is determined by the modular properties of the dual $\CFT_2$ partition function. This can be used to determine the measure and hence the one-loop determinants. Here we give a more refined version of that derivation and argue that the localization one-loop determinants are consistent with the Chern-Simons computation. 

In the four dimensional problem, the localization deformation was quadratic in the fields. As a consequence the one-loop determinants could not carry any dependence on the off-shell solution \footnote{In \cite{Murthy:2015yfa,Gupta:2015gga} the one-loop determinants depend explicitly on the localization manifold because the conformal factor of the metric is chosen to be field dependent. Nevertheless, such an approach requires the inclusion the Weyl multiplet in the localization determinants together with understanding the conformal gauge fixing procedure, features that are not clarified in those works.  }. In contrast, in five dimensions, the metric is fluctuating and so, on general grounds, we expect the one-loop determinant to be a function of the localization modes $\phi^0$, $\phi^a$ and the physical size $\vartheta(p)$. Furthermore, given the off-shell nature of the localization computation, the one-loop determinants can not depend on the couplings of the theory. In our problem these are effectively determined by the Chern-Simons couplings $\tilde{k}_L\propto p^3+c_2\cdot p$, $\tilde{k}_R\propto p^3+c_2\cdot p/2$ and $D_{ab}\sim p_a$, as we can see from the renormalized off-shell action. Another aspect to take into account is the fact that the localization deformation is Weyl invariant \footnote{The localization deformation $t QV$, with $V$ fermionic, is of the form $t\int \sqrt{g}\mathcal{L}_{\text{loc}}$. Since the path integral is $t$ invariant we must have that $\sqrt{g}\mathcal{L}_{\text{loc}}$ is Weyl invariant, otherwise, we could absorb the $t$ dependence in a Weyl rescaling of the metric. } and in odd dimensions Weyl invariance is preserved also at the quantum level. This means that the one-loop determinants, that is, the determinant over the non-zero modes, can not carry any dependence on $\vartheta(p)$, but only on the fields $\phi^0,\phi^a$ that are scale invariant. A dependence on $\vartheta(p)$ can arise due to zero modes, which we know are present in the $AdS_2$ path integral \cite{Sen:2009vz,Gomes:2015xcf,Banerjee:2009af}. Therefore, we can conclude that the one-loop contribution must have the form $Z^{\text{Loc}}_{\text{1-loop}}=\vartheta^{\alpha}f(\phi^0,\phi^a)$ with $\alpha$ determined by a zero mode counting. At this point we assume that $f(\phi^0,\phi^a)$ can  depend  on $\phi^0,\phi^a$ only polynomially. If this was not the case, the presence of exponential terms would correct the various terms  in the renormalized action (\ref{Ren 5D}) and change the saddle point equations. As a consequence, we would find that the on-shell values of $\phi^0,\phi^a$ were no longer the ones determined by the physical attractor geometry.  It would be important, nevertheless, to check by explicit computation that integration over the Kaluza-Klein modes on the circle does not give rise to such exponential terms in the one-loop determinants. Later we use the chiral primary picture of section \S\ref{sec polar states} to argue that this is the case. 

To determine $\vartheta^{\alpha}f(\phi^0,\phi^a)$ we can compare with the one-loop computation in the supersymmetric Chern-Simons theory \cite{Gomes:2015xcf}. For a Chern-Simons theory based on the gauge group $SL(2,\mathbb{R})_L\times SU(1,1|2)_R\times U(1)^{b_2}$ we compute the one-loop correction to the partition function as $Z=e^{\text{CS}(A)}Z^{\text{CS}}_{\text{1-loop}}$, where $\text{CS}(A)$ is the Chern-Simons action of a flat connection. The classical part given by the action of the flat connection $CS(A)$ can be shown to match with the renormalized entropy function (\ref{Ren 5D}) \cite{Dabholkar:2014ema,Gomes:2015xcf}. This gives
\begin{equation}\label{Z_CS one-loop}
Z^{\text{CS}}_{\text{1-loop}}=\vartheta\;\frac{(\phi^0)^{b_2/2+1/2}}{\sqrt{\tilde{k}_L\text{det}(D_{ab})}},
\end{equation}
where $\tilde{k}_L$ is the $SL(2,\mathbb{R})_L$ level and $D_{ab}$ parametrize the abelian $U(1)$ levels. On the other hand, from the localization computation we obtain after extremization of the finite dimensional integral, the following one-loop correction
\begin{equation}\label{Z saddle one-loop}
Z_{\text{1-loop}}=\vartheta^{\alpha}f(\phi^0,\phi^a)\frac{(\phi^0)^{b_2/2+3/2}}{\sqrt{\tilde{k}_L\text{det}(D_{ab})}}.
\end{equation}
The term $(\phi^0)^{b_2/2+3/2}/\sqrt{\tilde{k}_L\text{det}(D_{ab})}$ comes from evaluating the gaussian integrals at the saddle of the renormalized action (\ref{Ren 5D}). The saddle point approximation only requires that $|\hat{q}_0 p^3|\gg 1$, since we have $-\pi \hat{q}_0\phi^0 +\frac{\pi}{6}\frac{p^3}{\phi^0}\sim \sqrt{|\hat{q}_0 p^3|}(x+1/x)$, with $x\sim \phi^0 \sqrt{|\hat{q}_0|/p^3}$. At the extremum, we have $x\sim 1$ and thus the saddle value of $\phi^0$ can range from small values, for $|\hat{q}_0|\gg p^3$, to large values for $|\hat{q}_0|\ll p^3$, while keeping $|\hat{q}_0 p^3|\gg 1$. Since the Chern-Simons computation is valid for any value of $\phi^0$ \footnote{In the Chern-Simons computation the value of $\phi^0$ corresponds to a choice of metric, and hence it is equivalent to a choice of gauge.},  comparing the expressions (\ref{Z_CS one-loop}) and (\ref{Z saddle one-loop}), we must have that the localization one-loop determinant  is given by
\begin{equation}\label{1-loop localization}
Z^{\text{Loc}}_{\text{1-loop}}=\frac{\vartheta}{\phi^0}.
\end{equation}

We can check that the chiral primary computation of section \S\ref{sec polar states} reproduces the result (\ref{1-loop localization}). To do that, note that we have already included the effect of the massive hypermultiplets by means of the shift in the parameter $c_2$ of the five dimensional Lagrangian. Therefore, we only need to care about the supergravity modes, which include the effect of the graviton multiplet, $n_V$ vector multiplets and $n_H$ massless hypermultiplets. The idea is to repeat the computation of section \S\ref{sec polar states} for these modes. Since we have $n_V=n_H$ for the $\N=4$ theory, we can show that the supergravity modes cancel exactly for any value of $L_0>0$ in the trace (\ref{Ztop^2}). This is so because the canonical partition function for these modes is proportional to the MacMahon function to the power $\chi=-2(n_V-n_H)$ \cite{Gaiotto:2006ns}, which is also the Euler character of the Calabi-Yau. Therefore, for this supergravity theory we have only one polar term, that is, the trace over the chiral primaries is trivial. From this point of view, we also see that the five dimensional supergravity theory does not lead to problems related to the stringy exclusion principle. To obtain the black hole entropy we must perform a modular transformation as (\ref{modular transf polar states}). The degeneracy becomes
\begin{eqnarray}
d_{\text{BH}}(n,l_a)|_{\text{Sugra}}&=&\frac{c_R}{6}\int \prod_{a=1}^{b_2} dz^a d\tau \tau^{-\omega} \exp{\left(\pi i\frac{D_{ab} z^az^b}{\tau}+\frac{\pi i}{12}\frac{c_L}{\tau}-2\pi i\tau n-2\pi i z^al_a\right)}\nonumber,\\
{}
\end{eqnarray}
where $c_R$ and $c_L$ are respectively $c_R=p^3+\hat{c}_2\cdot p/2$ and $c_L=p^3+\hat{c}_2\cdot p$. As explained before, the $c_R$ factor in the measure comes from counting states in the angular momentum multiplet. Using the map
\begin{equation}
\tau=\frac{i}{2}\phi^0,\;z^a=\frac{\phi^a}{2i},\;n=-q_0,\,l_a=q_a,
\end{equation}
we can show that the integrand above is the exponential of the entropy function computed using localization. Moreover, from the analysis in section \S\ref{sec polar states} we have found $\omega=1$, and so we conclude that the one-loop determinant is $Z_{1-\text{loop}}^{\text{Loc}}=c_R/\phi^0$. From the analysis at the end of section \S \ref{sec Ideal sheaves}, we have found that the physical size $\vartheta$ of the geometry is proportional to $c_R$, which allows to reproduce the result (\ref{1-loop localization}), as we wanted to show.

The full partition function, including the one-loop determinants, is therefore
\begin{equation}\label{Z_5D localization}
\int_{\mathcal{C}} \prod^{b_2}_{I=0}d\phi^I\,\frac{\vartheta}{\phi^0}\, \exp{\left[-\pi \hat{q}_0\phi^0 +\frac{\pi}{6}\frac{p^3+c_2\cdot p}{\phi^0}-\frac{\pi}{2\phi^0}D_{ab}(\phi^a+q^a\phi^0)(\phi^b+q^b\phi^0)\right]}.
\end{equation}

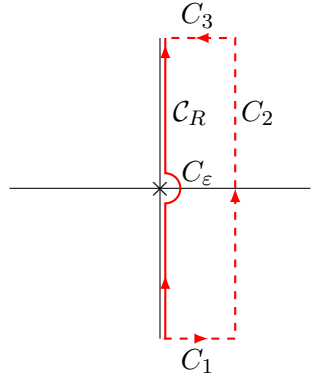
\begin{figure}
	\centering
	\begin{tikzpicture}
	\draw (0,-2) -- (0,2) ;  
	\draw (-2,0) -- (2,0) ;   
	
	\node at (0,0) {$\times$}; 
	
	\draw[thick,red,xshift=2pt,
	decoration={ markings,  
		mark=at position 0.2 with {\arrow{latex}}, 
		mark=at position 0.98 with {\arrow{latex}}
	}, 
	postaction={decorate}]
	(0,-2) -- (0,-0.2) arc (-90:90:.2) -- (0,2);
	
	\draw[red,thick,dashed,
	decoration={ markings,  
		mark=at position 0.1 with {\arrow{latex}}, 
		mark=at position 0.5 with {\arrow{latex}},
		mark=at position 0.93 with {\arrow{latex}}
	},
	postaction={decorate}]
	(0.05,-2)--(1,-2) -- (1,2) -- (0.05,2);
	
	\node at (0.5,0.2) {$C_{\varepsilon}$};
	\node at (0.5,2.3) {$C_3$};
	\node at (0.5,-2.3) {$C_1$};
	\node at (1.3,1) {$C_2$};
	\node at (0.4,1) {$\mathcal{C}_R$};
	
	\end{tikzpicture}
	\caption{Contour deformation for the integral $\int dz \frac{e^{az+b/z}}{z^c}$ with $a,b>0$ and $c>1$. $C_{\varepsilon}$ denotes the semicircle contour of radius $\epsilon$ and $\mathcal{C}_R$ is the original contour. It is easy to show that the integrals along the contours $C_1$ and $C_3$ vanish when we take $\text{Im}z(C_3)\rightarrow +\infty$ and $\text{Im}z(C_1)\rightarrow -\infty$. Since there is no pole inside the contour we must have that the integral along $\mathcal{C}_R$ equals the integral along $C_2$. In turn, the integral along $C_2$ is precisely the modified Bessel function of the first kind.} \label{fig:contour}
\end{figure}

To determine the integration contour $\mathcal{C}$ we proceed as follows. In the path integral, the measure includes an integration over the five dimensional metric and the gauge fields. Therefore, in the finite dimensional integral (\ref{Z_5D localization}) the appropriate variables of integration are the radius\footnote{To be more precise we are integrating over the vielbein $dR\sim de_u$. } $R\sim 1/\phi^0$ and the Wilson lines $A_{u}^a\sim \phi^a$, and so the integration measure must be proportional to $dR\,d\phi^a$. Furthermore, in the Euclidean four dimensional supergravity theory one has an $SO(1,1)$ R-symmetry instead of the usual $U(1)$ in the Minkowski theory. From a five dimensional point of view this effectively amounts to reduce the theory on a time-like circle \cite{Cortes:2003zd} instead of the euclidean circle. Looking at the geometry  (\ref{metric localization}) we see that we must integrate $R\sim 1/\phi^0$ over the imaginary axis, which determines the contour. To avoid the singularity at $R=0$, we take $R$ along the contour $\mathcal{C}_R=]-i\infty,-i\epsilon]\cup C_{\epsilon}\cup [i\epsilon,+i\infty]$ with $C_{\epsilon}$ a semicircle of radius $\epsilon\ll 1$  going around the origin in the anti-clockwise direction as depicted in figure (\ref{fig:contour}) (the integral would be zero if it circled the origin in the clockwise direction). Besides, one also has that the matrix $D_{ab}$ is not positive definite \footnote{The Hodge theorem ensures that for an $SU(3)$ Calabi-Yau the matrix $D_{ab}$ has exactly one negative eingenvalue \cite{Maldacena:1997de}, while for other Calabi-Yau there can be more than one.}, which renders the gaussian integral in (\ref{Z_5D localization}) ill-defined when $\text{Re}(R)>0$, or equivalently for $R\in C_{\epsilon}$. In a diagonal basis for $D_{ab}$, with $D_{ab}\phi^a\phi^a=\lambda_{+}\tilde{\phi}_{+}^2-\lambda_{-}\tilde{\phi}_{-}^2$ and $\lambda_{+},\lambda_{-}>0$, the solution is to analytically continue $\tilde{\phi}_{-}$ to imaginary values, making the gaussian integral  convergent. Deforming the contour, as described in figure (\ref{fig:contour}), the final integral is a modified Bessel function of the first kind, that is,
\begin{equation}
Z_{AdS_2\times S^1\times S^2}=\frac{\vartheta}{\sqrt{\text{det}(D)}}\int_{\epsilon'-i\infty}^{\epsilon'+i\infty} \frac{dR}{R^{1+b_2/2}}\exp{\left[-\pi \frac{\hat{q}_0}{R} +\frac{\pi}{6}(p^3+c_2\cdot p)R\right]}.
\end{equation}

\subsection{Subleading Bessel contributions}\label{sec subleading Bessels}

The global $AdS_3\times S^2$ solutions described in section \S\ref{sec Ideal sheaves} can be used to generate black hole $AdS_2\times S^1\times S^2$ geometries.  For example, start with the $AdS_3$ global metric
\begin{eqnarray}
ds^2_{AdS_3}=4\left(-\cosh^2\eta d\tau^2+d\eta^2+\sinh^2\eta d\sigma^2\right)\nonumber,
\end{eqnarray}
and write $y=\tau-\sigma$ and $\rho=2\eta$  such that, after some algebra,
\begin{equation}
ds^2_{AdS_3}=-4\left(d\tau +\frac{1}{2}(\cosh\rho-1)dy\right)^2+d\rho^2+\sinh^2\rho dy^2.
\end{equation}
For the purpose of computing the euclidean path integral we can declare $y$ to be the euclidean time and $\tau$ the spatial circle instead. Then, the $AdS_2$ factor $d\rho^2+\sinh^2\rho dy^2$ together with the sphere becomes the euclidean near-horizon geometry of the four dimensional black hole \cite{Sen:2008vm}. The spatial circle has nevertheless, a time-like signature and so we have to consider its Wick rotation. In particular, we consider the identification $i\tau\sim i\tau+ 2\pi/\phi^0$- similar details concerning the analytic continuation were considered  in the context of string theory on $\mathbb{R}^4\times S^1$ in \cite{Dedushenko:2014nya}. The metric becomes
\begin{equation}
ds^2_{AdS_2\times S^1}=\frac{4}{(\phi^0)^2}\left(du +\frac{i}{2}\phi^0(\cosh\rho-1)dy\right)^2+ds^2_{AdS_2},
\end{equation}
where we have defined $u=i\tau \phi^0$ with periodicity $u\sim u+2\pi$. This metric can also be seen as the extremal limit of a BTZ black hole in $AdS_3$. Physical considerations for this spacetime in the context of the entropy function were made in \cite{Gupta:2008ki,Murthy:2009dq}.

Furthermore, on top of the geometry we need to consider the effect of the singular gauge field configurations, which generate spectral flow transformations. This is motivated from the gauge field configuration (\ref{attractor fields}), that one obtains in the decoupling limit of the polar configurations. Asymptotically we have 
\begin{equation}\label{5d connection after flux}
A^{a}_{5D}\simeq  -p^a \cos\theta (d\phi+\tilde{A})-i2D^{ab}\frac{\Delta\beta_b}{\phi^0} du -2D^{ab}\Delta\beta_b dy,
\end{equation}
after analytically continuing the time-like circle,  as described above. We see that the total M2 charge contribution $\Delta\beta=\beta-\bar{\beta}$ induces the gauge transformation 
\begin{equation}\label{M2 singular connection}
-2D^{ab}\Delta\beta_a dy.
\end{equation}
Nevertheless, this gauge transformation is not well defined because the circle parametrized by $y$ is contractible at the origin $\rho=0$, which gives rise to a delta function singularity in its field strength. Similar gauge field configurations were discussed in \cite{Dijkgraaf:2000fq}.

Note that $\Delta\beta^a\equiv D^{ab}\Delta\beta_a$ lives in $\Lambda^*/\Lambda$, where $\Lambda$ corresponds to the lattice $k_a \in \mathbb{Z}$, while $\Lambda^*$ is the dual lattice. As $D_{ab}$ is not unimodular (its determinant is not one), $\Delta\beta^a$ is not necessarily an integer, and so the holonomies $\exp{\oint A_{5D}}$ around the contractible cycle can be non-trivial.

The connection (\ref{M2 singular connection}) is thus a large gauge transformation and we want to understand its contribution to the localization computation.  We will argue that its contribution to the renormalized action comes solely from the abelian Chern-Simons terms, since these are the only terms in the Lagrangian that are not gauge invariant. To deal with the delta function singularity we remove a disk of radius $\epsilon$ from the origin \footnote{The space has topology $D\times S^1$ where $D$ is a disk.} and at the end of the computation we consider the limit when $\epsilon$ goes to zero. This way we ensure that the localization equations are left unchanged, as they depend only on the field strengths. There is a subtlety in this procedure. It turns out, that we also need to supplement the Chern-Simons integral with a boundary term at $r=\epsilon$, which is the inner boundary disk. This term is necessary to correctly take into account the delta function singularity. To exemplify this, consider the integral on the disk $D$
\begin{equation}
I=\int_{D} f(r) F,
\end{equation}
where $F=d(d\theta)$, with $\theta$ is the angle on the disk $D$, and $f(r)$ is a test function. Since we have $F=\delta(r)dr\wedge d\theta$, the integral gives $I=2\pi f(0)$. In contrast, if we put a regulator at $r=\epsilon$ we would find zero since in that case we have $F=0$ everywhere outside the origin. Instead the regulated integral should have the form
\begin{equation}\label{reg at the origin}
I_{\epsilon}=\int_{D_{\epsilon}} f(r) F+\int_{\partial D_{\epsilon} }f(\epsilon)A,
\end{equation}
with $D_{\epsilon}$ the regulated disk and $\partial D_{\epsilon}$ being the boundary of the inner disk. In the limit when $\epsilon\rightarrow 0$ we recover the result $I=2\pi f(0)$. This example can be easily adapted to the Chern-Simons form $A\wedge F$, in which case the inner boundary term becomes $\int A_{\theta}A_{u}$, with $A_{u}$  the component along the circle $S^1$.  

We now discuss the contribution of large gauge transformation to the renormalized action coming from the Chern-Simons action. To simplify the problem, we first integrate on the sphere. The five dimensional Chern-Simons terms give rise to three dimensional abelian Chern-Simons terms
\begin{equation}\label{Chern-Simons terms}
\frac{i}{192 \pi^2}D_{abc}\int A^a\wedge F^b\wedge F^c\rightarrow \frac{i}{16\pi}D_{abc}p^c\int A^a\wedge F^b.
\end{equation} 
By the localization equations, the five dimensional gauge field has the form  (\ref{localization 5D gauge field})
\begin{eqnarray}\label{localization sol}
A^{a}_{5D}= -2\frac{\phi^a(\rho)}{\phi^0(\rho)}(du +A^{*0}) +A^{*a}_{4D}-2D^{ab}\Delta\beta_b dy,
\end{eqnarray}
where $A^{*0}$ and $A^{*a}_{4D}$ have the attractor values of the unperturbed solution. In contrast the fields $\phi^a$ and $\phi^0$ have non-trivial radial profiles. The boundary conditions are such that at infinity one has  
\begin{equation}\label{boundary phi^a}
\lim_{\rho\rightarrow\infty}\frac{\phi^a}{\phi^0}=-q^a,
\end{equation}
which follows from the equations of motion. Note that due to the Chern-Simons terms only the component of the gauge field along $du$, which is proportional to $q^a$ at infinity, is kept fixed while the component $dy$ is allowed to fluctuate \cite{Dabholkar:2014ema}. To simplify the discussion we consider the case of only one gauge field.  Due to the boundary conditions the Chern-Simons action has to be supplemented with a boundary action, that is,
\begin{equation}\label{simple CS action}
\int_{D\times S^1} A\wedge F+\int_{T^2}A_y A_u,
\end{equation}
with $A_u$ fixed but $A_y$ can fluctuate. We write the gauge field as a non singular part $A_{\text{n.s.}}$ plus a singular part, which is the gauge transformation $d\Lambda=d\Lambda_y\, dy$. Then the action (\ref{simple CS action}) splits into a term that depends only on $A_{\text{n.s.}}$ and a term linear in $d\Lambda$. The first joins the remaining terms in the Lagrangian to give the renormalized action (\ref{Ren 5D}). On the other hand, the term linear in the gauge transformation is
\begin{equation}\label{large gauge transf action}
2\int_{D_{\epsilon}\times S^1} d\Lambda\wedge F_{\text{n.s}} +\int_{D_{\epsilon}\times S^1} d(d\Lambda\wedge A_{\text{n.s}})+ \int_{T^2} (d\Lambda)_y A^{\text{n.s }}_u.
\end{equation}
The term $\int d(d\Lambda\wedge A_{\text{n.s}})$ gives rise to the integral of a delta function and so we need the regularization term at the inner boundary of $D_{\epsilon}$ as in (\ref{reg at the origin}). Being a total derivative, it gives rise to two contributions. The first is a boundary term at infinity that cancels the third term. The second is a term at the inner boundary $\partial D_{\epsilon}$, but this cancels against a similar term in the regularization (\ref{reg at the origin}). In the present problem the components of $F$ which are relevant are those coming from the term $-2\partial_{\rho}(\phi^a/\phi^0)d\rho\wedge du$. Then the term $d\Lambda\wedge F$ gives rise to a total derivative that we can calculate. Specializing for the Chern-Simons terms (\ref{Chern-Simons terms}), (\ref{large gauge transf action}) gives
\begin{equation}
2\pi i \left.\frac{\phi^a}{\phi^0}\right|^{\rho=\infty}_{\rho=0}\Delta\beta_a=-2\pi iq^a\Delta\beta_a-2\pi i\left.\frac{\phi^a}{\phi^0}\right|_{\rho=0}\Delta\beta_a.
\end{equation}
where $\left.\frac{\phi^a}{\phi^0}\right|_{\rho=0}$ is the field computed at the origin, which coincides with the variables in the renormalized action. The final result for the renormalized action including the effect of the large gauge transformation is, therefore,
\begin{eqnarray}\label{full renormalized S}
\text{Ren}(S_{5D})|_{ A+d\Lambda}=-\pi q_I\phi^I+\frac{\pi}{6}\frac{p^3+\hat{c}_{2}\cdot p}{\phi^0}-\frac{\pi}{2}\frac{D_{ab}\phi^a\phi^b}{\phi^0}-2\pi i\frac{\phi^a}{\phi^0}\Delta\beta_a-2\pi iq^a\Delta\beta_a,
\end{eqnarray}
where we have included the effect of the shift in $c_2\rightarrow c_2-12(\beta+\bar{\beta})$ in $\hat{c}_2$. After some algebra this can be written as
\begin{eqnarray}\label{renorm action w large gauge transf}
-\pi \hat{q}_0\phi^0 +\frac{\pi}{6}\frac{p^3+\hat{c}_{2}\cdot p-12D_{ab}\Delta\beta^a\Delta\beta^b}{\phi^0}
-\frac{\pi}{2\phi^0}D_{ab}(\phi^a+q^a\phi^0+2i\Delta\beta^a)(\phi^b+q^b\phi^0+2i\Delta\beta^b)\nonumber,
\end{eqnarray}
with $\hat{q}_0=q_0-D_{ab}q^aq^b/2$. 

Note that by extremezing $\text{Ren}(S_{5D})$, the saddle values for $\phi^0,\phi^a$ are different from the ones defined by the attractor values (\ref{attractor equations}). The difference is due to the terms proportional to $\Delta\beta$, which do not appear in the classical prepotential for computing the on-shell values. At first site, this could have looked puzzling. The fact is that the fields $\phi^0,\phi^a$ in the renormalized action (\ref{full renormalized S}) are the values of the fields computed at the origin, which can be different from the values at the boundary. The mismatch is due to the singular gauge transformation, whose contribution to the action  behaves more like a non-local term. As a side comment, note that if (\ref{renorm action w large gauge transf}) was the result of a local contribution, then by the equations of motion we would find the attractor value $\phi^a=-2i\Delta\beta^a$, for $q^a=0$. We would recover precisely the asymptotic value of the five dimensional gauge field given in (\ref{5d connection after flux}).

The localization one-loop determinant works much the same way as discussed in the previous section. The only difference is the singular gauge transformation. Nevertheless, since the localization action depends only on the field strengths, given the regularization procedure,  the determinant is not expected to depend on the effect of the large gauge transformation. Therefore for the purpose of computing the one-loop determinant we can set $\Delta\beta_a=0$. Hence, using  (\ref{1-loop localization}) we obtain
\begin{equation}
Z^{\text{Loc}}_{1-\text{loop}}=\frac{\vartheta(p,\beta,\bar{\beta})}{\phi^0}=\frac{1}{\phi^0}\left(\frac{p^3}{6}+\frac{\hat{c}_2\cdot p}{12}\right),
\end{equation} 
where $\vartheta$ is the physical size of the geometry (\ref{condition on the size}), which contains the backreaction of the fluxes. Note that this expression is precisely the K\"{a}hler potential of the four dimensional theory, computed with the classical prepotential (\ref{classical prepotential reduction}). 

We are ready to assemble the $\mathcal{N}=4$ answer. In both the $K3\times T^2$ and $T^4\times T^2$  CHL  compactifications we have $\beta_a=\beta\delta_{1a}$ and $\bar{\beta}_a=\bar{\beta}\delta_{1a}$, following the description in terms of M2 and anti-M2 branes of section \S \ref{sec polar states}. For the $\N=8$ compactification one is effectively turning off the fluxes, since the index vanishes otherwise. Furthermore, the intersection matrix is $D_{ab1}=D_{a1b}=D_{1ab}=C_{ab}$ with the other components zero, and the second Chern-class is $c_{2a}=24 n_p\delta_{1a}$ with $n_p=1,0$ for $K3$ and $T^4$  respectively. We compute
\begin{eqnarray}
&&\frac{p^3}{6}=\frac{p^1}{2}P^2,\;\;\frac{\hat{c}_2\cdot p}{12}=p^1\left(2-(\beta+\bar{\beta})\right),\;\;D_{ab}\Delta\beta^a\Delta\beta^b=\Delta\beta^1\Delta\beta_1=-\frac{p^1}{P^2}(\beta-\bar{\beta})^2\nonumber,
\end{eqnarray}
where we defined $P^2=C_{ab}p^ap^b$. This gives
\begin{eqnarray}
\frac{1}{6}\left(p^3+\hat{c}_{2}\cdot p-12D_{ab}\Delta\beta^a\Delta\beta^b\right)=4p^1\left[\frac{(P^2/2-(\beta-\bar{\beta}))^2}{2P^2}-\bar{\beta}+n_p\right],
\end{eqnarray}
and 
\begin{equation}
\vartheta(p,\beta,\bar{\beta})=p^1(P^2/2+2n_p-(\beta+\bar{\beta})).
\end{equation}

The full $\mathcal{N}=4$ answer is therefore the sum of the five dimensional partition function for each of the saddles parametrized by $\beta,\bar{\beta}$, that is,
\begin{equation}
Z_{\N=4}(q_I,p^I)=\sum_{\beta,\bar{\beta}\in \vartheta(p,\beta,\bar{\beta})>0} \mathcal{D}(\beta)\mathcal{D}(\bar{\beta})Z_{5D}(q_I,p^I,\beta,\bar{\beta}),
\end{equation}
where $\mathcal{D}(\beta),\mathcal{D}(\bar{\beta})$ represent a measure, or better an Euler characteristic, for the ideal sheaves contribution; in the M2 brane picture this measure is given by an index. It would be important to understand more clearly how these Euler characteristics are computed from the full M-theory path integral. Nevertheless, if we use the M2-brane counting we find that $\mathcal{D}(\beta)=\oint \frac{dq}{q^{\beta+1}}\frac{1}{g(q)}$ \footnote{It is possible that the ideal sheaves counting differs from the M2-brane index. The reason is that while the index is computed in flat spacetime transverse to the Calabi-Yau, the ideal sheaves moduli space might carry a $AdS_2\times S^2$ factor instead of $\mathbb{R}^4$. The instanton computation on $AdS_2\times S^2$ of \cite{Beasley:2006us} using string worldsheet methods seems to point in that direction. It is found that a string wrapping a two-cycle in the Calabi-Yau, has both bosonic and fermionic collective coordinates on $AdS_2\times S^2$.},where $g(q)$ is the worldsheet instanton partition function. For example, for $K3$ one has $g(q)=\eta^{24}(q)$. Assembling all the pieces in the formula above we obtain 
\begin{eqnarray}\label{non-perturbative entropy}
d(q_I,p^I)=&&\sum_{\beta,\bar{\beta}>0}^{\text{Max}(\beta,\bar{\beta})}p^1\left(\frac{P^2}{2}+2n_p-(\beta+\bar{\beta})\right)\times \mathcal{D}(\beta) \mathcal{D}(\bar{\beta}) \nonumber\\
&&\times \frac{1}{\sqrt{\text{det}(D_{ab})}}\int_{\epsilon'-i\infty}^{\epsilon'+i\infty} \frac{dR}{R^{1+b_2/2}} \exp{\left[-\pi \frac{\hat{q}_0}{R} +4\pi p^1\left(\frac{(P^2/2-(\beta-\bar{\beta}))^2}{2P^2}-\bar{\beta}+n_p\right) R\right]}.\nonumber\\
{}
\end{eqnarray}
Furthermore, the Bessel contour forces the condition
\begin{equation}
\frac{(P^2/2-(\beta-\bar{\beta}))^2}{2P^2}-\bar{\beta}+n_p>0,
\end{equation}
otherwise the integral vanishes. That is, if the condition is non-positive we can close the contour on the right hand side by an infinite semicircle, because the integral along this arc vanishes. Since there is no pole inside the contour, the Bessel integral must vanish too.

Putting all factors together we find that the macroscopic answer (\ref{non-perturbative entropy}) is in perfect agreement with the microscopic answer (\ref{finite sum Bessels}) up to a phase involving the charges, and an overall $p^1$ dependence. The phase $\exp{[\pi i (r-2s)l/m]}$ is a Kloosterman sum. In a related work  \cite{Dabholkar:2014ema}, Kloosterman sums were shown to  be related to  a sum over flat connections in Chern-Simons theory on a Dhen filled solid torus. It would be important to check if the missing phase arises in the same way \cite{Gomes17-2}. The correct $p^1$ dependence can be incorporated by taking into account a redefinition of the Chern-Simons couplings \cite{Gomes:2015xcf}.

We end this section by making a comment about the mock-modular nature of the $\N=4$ microscopic answer \cite{Dabholkar:2012nd}. The microscopic degeneracy of one-quarter BPS dyons still admits a Rademacher expansion but with some important differences, which result essentially from the meromorphicity of the Jacobi form. As expected from the usual Rademacher expansion, one finds a sum over Bessel functions of index $-\omega+3/2$, with $\omega$ the weight of the meromorphic Jacobi form \cite{Murthy:2015zzy}, but due to its mock-modular nature, the degeneracy formula contains in addition Bessel functions of index $-\omega+2$ and  an integral over a Bessel of index $-\omega+5/2$ \cite{Ferrari:2017msn} (they have used $\omega=-10$). The argument of the unusual Bessel functions lies precisely at the lower boundary of the polarity. 

We can see that our construction breaks down precisely when such deviations of the usual Rademacher expansion are expected. This happens when the geometry attains its possible minimal size, that is, when
\begin{equation}
\beta+\bar{\beta}=\frac{P^2}{2}
\end{equation}
and the spectral flow invariant combination that appears in the argument of the Bessel function, becomes zero, that is
\begin{equation}
\frac{(P^2/2-(\beta-\bar{\beta}))^2}{2P^2}-\bar{\beta}=0.
\end{equation}
This last condition signals a breakdown of the Rademacher expansion for holomorphic Jacobi forms. The solution to both equations is
\begin{equation}
\beta=\frac{P^2}{2},\;\bar{\beta}=0,\vee\; \beta=0,\;\bar{\beta}=\frac{P^2}{2},
\end{equation}
and  the contribution to the full partition function is 
\begin{equation}\label{mock Bessel}
4p^1 \mathcal{D}(0) \mathcal{D}(P^2/2)  \frac{1}{\sqrt{\text{det}(D_{ab})}}\int_{\epsilon'-i\infty}^{\epsilon'+i\infty} \frac{dR}{R^{1+b_2/2}} \exp{\left[-\pi \frac{\hat{q}_0}{R} +4\pi p^1 n_p R\right]}
\end{equation}
Even though the index of this Bessel function does not match with the prediction coming from the mixed Rademacher expansion \cite{Ferrari:2017msn}, we can easily check that the argument of the Bessel (\ref{mock Bessel}) is in perfect agreement (compare with equation 3.42 of \cite{Ferrari:2017msn}). 

The origin of the mock-modular nature may be related to the fact that in the canonical ensemble there is a configuration which competes with a similar growth of (\ref{mock Bessel}). This happens when $P^2=0$ and $\beta=\bar{\beta}=0$. In this case $\text{det}(D_{ab})$ in (\ref{mock Bessel}) is zero and so we need to reconsider the computation of the one-loop determinants. Following the Chern-Simons formulation, we see that the $U(1)_L$ factor has level $k_L=P^2/2=0$. This is consistent with the fact that at the level of the renormalized action the coefficient of $(\phi^1)^2$ is zero. We have to truncate the integration over $\phi^1$ to a finite interval. Since the combination $\tau=(\phi^1+ip^1)/\phi^0$ transforms under electric-magnetic duality as $\tau\rightarrow \tau+1$, it is natural to truncate $\phi^1$ to the interval $[0,\phi^0]$. Therefore, integration over $\phi^1$ gives a factor of $\phi^0$ in the measure, which corresponds to an additional power of $(\phi^0)^{1/2}=R^{-1/2}$ relative to the Bessel (\ref{mock Bessel}). For the $K3\times T^2$ compactification we have $b_2=23$ and so we find a Bessel function with the same argument as (\ref{mock Bessel}) but with index $12$.

\subsection{4D quantum effective action and the topological string}\label{sec effective action}

In this section, we make a comparison with four dimensional string theory. To do that, we consider the limit when the M-theory circle radius becomes parametrically smaller than the size of $AdS_2\times S^2$. This amounts to take $1/\phi^0\ll 1$, which is equivalent to the regime of weak topological string coupling. We show that in this limit the number of Bessel functions grows exponentially, allowing to perform the sum by a saddle approximation. As a consequence, we can expand the entropy $\ln d(q,p)$ in a perturbative expansion in powers of $g_{\text{top}}=1/\phi^0$. We can then identify some of the terms in the expansion with the topological string free energies (holomorphic part), as expected from the OSV proposal \cite{Ooguri:2004zv}. Nevertheless, we will encounter other terms  such as logarithmic corrections, which signal a departure from the Wilsonian action point of view. This may be regarded as a way to derive the non-holomorphic corrections proposed in \cite{LopesCardoso:2006bg,Cardoso:2008fr,Cardoso:2010gc,Cardoso:2012nh,Cardoso:2014kwa}.

 In the regime of charges
\begin{equation}\label{scale charges}
q_I\rightarrow \lambda q_I,\;p^I\rightarrow \lambda p^I,\;\lambda\gg 1,
\end{equation}
the attractor values scale as
\begin{equation}\label{scale phi}
\phi^0\rightarrow \lambda \phi^0,\;\phi^a\rightarrow  \lambda\phi^a,
\end{equation}
and thus the M-theory circle, which is proportional to $1/\phi^0$, becomes very small for $\lambda\gg 1$. Moreover, in this limit we keep fixed the K\"{a}hler class of the Calabi-Yau, which is proportional to $p/\phi^0$ \cite{Beasley:2006us}, while taking the size $L^2\propto p^3$ of $AdS_2\times S^2$ to large values. 

However, the scaling limits (\ref{scale charges}) and (\ref{scale phi}) are valid only for the solution without fluxes. For $p^I\gg 1$ the number of  fluxes increases because of the condition $p^3/6+c_2\cdot p/12-(\beta+\bar{\beta})\cdot p>0$. At the order $\beta,\bar{\beta}\sim \lambda^2$  we can expect a breakdown of the scaling limits because the factors $\mathcal{D}(\beta\sim \lambda^2)$ in (\ref{non-perturbative entropy}) will have exponential growth. In the case of $K3$ for example, we have
\begin{equation}
\mathcal{D}(\beta)\simeq e^{4\pi\sqrt{\beta}},\;\beta\gg 1.
\end{equation}
 We can use this to obtain an approximate formula for the degeneracy (\ref{non-perturbative entropy}) in the limit (\ref{scale charges}). Using a saddle point approximation for each of the Bessel functions, we obtain
\begin{equation}
d(q_I,p^I)\sim \sum_{\beta,\bar{\beta}\gg 1}^{\lambda^2} \exp{\left[4\pi\sqrt{\beta}+4\pi\sqrt{\bar{\beta}}+4\pi\sqrt{p^1|\hat{q}_0|\left(\frac{(P^2/2-(\beta-\bar{\beta}))^2}{2P^2}-\bar{\beta}+n_p\right)}\right]},\;p\gg 1.
\end{equation}
Next, we approximate the sum over fluxes by a continuum, which allows to make a new saddle point approximation with respect to $\beta,\bar{\beta}$. After some algebra, we find that the saddle is at $\beta=\bar{\beta}$ with $\beta$ finite of order $\sim P^2/(p^1\hat{q}_0)\sim \mathcal{O}(\lambda^{0})$. Therefore, we see that the saddle point approximation is not consistent with $\beta,\bar{\beta}\gg 1$, and thus the leading contribution must come, instead, from small values of $\beta,\bar{\beta}$. In this case the dominant contribution comes from the term with $\beta=\bar{\beta}=0$, which is the Bessel function of maximal polarity.

Therefore for large $\lambda$ we approximate
\begin{eqnarray}
&&d(q_I,p^I)\sim\nonumber\\ &&\sim \sum_{\beta,\bar{\beta}\lesssim \lambda^2}\mathcal{D}(\beta)\mathcal{D}(\bar{\beta})e^{\left[-\pi \hat{q}_0\phi^{*0}+\frac{\pi p^1 P^2}{2\phi^{*0}}+\frac{\pi P^2}{2p^1\phi^{*0}}(\phi^{*1}+q^1\phi^{*0})^2-2\pi \frac{p^1}{\phi^{*0}}(\beta+\bar{\beta})+2\pi i(\beta-\bar{\beta})\frac{\phi^{*1}}{\phi^{*0}}\right]}\nonumber\\
&&=e^{\left[-\pi \hat{q}_0\phi^{*0}+\frac{\pi p^1 P^2}{2\phi^{*0}}+\frac{\pi P^2}{2p^1\phi^{*0}}(\phi^{*1}+q^1\phi^{*0})^2-\ln\left(\frac{p^1}{\phi^{*0}}\right)^{12}-\ln\Big|\eta\left(\frac{\phi^{*1}+ip^1}{\phi^{*0}}\right)\Big|^{48}\right]}+\mathcal{O}(1/\lambda^2),\nonumber\\
{}\label{on-shell entropy function}
\end{eqnarray}
where $\phi^{*0}$ and $\phi^{*1}$ are the on-shell values determined from the Bessel of maximal polarity. To arrive at the formula above we approximated the subleading Bessel contributions by their saddle point value,  and then expanded their growth formula for $\beta,\bar{\beta}\ll p^3$. The measure factor $\sim P^2-4(\beta+\bar{\beta})$ in (\ref{non-perturbative entropy}), is offset by the saddle point gaussian integrals, which give an overall factor of $1/P^2$, and hence in the limit $P^2\sim \lambda^2$, they are of the same order. In the second line, we have extended the range of $\beta$ to infinity which is justified for $\lambda\gg 1$. This allowed to resum the contributions of $\mathcal{D}(\beta)$ into their generating function $\eta^{24}(\beta)$. Similarly, we could have repeated the same exercise for CHL models, with $\eta^{24}(\beta)$ being replaced by $g(\beta)$, the worldsheet instanton partition function.

From the four dimensional point of view, the exponential in (\ref{on-shell entropy function}) can be interpreted as the renormalized 1PI effective action computed on the near-horizon geometry $AdS_2\times S^2$ \cite{Sen:2008vm,Sen:2007qy}.  In particular, the expression
\begin{equation}\label{2der leading}
-\pi \hat{q}_0\phi^{*0}+\frac{\pi p^1 P^2}{2\phi^{*0}}+\frac{\pi P^2}{2p^1\phi^{*0}}(\phi^{*1}+q^1\phi^{*0})^2,
\end{equation}
 can be identified with the two derivative Lagrangian contribution, which gives the dominant contribution to the entropy since it scales as $\lambda^2$. On the other hand, the logarithmic part 
 \begin{equation}\label{lambda^zero}
 -\ln\left(\frac{p^1}{\phi^{*0}}\right)^{12}-\ln\Big|\eta\left(\frac{\phi^{*1}+ip^1}{\phi^{*0}}\right)\Big|^{48},
 \end{equation}
 grows as $\lambda^{0}$. It can be computed by evaluating the contribution of the Gauss-Bonnet $R^2$ corrections  on $AdS_2\times S^2$ \cite{Sen:2007qy}. Furthermore, it is modular invariant in the variable $\tau=\frac{\phi^{*1}+ip^1}{\phi^{*0}}$, as expected from the four dimensional electric-magnetic duality of string theory on $K3\times T^2$. Finally, there is no term of order $\ln\lambda$, which is in agreement with the logarithmic correction computed in the four dimensional $\N=4$  supergravity theory \cite{Banerjee:2010qc,Banerjee:2011jp}.

From the topological string point of view we can identify the different corrections as an expansion in $g_{\text{top}}\sim 1/\lambda\ll 1$. For example, the leading contribution (\ref{2der leading}) corresponds to the real part of the tree level free energy $F_0(t)$ multiplied by $1/g_{\text{top}}^2$, while the term of order $\lambda^0$ (\ref{lambda^zero}) can be identified with the one-loop contribution, with the complexified K\"{a}hler class $t$ being $\tau$. Such an expansion is precisely the conjectured OSV formula \cite{Ooguri:2004zv}. However, we must stress that further corrections can carry the imprint of the $AdS_2\times S^2$ physics and deviate from the topological string free energies that we obtain from the $\mathbb{R}^4$ computation. 

\section{Quantum Foam and Non-perturbative topological string}\label{sec Foam}

In this work, we have argued that the path integral of M-theory should include the contribution of singular gauge field configurations. Their effect was to produce a finite renormalization of the parameter $c_a$ of the five dimensional Lagrangian that parametrizes the mixed gauge-gravitational Chern-Simons terms. If this idea is correct, this would imply that one is effectively summing over different topologies of the internal manifold, since the parameter $c_a$ descends from the second Chern-class $c_2(X)$ of the Calabi-Yau $X$. In a way, this is reminiscent of the idea of quantum foam and melting crystals discussed in \cite{Iqbal:2003ds}, which we briefly review  now.

The  goal of  \cite{Iqbal:2003ds} was to provide with a non-perturbative definition for the topological string.  The example under discussion was the case of the A-model. From the target space perspective, the A-model can be described by a theory known as K\"{a}hler gravity \cite{Bershadsky:1994sr}, and the classical solutions of this theory are given by K\"{a}hler forms $k$, with action proportional to the volume form 
\begin{equation}
S=\frac{1}{g^2\,3!}\int_{X} k\wedge k\wedge k,
\end{equation}
with $g$ the topological string coupling constant. We can consider higher derivatives in this action by adding the term $\frac{1}{24}\int k\wedge c_2(X)$.

In the quantum problem we consider fluctuations of the macroscopic solution $k_0$ as
\begin{equation}
k=k_0+g\,F,
\end{equation}
with $F$ the fluctuation. Since it obeys $dF=0$ due to the K\"{a}ler condition, $F$ can be seen as the field strength of a gauge field. In addition, we want to preserve the macroscopic K\"{a}hler form, that is, we need $\int_{\alpha} F=0$
for any two-cycle $\alpha\in H_2(\mathbb{Z},X)$. As explained before, if we require $F$ to have non-trivial higher Chern-classes, then it must be the field strength of a singular gauge connection, or in the appropriate bundle generalization an ideal sheaf. It is argued in \cite{Iqbal:2003ds} that these singular fluctuations  lead to a foamy description of quantum gravity characterized by wild changes of the geometry and the topology. Instead of dealing directly with the quantum gravity picture, which may lead to puzzles related to black hole formation, they propose that the same physics should be  described in terms of the topologically twisted maximally supersymmetric $U(1)$ theory, that is, the $\D6$ brane worldvolume theory. 

After some algebra, the quantum action for $k=k_0+gF$ becomes
\begin{equation}\label{Kahler grav action}
S_{\text{Quantum}}=\frac{1}{g^2\,3!}\int_{X} k_0\wedge k_0\wedge k_0+\int k_0\wedge c_2(X)+\frac{1}{2}\int k_0\wedge F\wedge F+\frac{g}{3!}\int F\wedge F\wedge F.
\end{equation}
We have included the effect of higher derivative corrections proportional to $c_2\cdot k_0$. Using localization, they show that the partition function of the $\D6$ theory on $S^1\times X$, with periodic boundary conditions on $S^1$ for the fermions, reproduces the gravity path integral.

Our problem is slightly different because in this case we have a $\D6-\aD6$ configuration, but we can easily mimic many of the quantum foam features. We  thus expect fluctuations $F$ and $\overline{F}$ of the K\"{a}hler form, coming from each of the centers. The total quantum action receives the contribution from both the D-branes, that is, $S_{\text{Quantum}}=S_{\D6}+S_{\aD 6}$ with
\begin{eqnarray}
&&S_{\D6}=\frac{1}{g^2\,3!}k_0^3+c_2\cdot k_0-\beta\cdot k_0-g n,\\
&&S_{\aD 6}=\frac{1}{g^2\,3!}k_0^3+c_2\cdot k_0-\bar{\beta}\cdot k_0-g \bar{n},
\end{eqnarray} 
where we have defined the second and third Chern-classes of the bundles $F,\,\overline{F}$ by $-\beta,-\bar{\beta}$ and $-n,-\bar{n}$ respectively. We have $g=1/\phi^0$, and $k_0^a=g p^a/2$ so that the total K\"{a}hler class is $2k_0=p/\phi^0$ in agreement with the attractor geometry.  Note that the quantum K\"{a}hler class $k_0+gF$ is equivalent to the flux $p/2+F$ that we turn in M-theory, as discussed in the section \S\ref{sec Ideal sheaves}. From the K\"{a}hler gravity point of view it becomes clear that the effect of the M-theory fluctuations parametrized by $\beta,\bar{\beta}$ is to renormalize the second Chern-class $c_2(X)$, as we see from (\ref{Kahler grav action}). The final result is 
\begin{equation}\label{quantum action Sq}
S_{\text{Quantum}}=\frac{p^3+c_2\cdot p}{24\phi^0}-\frac{(\beta+\bar{\beta})\cdot p}{2\phi^0}-\frac{n+\bar{n}}{\phi^0}.
\end{equation}
For $n,\bar{n}=0$, we recognize the action $2\pi S_{\text{Quantum}}$ as the on-shell renormalized physical action on the near-horizon geometry (\ref{Ren 5D}), before including the electric charges and the effect of the large gauge transformation. 

We thus see that there are striking similarities between our problem and the quantum foam description, with the topological string playing a special role. It would be important to make more precise the connection between quantum black hole entropy, K\"{a}hler gravity and the $\D 6$ brane theory.

\section{Discussion and Conclusion}

In this work we have discussed a proposal for explaining the non-perturbative corrections to black hole entropy, related to the polar subleading Bessel contributions in the Rademacher expansion. In summary, the main results are:
\begin{itemize}
	\item \emph{New family of saddle geometries and the stringy exclusion principle}: we have argued that the path integral of M-theory on the near-horizon geometry of the black hole receives the contribution of a new family of saddle geometries, whose contribution is related to the polar terms in the Rademacher expansion. We discussed the possibility that these saddles arise after turning on ideal sheaves fluxes on the Calabi-Yau, which in turn induce corrections on the parameters that define the effective five dimensional Lagrangian. This picture has an alternative description in terms of M2 and anti-M2 branes wrapping holomorphic cycles on the Calabi-Yau. However, in previous works such as \cite{Simons:2004nm,Gaiotto:2006ns},  the backreaction of the M2 branes on the geometry is not taken into account. Instead, our proposal considers backreaction, which allows to solve many puzzles. In fact, one can show that effective field theory on a fixed background is a good approximation for very large central charge, as expected. But for finite central charge, which is the main goal of this work, backreaction becomes important and we find that a finite number of geometries contributes.  The bound on the number of geometries is precisely the bound imposed by the stringy exclusion principle \cite{Maldacena:1998bw}. 
	
	\item \emph{5D Supersymmetric localization}: as an intermediate problem, we have considered supersymmetric localization at the level of five dimensional supergravity on the $AdS_2\times S^1\times S^2$ background, generalizing the results found in \cite{Dabholkar:2010uh}. Two main results stand out in this computation. The first is that the solutions of the localization equations are an uplift of the four dimensional solutions found in \cite{Dabholkar:2010uh}. The second, and most important, is that the five dimensional supergravity action computed on the localization locus, which includes the contribution from the supersymmetrization of the gauge-gravitational Chern-Simons term, gives precisely the renormalized four dimensional result using the classical holomorphic prepotential (the one that does not contain the contribution from instantons). This result elevates the non-renormalization theorem, pointed out recently in \cite{Butter:2014iwa}, to the quantum level. The finite dimensional integral obtained using localization, is exactly the modified Bessel function that appears in the Rademacher expansion, including the exact spectrum of the polar terms.
	
	\item \emph{Quantum effective action and the topological string}: from our formulas, it becomes clear that the effect of backreaction is relevant for small central charge, because it induces only a small number of Bessel contributions. Whereas for large central charge, the number of Bessel functions grows exponentially, which allows for a saddle point approximation on the sum over Bessels. This limit is equivalent to take the M-theory circle parametrically much smaller than the size of $AdS_2$. We find that the final result for the degeneracy, after integrating out the contribution from the fluxes, matches with the four dimensional quantum effective action computed on the near-horizon geometry of the black hole, in agreement with previous results. From this effective action we can read the topological string contributions as an expansion in powers of $g_{\text{top}}=1/\phi^0 \ll 1$, in agreement with the OSV proposal \cite{Ooguri:2004zv}.
	
\end{itemize}

There are many features in our construction that are similar to the work of Denef and Moore \cite{Denef:2007vg}. Many of those have been discussed in previous sections. The differences though, are essential for deriving an exact formula for the entropy of four dimensional $\N=2$ black holes. In the following we discuss some of them.  First of all, we consider a path integral formulation directly in the near-horizon geometry, without relying on the properties of the dual CFT partition function. Furthermore, our construction avoids the enigmatic multi-center configurations discussed in \cite{Denef:2007vg}. The reason is that localization  only allows for geometries of the type $AdS_2\times S^1\times S^2$, but the decoupling limit of the enigmatic configuration, discussed in \cite{deBoer:2008fk}, contains a black hole localized on the sphere, which is, thus, physically very different.  Without entering in many details about a  derivation of $\N=2$ black hole entropy, which we leave for future work \cite{Gomes17-1}, we can already see a few advantages of our construction in comparison to \cite{Denef:2007vg}. First, our degeneracy formula is always finite, without the need of including any cutoff. This solves many issues related to the original OSV proposal \cite{Ooguri:2004zv}. That is, the topological string partition function is defined formally in the form of the Gopakumar-Vafa infinite products, which in general is not convergent. In our work, finiteness of the degeneracy follows essentially from the microcanonical ensemble and the bound on the number of geometries. Another key aspect of our construction, is that it can be extended to the regime of weak topological string coupling,  in contrast with Denef and Moore's work. This is possible because the localization computation provides with a result that is valid for any value of the charges, which obviously includes the regime $g_{\text{top}}=1/\phi^0\ll 1$, and moreover is finite.

Finally we comment on three ongoing projects. These are an application of the ideas proposed in this work and provide with further support of our claims, by testing our proposal against microscopic and macroscopic computations. These projects extend through four dimensional  black holes in Calabi-Yau compactifications, including a study of the logarithmic corrections following \cite{Belin:2016knb}, and the study of number theoretic properties of the black hole degeneracy related to Kloosterman sums and U-duality invariants.   Succinctly, these projects can be summarized as:
\begin{itemize}
	\item \emph{$\mathcal{N}=2$ black hole entropy }\cite{Gomes17-1}: we will extend the present results to black holes in $\mathcal{N}=2$ compactifications \cite{Maldacena:1997de}. In particular, the goal is to derive an exact formula for the entropy. There are two important requirements. First, the entropy formula must agree with Denef and Moore's formula \cite{Denef:2007vg} in the regime of strong topological string coupling. Second, it should reproduce the logarithmic corrections computed using the quantum entropy formalism in $\mathcal{N}=2$ supergravity \cite{Keeler:2014bra,Sen:2011ba}. In particular, this is the regime of weak topological string coupling.
	
	\item \emph{Generalized Kloosterman Sums from M2-branes }\cite{Gomes17-2} : in this project we will consider the contribution of  smooth $AdS_2\times S^1\times S^2/\mathbb{Z}_c$ orbifolds to the path integral in $\mathcal{N}=4$ compactifications, following previous work \cite{Dabholkar:2014ema}. The goal is to reproduce the structure of the Rademacher expansion, and in particular, to derive expressions for the Kloosterman sums that can be compared with arbitrary level (mock) Jacobi forms \cite{Ferrari:2017msn}. These orbifolds result from an $SL(2,\mathbb{Z})$ Dehn filling of the bulk solid torus and are thus topology changing.  We will follow \cite{Dabholkar:2014ema} and compute the contribution of flat connections to the Chern-Simons path integral.
	
    \item \emph{Quantum entropy, Koosterman Sums and U-duality invariance }\cite{Gomes17-3}: we will consider the contribution of  Eguchi-Hanson space with GH charges $q$ and $-q$ (\ref{Eguchi-Hanson}), which are  $AdS_2\times S^1\times S^2$ geometries with $\mathbb{Z}_q$ orbifold singularities. We argue that these are the geometries corresponding to a configuration of $q\,\D6$ and $q\,\aD6$. We will study the dependence of the exact entropy on arithmetic invariants and compare with microscopic formulas for both the $\mathcal{N}=8$ \cite{Sen:2008sp} and $\mathcal{N}=4$ \cite{Banerjee:2008pu,Dabholkar:2008zy} examples. In a sense, we want to extend the results \cite{Sen:2009vz,Sen:2009gy} to the quantum theory.
    
\end{itemize}

\subsection*{Acknowledgments}

We would like to thank Jan de Boer, Frederik Denef, Alejandra Castro, Bernard de Wit and Atish Dabholkar for discussions on related topics and for comments on the draft. This work is part of the Delta ITP consortium, a program of the Netherlands Organisation for Scientific Research (NWO) that is funded by the Dutch Ministry of Education, Culture and Science (OCW).

\bibliographystyle{JHEP}
\bibliography{measure3}

\providecommand{\href}[2]{#2}\begingroup\raggedright\begin{thebibliography}{10}

\bibitem{Banerjee:2009af}
N.~Banerjee, S.~Banerjee, R.~K. Gupta, I.~Mandal, and A.~Sen, {\it
  {Supersymmetry, Localization and Quantum Entropy Function}},  {\em JHEP} {\bf
  02} (2010) 091, [\href{http://xxx.lanl.gov/abs/0905.2686}{{\tt
  arXiv:0905.2686}}].

\bibitem{Dabholkar:2010uh}
A.~Dabholkar, J.~Gomes, and S.~Murthy, {\it {Quantum black holes, localization
  and the topological string}},  \href{http://xxx.lanl.gov/abs/1012.0265}{{\tt
  arXiv:1012.0265}}.

\bibitem{Sen:2008vm}
A.~Sen, {\it {Quantum Entropy Function from AdS(2)/CFT(1) Correspondence}},
  \href{http://xxx.lanl.gov/abs/0809.3304}{{\tt arXiv:0809.3304}}.

\bibitem{Dabholkar:2011ec}
A.~Dabholkar, J.~Gomes, and S.~Murthy, {\it {Localization \&; Exact
  Holography}},  {\em JHEP} {\bf 1304} (2013) 062,
  [\href{http://xxx.lanl.gov/abs/1111.1161}{{\tt arXiv:1111.1161}}].

\bibitem{Ooguri:2004zv}
H.~Ooguri, A.~Strominger, and C.~Vafa, {\it {Black hole attractors and the
  topological string}},  \href{http://xxx.lanl.gov/abs/hep-th/0405146}{{\tt
  hep-th/0405146}}.

\bibitem{Gopakumar:1998ii}
R.~Gopakumar and C.~Vafa, {\it {M-theory and topological strings. I}},
  \href{http://xxx.lanl.gov/abs/hep-th/9809187}{{\tt hep-th/9809187}}.

\bibitem{Gopakumar:1998jq}
R.~Gopakumar and C.~Vafa, {\it {M-theory and topological strings. II}},
  \href{http://xxx.lanl.gov/abs/hep-th/9812127}{{\tt hep-th/9812127}}.

\bibitem{Dabholkar:2014ema}
A.~Dabholkar, J.~Gomes, and S.~Murthy, {\it {Nonperturbative black hole entropy
  and Kloosterman sums}},  {\em JHEP} {\bf 1503} (2015) 074,
  [\href{http://xxx.lanl.gov/abs/1404.0033}{{\tt arXiv:1404.0033}}].

\bibitem{Murthy:2015yfa}
S.~Murthy and V.~Reys, {\it {Functional determinants, index theorems, and exact
  quantum black hole entropy}},  \href{http://xxx.lanl.gov/abs/1504.0140}{{\tt
  arXiv:1504.0140}}.

\bibitem{Gupta:2015gga}
R.~K. Gupta, Y.~Ito, and I.~Jeon, {\it {Supersymmetric Localization for BPS
  Black Hole Entropy: 1-loop Partition Function from Vector Multiplets}},
  \href{http://xxx.lanl.gov/abs/1504.0170}{{\tt arXiv:1504.0170}}.

\bibitem{Shih:2005he}
D.~Shih and X.~Yin, {\it {Exact black hole degeneracies and the topological
  string}},  {\em JHEP} {\bf 04} (2006) 034,
  [\href{http://xxx.lanl.gov/abs/hep-th/0508174}{{\tt hep-th/0508174}}].

\bibitem{Gomes:2015xcf}
J.~Gomes, {\it {Exact Holography and Black Hole Entropy in N=8 and N=4 String
  Theory}},  \href{http://xxx.lanl.gov/abs/1511.0706}{{\tt arXiv:1511.0706}}.

\bibitem{Murthy:2015zzy}
S.~Murthy and V.~Reys, {\it {Single-centered black hole microstate degeneracies
  from instantons in supergravity}},
  \href{http://xxx.lanl.gov/abs/1512.0155}{{\tt arXiv:1512.0155}}.

\bibitem{Harvey:1996ir}
J.~A. Harvey and G.~W. Moore, {\it {Fivebrane instantons and R**2 couplings in
  N = 4 string theory}},  {\em Phys. Rev.} {\bf D57} (1998) 2323--2328,
  [\href{http://xxx.lanl.gov/abs/hep-th/9610237}{{\tt hep-th/9610237}}].

\bibitem{Beasley:2006us}
C.~Beasley, D.~Gaiotto, M.~Guica, L.~Huang, A.~Strominger, {\em et.~al.}, {\it
  {Why Z(BH) = |Z(top)|**2}},
  \href{http://xxx.lanl.gov/abs/hep-th/0608021}{{\tt hep-th/0608021}}.

\bibitem{Gomes:2013cca}
J.~Gomes, {\it {Quantum entropy and exact 4d/5d connection}},
  \href{http://xxx.lanl.gov/abs/1305.2849}{{\tt arXiv:1305.2849}}.

\bibitem{Rademacher:1964ra}
H.~Rademacher, {\em {Lectures on Elementary Number Theory}}.
\newblock Robert E. Krieger Publishing Co., 1964.

\bibitem{Dabholkar:2012nd}
A.~Dabholkar, S.~Murthy, and D.~Zagier, {\it {Quantum Black Holes, Wall
  Crossing, and Mock Modular Forms}},
  \href{http://xxx.lanl.gov/abs/1208.4074}{{\tt arXiv:1208.4074}}.

\bibitem{Gupta:2012cy}
R.~K. Gupta and S.~Murthy, {\it {All solutions of the localization equations
  for N=2 quantum black hole entropy}},  {\em JHEP} {\bf 1302} (2013) 141,
  [\href{http://xxx.lanl.gov/abs/1208.6221}{{\tt arXiv:1208.6221}}].

\bibitem{Maldacena:1998bw}
J.~M. Maldacena and A.~Strominger, {\it {AdS(3) black holes and a stringy
  exclusion principle}},  {\em JHEP} {\bf 12} (1998) 005,
  [\href{http://xxx.lanl.gov/abs/hep-th/9804085}{{\tt hep-th/9804085}}].

\bibitem{Gaiotto:2006ns}
D.~Gaiotto, A.~Strominger, and X.~Yin, {\it {From AdS(3)/CFT(2) to black holes
  / topological strings}},  {\em JHEP} {\bf 09} (2007) 050,
  [\href{http://xxx.lanl.gov/abs/hep-th/0602046}{{\tt hep-th/0602046}}].

\bibitem{Dedushenko:2014nya}
M.~Dedushenko and E.~Witten, {\it {Some Details On The Gopakumar-Vafa and
  Ooguri-Vafa Formulas}},  {\em Adv. Theor. Math. Phys.} {\bf 20} (2016)
  1--133, [\href{http://xxx.lanl.gov/abs/1411.7108}{{\tt arXiv:1411.7108}}].

\bibitem{Castro:2011ui}
A.~Castro, T.~Hartman, and A.~Maloney, {\it {The Gravitational Exclusion
  Principle and Null States in Anti-de Sitter Space}},  {\em Class. Quant.
  Grav.} {\bf 28} (2011) 195012, [\href{http://xxx.lanl.gov/abs/1107.5098}{{\tt
  arXiv:1107.5098}}].

\bibitem{Banerjee:2010qc}
S.~Banerjee, R.~K. Gupta, and A.~Sen, {\it {Logarithmic Corrections to Extremal
  Black Hole Entropy from Quantum Entropy Function}},
  \href{http://xxx.lanl.gov/abs/1005.3044}{{\tt arXiv:1005.3044}}.

\bibitem{Banerjee:2011jp}
S.~Banerjee, R.~K. Gupta, I.~Mandal, and A.~Sen, {\it {Logarithmic Corrections
  to N=4 and N=8 Black Hole Entropy: A One Loop Test of Quantum Gravity}},
  {\em JHEP} {\bf 11} (2011) 143,
  [\href{http://xxx.lanl.gov/abs/1106.0080}{{\tt arXiv:1106.0080}}].

\bibitem{Gaiotto:2006wm}
D.~Gaiotto, A.~Strominger, and X.~Yin, {\it {The M5-brane elliptic genus:
  Modularity and BPS states}},  {\em JHEP} {\bf 08} (2007) 070,
  [\href{http://xxx.lanl.gov/abs/hep-th/0607010}{{\tt hep-th/0607010}}].

\bibitem{Denef:2007vg}
F.~Denef and G.~W. Moore, {\it {Split states, entropy enigmas, holes and
  halos}},  \href{http://xxx.lanl.gov/abs/hep-th/0702146}{{\tt
  hep-th/0702146}}.

\bibitem{Denef:2000nb}
F.~Denef, {\it {Supergravity flows and D-brane stability}},  {\em JHEP} {\bf
  08} (2000) 050, [\href{http://xxx.lanl.gov/abs/hep-th/0005049}{{\tt
  hep-th/0005049}}].

\bibitem{Denef:2007yt}
F.~Denef, D.~Gaiotto, A.~Strominger, D.~Van~den Bleeken, and X.~Yin, {\it
  {Black Hole Deconstruction}},  {\em JHEP} {\bf 03} (2012) 071,
  [\href{http://xxx.lanl.gov/abs/hep-th/0703252}{{\tt hep-th/0703252}}].

\bibitem{deBoer:2008fk}
J.~{de Boer}, F.~Denef, S.~El-Showk, I.~Messamah, and D.~{Van den Bleeken},
  {\it {Black hole bound states in AdS(3) x S**2}},  {\em JHEP} {\bf 11} (2008)
  050, [\href{http://xxx.lanl.gov/abs/0802.2257}{{\tt arXiv:0802.2257}}].

\bibitem{deBoer:2006vg}
J.~{de Boer}, M.~C.~N. Cheng, R.~Dijkgraaf, J.~Manschot, and E.~Verlinde, {\it
  {A farey tail for attractor black holes}},  {\em JHEP} {\bf 11} (2006) 024,
  [\href{http://xxx.lanl.gov/abs/hep-th/0608059}{{\tt hep-th/0608059}}].

\bibitem{Dijkgraaf:2000fq}
R.~Dijkgraaf, J.~M. Maldacena, G.~W. Moore, and E.~P. Verlinde, {\it {A Black
  hole Farey tail}},  \href{http://xxx.lanl.gov/abs/hep-th/0005003}{{\tt
  hep-th/0005003}}.

\bibitem{Murthy:2009dq}
S.~Murthy and B.~Pioline, {\it {A Farey tale for N=4 dyons}},  {\em JHEP} {\bf
  09} (2009) 022, [\href{http://xxx.lanl.gov/abs/0904.4253}{{\tt
  arXiv:0904.4253}}].

\bibitem{Iqbal:2003ds}
A.~Iqbal, N.~Nekrasov, A.~Okounkov, and C.~Vafa, {\it {Quantum foam and
  topological strings}},  {\em JHEP} {\bf 04} (2008) 011,
  [\href{http://xxx.lanl.gov/abs/hep-th/0312022}{{\tt hep-th/0312022}}].

\bibitem{deBoer:1998ip}
J.~{de Boer}, {\it {Six-dimensional supergravity on S**3 x AdS(3) and 2-D
  conformal field theory}},  {\em Nucl. Phys.} {\bf B548} (1999) 139--166,
  [\href{http://xxx.lanl.gov/abs/hep-th/9806104}{{\tt hep-th/9806104}}].

\bibitem{deBoer:1998us}
J.~{de Boer}, {\it {Large N elliptic genus and AdS / CFT correspondence}},
  {\em JHEP} {\bf 05} (1999) 017,
  [\href{http://xxx.lanl.gov/abs/hep-th/9812240}{{\tt hep-th/9812240}}].

\bibitem{10.2307/2371313}
H.~Rademacher, {\it The fourier coefficients of the modular invariant
  {$J(\tau)$}},  {\em American Journal of Mathematics} {\bf 60} (1938), no.~2
  501--512.

\bibitem{Kloos}
H.~Kloosterman, {\it {On the representation of numbers in the form
  {$ax^2+by^2+cw^2+dz^2$}}},  {\em Acta Math.} {\bf 49} (1926) 407--464.

\bibitem{Sen:2009vz}
A.~Sen, {\it {Arithmetic of Quantum Entropy Function}},  {\em JHEP} {\bf 0908}
  (2009) 068, [\href{http://xxx.lanl.gov/abs/0903.1477}{{\tt
  arXiv:0903.1477}}].

\bibitem{Sen:2007qy}
A.~Sen, {\it {Black Hole Entropy Function, Attractors and Precision Counting of
  Microstates}},  \href{http://xxx.lanl.gov/abs/0708.1270}{{\tt 0708.1270}}.

\bibitem{Simons:2004nm}
A.~Simons, A.~Strominger, D.~M. Thompson, and X.~Yin, {\it {Supersymmetric
  branes in AdS(2) x S**2 x CY(3)}},  {\em Phys.Rev.} {\bf D71} (2005) 066008,
  [\href{http://xxx.lanl.gov/abs/hep-th/0406121}{{\tt hep-th/0406121}}].

\bibitem{Maldacena:1997de}
J.~M. Maldacena, A.~Strominger, and E.~Witten, {\it {Black hole entropy in
  M-theory}},  {\em JHEP} {\bf 12} (1997) 002,
  [\href{http://xxx.lanl.gov/abs/hep-th/9711053}{{\tt hep-th/9711053}}].

\bibitem{Sen:1997js}
A.~Sen, {\it {Dynamics of multiple Kaluza-Klein monopoles in M and string
  theory}},  {\em Adv. Theor. Math. Phys.} {\bf 1} (1998) 115--126,
  [\href{http://xxx.lanl.gov/abs/hep-th/9707042}{{\tt hep-th/9707042}}].

\bibitem{Kraus:2006nb}
P.~Kraus and F.~Larsen, {\it {Partition functions and elliptic genera from
  supergravity}},  \href{http://xxx.lanl.gov/abs/hep-th/0607138}{{\tt
  hep-th/0607138}}.

\bibitem{Bena:2010gg}
I.~Bena, N.~Bobev, S.~Giusto, C.~Ruef, and N.~P. Warner, {\it {An
  Infinite-Dimensional Family of Black-Hole Microstate Geometries}},  {\em
  JHEP} {\bf 03} (2011) 022, [\href{http://xxx.lanl.gov/abs/1006.3497}{{\tt
  arXiv:1006.3497}}]. [Erratum: JHEP04,059(2011)].

\bibitem{Castro:2008ne}
A.~Castro, J.~L. Davis, P.~Kraus, and F.~Larsen, {\it {String Theory Effects on
  Five-Dimensional Black Hole Physics}},  {\em Int. J. Mod. Phys.} {\bf A23}
  (2008) 613--691, [\href{http://xxx.lanl.gov/abs/0801.1863}{{\tt
  arXiv:0801.1863}}].

\bibitem{Castro:2007hc}
A.~Castro, J.~L. Davis, P.~Kraus, and F.~Larsen, {\it {5D Black Holes and
  Strings with Higher Derivatives}},  {\em JHEP} {\bf 06} (2007) 007,
  [\href{http://xxx.lanl.gov/abs/hep-th/0703087}{{\tt hep-th/0703087}}].

\bibitem{Hansen:2006wu}
J.~Hansen and P.~Kraus, {\it {Generating charge from diffeomorphisms}},  {\em
  JHEP} {\bf 0612} (2006) 009,
  [\href{http://xxx.lanl.gov/abs/hep-th/0606230}{{\tt hep-th/0606230}}].

\bibitem{Kraus:2005vz}
P.~Kraus and F.~Larsen, {\it {Microscopic black hole entropy in theories with
  higher derivatives}},  {\em JHEP} {\bf 09} (2005) 034,
  [\href{http://xxx.lanl.gov/abs/hep-th/0506176}{{\tt hep-th/0506176}}].

\bibitem{deBoer:2009un}
J.~de~Boer, S.~El-Showk, I.~Messamah, and D.~Van~den Bleeken, {\it {A Bound on
  the entropy of supergravity?}},  {\em JHEP} {\bf 02} (2010) 062,
  [\href{http://xxx.lanl.gov/abs/0906.0011}{{\tt arXiv:0906.0011}}].

\bibitem{Freed:1998tg}
D.~Freed, J.~A. Harvey, R.~Minasian, and G.~W. Moore.

\bibitem{Mohaupt:2000mj}
T.~Mohaupt, {\it {Black hole entropy, special geometry and strings}},  {\em
  Fortsch. Phys.} {\bf 49} (2001) 3--161,
  [\href{http://xxx.lanl.gov/abs/hep-th/0007195}{{\tt hep-th/0007195}}].

\bibitem{Gupta:2008ki}
R.~K. Gupta and A.~Sen, {\it {Ads(3)/CFT(2) to Ads(2)/CFT(1)}},  {\em JHEP}
  {\bf 04} (2009) 034, [\href{http://xxx.lanl.gov/abs/0806.0053}{{\tt
  arXiv:0806.0053}}].

\bibitem{Dabholkar:2010rm}
A.~Dabholkar, J.~Gomes, S.~Murthy, and A.~Sen, {\it {Supersymmetric Index from
  Black Hole Entropy}},  \href{http://xxx.lanl.gov/abs/1009.3226}{{\tt
  arXiv:1009.3226}}.

\bibitem{Sen:2005iz}
A.~Sen, {\it {Entropy function for heterotic black holes}},
  \href{http://xxx.lanl.gov/abs/hep-th/0508042}{{\tt hep-th/0508042}}.

\bibitem{Murthy:2015yfa-1}
S.~Murthy and V.~Reys, {\it {Functional determinants, index theorems, and exact
  quantum black hole entropy}},  \href{http://xxx.lanl.gov/abs/1504.0140}{{\tt
  arXiv:1504.0140}}.

\bibitem{Banerjee:2011ts}
N.~Banerjee, B.~{de Wit}, and S.~Katmadas, {\it {The Off-Shell 4D/5D
  Connection}},  {\em JHEP} {\bf 1203} (2012) 061,
  [\href{http://xxx.lanl.gov/abs/1112.5371}{{\tt arXiv:1112.5371}}].

\bibitem{deWit:2009de}
B.~{de Wit} and S.~Katmadas, {\it {Near-Horizon Analysis of D=5 BPS Black Holes
  and Rings}},  {\em JHEP} {\bf 1002} (2010) 056,
  [\href{http://xxx.lanl.gov/abs/0910.4907}{{\tt arXiv:0910.4907}}].

\bibitem{Elitzur:1989nr}
S.~Elitzur, G.~W. Moore, A.~Schwimmer, and N.~Seiberg, {\it {Remarks on the
  Canonical Quantization of the Chern-Simons-Witten Theory}},  {\em Nucl.Phys.}
  {\bf B326} (1989) 108.

\bibitem{Butter:2013lta}
D.~Butter, B.~de~Wit, S.~M. Kuzenko, and I.~Lodato, {\it {New higher-derivative
  invariants in N=2 supergravity and the Gauss-Bonnet term}},  {\em JHEP} {\bf
  12} (2013) 062, [\href{http://xxx.lanl.gov/abs/1307.6546}{{\tt
  arXiv:1307.6546}}].

\bibitem{Butter:2014iwa}
D.~Butter, B.~de~Wit, and I.~Lodato, {\it {Non-renormalization theorems and N=2
  supersymmetric backgrounds}},  {\em JHEP} {\bf 03} (2014) 131,
  [\href{http://xxx.lanl.gov/abs/1401.6591}{{\tt arXiv:1401.6591}}].

\bibitem{deWit:2010za}
B.~{de Wit}, S.~Katmadas, and M.~{van Zalk}, {\it {New supersymmetric
  higher-derivative couplings: Full N=2 superspace does not count!}},  {\em
  JHEP} {\bf 1101} (2011) 007, [\href{http://xxx.lanl.gov/abs/1010.2150}{{\tt
  arXiv:1010.2150}}].

\bibitem{Murthy:2013xpa}
S.~Murthy and V.~Reys, {\it {Quantum black hole entropy and the holomorphic
  prepotential of N=2 supergravity}},
  \href{http://xxx.lanl.gov/abs/1306.3796}{{\tt arXiv:1306.3796}}.

\bibitem{Cortes:2003zd}
V.~Cortes, C.~Mayer, T.~Mohaupt, and F.~Saueressig, {\it {Special geometry of
  Euclidean supersymmetry. I: Vector multiplets}},  {\em JHEP} {\bf 03} (2004)
  028, [\href{http://xxx.lanl.gov/abs/hep-th/0312001}{{\tt hep-th/0312001}}].

\bibitem{Gomes17-2}
J.~Gomes, ``{Generalized Kloosterman Sums from $M2$-branes}.'' work in
  progress, 2017.

\bibitem{Ferrari:2017msn}
F.~Ferrari and V.~Reys, {\it {Mixed Rademacher and BPS Black Holes}},
  \href{http://xxx.lanl.gov/abs/1702.0275}{{\tt arXiv:1702.0275}}.

\bibitem{LopesCardoso:2006bg}
G.~{Lopes Cardoso}, B.~{de Wit}, J.~Kappeli, and T.~Mohaupt, {\it {Black hole
  partition functions and duality}},  {\em JHEP} {\bf 03} (2006) 074,
  [\href{http://xxx.lanl.gov/abs/hep-th/0601108}{{\tt hep-th/0601108}}].

\bibitem{Cardoso:2008fr}
G.~L. Cardoso, B.~{de Wit}, and S.~Mahapatra, {\it {Subleading and
  non-holomorphic corrections to N=2 BPS black hole entropy}},  {\em JHEP} {\bf
  02} (2009) 006, [\href{http://xxx.lanl.gov/abs/0808.2627}{{\tt
  arXiv:0808.2627}}].

\bibitem{Cardoso:2010gc}
G.~L. Cardoso, B.~{de Wit}, and S.~Mahapatra, {\it {BPS black holes, the Hesse
  potential, and the topological string}},  {\em JHEP} {\bf 06} (2010) 052,
  [\href{http://xxx.lanl.gov/abs/1003.1970}{{\tt arXiv:1003.1970}}].

\bibitem{Cardoso:2012nh}
G.~Lopes~Cardoso, B.~de~Wit, and S.~Mahapatra, {\it {Non-holomorphic
  deformations of special geometry and their applications}},  {\em Springer
  Proc. Phys.} {\bf 144} (2013) 1--58,
  [\href{http://xxx.lanl.gov/abs/1206.0577}{{\tt arXiv:1206.0577}}].

\bibitem{Cardoso:2014kwa}
G.~L. Cardoso, B.~de~Wit, and S.~Mahapatra, {\it {Deformations of special
  geometry: in search of the topological string}},  {\em JHEP} {\bf 09} (2014)
  096, [\href{http://xxx.lanl.gov/abs/1406.5478}{{\tt arXiv:1406.5478}}].

\bibitem{Bershadsky:1994sr}
M.~Bershadsky and V.~Sadov, {\it {Theory of Kahler gravity}},  {\em Int. J.
  Mod. Phys.} {\bf A11} (1996) 4689--4730,
  [\href{http://xxx.lanl.gov/abs/hep-th/9410011}{{\tt hep-th/9410011}}].

\bibitem{Gomes17-1}
J.~Gomes, ``{Exact $\mathcal{N}=2$ Black Hole Entropy}.'' work in progress,
  2017.

\bibitem{Belin:2016knb}
A.~Belin, A.~Castro, J.~Gomes, and C.~A. Keller, {\it {Siegel Modular Forms and
  Black Hole Entropy}},  {\em JHEP} {\bf 04} (2017) 057,
  [\href{http://xxx.lanl.gov/abs/1611.0458}{{\tt arXiv:1611.0458}}].

\bibitem{Keeler:2014bra}
C.~Keeler, F.~Larsen, and P.~Lisbao, {\it {Logarithmic Corrections to $N \geq
  2$ Black Hole Entropy}},  {\em Phys. Rev.} {\bf D90} (2014), no.~4 043011,
  [\href{http://xxx.lanl.gov/abs/1404.1379}{{\tt arXiv:1404.1379}}].

\bibitem{Sen:2011ba}
A.~Sen, {\it {Logarithmic Corrections to N=2 Black Hole Entropy: An Infrared
  Window into the Microstates}},  {\em Gen. Rel. Grav.} {\bf 44} (2012), no.~5
  1207--1266, [\href{http://xxx.lanl.gov/abs/1108.3842}{{\tt
  arXiv:1108.3842}}].

\bibitem{Gomes17-3}
J.~Gomes, ``{Quantum Entropy, Koosterman Sums and U-duality Invariance}.'' work
  in progress, 2017.

\bibitem{Sen:2008sp}
A.~Sen, {\it {U-duality Invariant Dyon Spectrum in type II on T**6}},  {\em
  JHEP} {\bf 0808} (2008) 037, [\href{http://xxx.lanl.gov/abs/0804.0651}{{\tt
  arXiv:0804.0651}}].

\bibitem{Banerjee:2008pu}
S.~Banerjee, A.~Sen, and Y.~K. Srivastava, {\it {Partition Functions of Torsion
  $ > 1$ Dyons in Heterotic String Theory on {$T^6$}}},
  \href{http://xxx.lanl.gov/abs/0802.1556}{{\tt 0802.1556}}.

\bibitem{Dabholkar:2008zy}
A.~Dabholkar, J.~Gomes, and S.~Murthy, {\it {Counting all dyons in N =4 string
  theory}},  \href{http://xxx.lanl.gov/abs/0803.2692}{{\tt arXiv:0803.2692}}.

\bibitem{Sen:2009gy}
A.~Sen, {\it {Arithmetic of N=8 Black Holes}},  {\em JHEP} {\bf 02} (2010) 090,
  [\href{http://xxx.lanl.gov/abs/0908.0039}{{\tt arXiv:0908.0039}}].

\end{thebibliography}\endgroup
\end{document}